\documentclass[%
preprintnumbers,
nofootinbib,
reprint,
amsmath,amssymb,
aps, prl,
]{revtex4-2}
\usepackage{aas_macros}
\usepackage{graphicx}
\usepackage{dcolumn}
\usepackage{bm}
\usepackage{xcolor}
\usepackage[normalem]{ulem} 

\usepackage{hyperref}

\begin{document}
	
	\preprint{MIT-CTP/5952}
	
	\title{Boosting the cosmic 21-cm signal with exotic Lyman-\texorpdfstring{$\alpha$}{alpha} from dark matter}

	\author{Dominic Agius}
	\email{dagius@ific.uv.es}
	\affiliation{Instituto de Física Corpuscular (IFIC), CSIC‐-Universitat de València, Valencia 46071, Spain}
	
	\author{Tracy R. Slatyer}%
	\email{tslatyer@mit.edu}
	\affiliation{%
		MIT Center for Theoretical Physics -- a Leinweber Institute, Massachusetts Institute of Technology, Cambridge, Massachusetts 02139, U.S.A
	}%

	\date{\today}
	
	\begin{abstract}
		The 21-cm signal from the epoch of cosmic dawn ($z \sim 10\textup{--}30$) offers a powerful probe of new physics. One standard mechanism for constraining decaying dark matter from 21-cm observations relies on heating of the intergalactic medium by the decay products, an effect whose observability is entangled with the uncertain Lyman-$\alpha$ fluxes and X-ray heating from the first stars. In this \textit{Letter}, we explore a novel mechanism, where the Lyman-$\alpha$ photons produced from dark matter decay initiate early Wouthuysen--Field coupling of the spin temperature to the gas temperature, thereby boosting the 21-cm signal. This mechanism provides constraints on dark matter that are less dependent on uncertainties associated with star formation than constraints on exotic heating. We study this effect for decaying dark matter with masses $m_{\chi}\sim20.4 \textup{--}27.2$ eV, where diphoton decay efficiently produces Lyman-series photons.  We present forecasts for the Hydrogen Epoch of Reionization Array and the Square Kilometre Array, showing their potential to probe an unconstrained parameter space for light decaying DM, including axion-like particles. 
	\end{abstract}

	\maketitle

	The redshifted 21-cm line, arising from spin-flip transitions between the singlet and triplet states in neutral hydrogen, is a promising new observable capable of probing the properties of the Universe at epochs spanning the late dark ages through to the cosmic dawn (e.g.~\cite{Barkana:2000fd,Pritchard:2011xb}). The 21-cm signal is governed by the difference between the spin temperature $T_S$ and the radiation temperature, with the latter typically given by the cosmic microwave background (CMB) temperature $T_{\rm CMB}$. A sufficient flux of Lyman-$\alpha$ photons will couple the spin temperature to the gas kinetic temperature $T_k$ during the cosmic dawn via the Wouthuysen--Field (WF) effect \cite{Wouthuysen:1952,Field:1958}, leading to a difference between $T_S$ and $T_{\rm CMB}$ that induces a nonzero 21-cm signal.

    As a consequence, the 21-cm signal provides a probe of the properties of the Universe across a wide range of redshifts. Next-generation radio telescopes such as the {\tt HERA}~\cite{HERA:2016} and {\tt SKA}~\cite{SKA:2012} are poised to measure the power spectrum associated with this signal at redshifts $z \sim 6\textup{--}25$, and thus provide a window into the thermal and ionization history of the Universe at the Epoch of Reionization (EoR) and the cosmic dawn. In this way, the 21-cm signal will provide an excellent probe of properties of the first stars~\cite{Greig:2015qca, Greig:2016wjs, Park:2018ljd, Pochinda:2023uom,Katz:2024ayw, Katz:2025sie,Dhandha:2025dtn} and further provide an insight into any additional exotic sources presence at these times~\cite{Furlanetto:2006wp, Valdes:2007cu, Evoli:2014pva, Lopez-Honorez:2016sur, Poulin:2016anj, Clark:2018ghm, DAmico:2018sxd, Hektor:2018qqw, Liu:2018uzy, Mitridate:2018iag, Mena:2019nhm, Yang:2020egn, Cang:2021owu, Halder:2021jiv, Halder:2021rbq, Mittal:2021egv, Natwariya:2021xki, Saha:2021pqf, Yang:2021idt, Mukhopadhyay:2022jqc, Yang:2022puh, Facchinetti:2023slb, Qin:2023kkk, Sun:2023acy, Novosyadlyj:2024bie, Zhao:2024jad, Bae:2025uqa, Natwariya:2025jlw, Sun:2025ksr, Zhao:2025ddy,Agius:2025xbj}.

	Many previous studies have considered Dark Matter (DM) constraints using the 21-cm signal as a probe. In particular, if DM decays or annihilates to visible particles, these particles can induce changes to the gas temperature compared to the fiducial astrophysical scenario and hence enable strong forecast constraints on DM~\cite{Poulin:2016anj,Mena:2019nhm,Facchinetti:2023slb,Sun:2023acy,Clark:2018ghm,DAmico:2018sxd,Fraser:2018acy,Barkana:2018qrx,Fialkov:2018xre,Berlin:2018sjs}. However, such constraints have generally assumed the presence of a sufficient flux of Lyman-$\alpha$ photons originating from the first stars and hence an  efficient WF coupling. Furthermore, there is a strong degeneracy with the heating from DM and that from the first stars. Recent lower limits on early astrophysical heating from the first sources~\cite{HERA:2021noe,HERA:2022wmy} thus imply correspondingly weaker forecast constraints on DM scenarios, since stronger astrophysical heating reduces the relative imprint of DM-induced heating effects~\cite{Agius:2025xbj}. 
	
	In this \textit{Letter}, we consider the possibility of DM providing sufficient Lyman-$\alpha$ flux for efficient WF coupling, independent of the ultraviolet properties of the first stars. With this possibility in mind, we study decaying DM in the mass range $20\textup{--}50 $ eV, where masses below $\sim$ twice the ionization threshold can efficiently deposit energy into the Lyman-$\alpha$ band.  Doing so, we derive forecast constraints in the $ 20.4\textup{--}27.2$ eV mass range which are orders of magnitude stronger than previous constraints, assuming typical fiducial astrophysics scenarios, and we extend our constraints up to $50$ eV for completeness. We further show that our results remain relatively robust to variations in the X-ray and ultraviolet properties of the first stars, particularly when compared to heating based constraints, due to the alternative physical mechanism underlying our constraints. 
	
	\textbf{Efficient Lyman-$\alpha$ coupling from Dark Matter.}---
    The key observable is the 21-cm signal's differential brightness temperature $\delta T_{21}$, governed by the spin temperature through
	\begin{equation}\label{eqn:T21}
		\delta T_{21} \propto x_H  \left(1 - \frac{T_{\rm CMB}}{T_S}\right) \left(\frac{1 + z}{10}\right)^{1/2}  \, \text{mK},
	\end{equation}
	where $x_H$ is the fraction of neutral hydrogen. The sky average of this signal, $ \overline{\delta T}_{21}$, is known as the global 21-cm signal. Spatial and temporal fluctuations can be characterized by the power spectrum of this signal, which is defined via the two-point correlation of the signal in Fourier space, defined in  \textit{Appendix~A}. We use the power spectrum as our observable in this work. 
    
 The evolution of the spin temperature can be understood through 
	\begin{equation}\label{eqn:Ts}
		T_S^{-1}  = \frac{T_{\text{CMB}}^{-1} + x_{\alpha} T_{c}^{-1} + x_c T_k^{-1}}{1+x_{\alpha} + x_c}, 
	\end{equation}
	where $x_c$ and $x_{\alpha}$ are the respective collisional and WF coupling coefficients. In the epoch of interest ($z \sim 5\textup{--}30$), the color temperature $T_c \simeq T_k$, within 5\%~\cite{Hirata:2005mz}.  The WF coupling $x_\alpha $ is proportional to the Lyman-$\alpha$ flux, and during the cosmic dawn the collisional coupling $x_c$ is inefficient due to the Universe's expansion. Hence, a sufficiently large Lyman-$\alpha$ flux makes $x_\alpha$ the dominant term, driving $T_S \to T_k$. In the absence of extreme X-ray heating at early times, $ T_k < T_{\rm CMB}$ during this epoch, and so a strong absorption signal is produced during these times.  In the standard cosmological picture, this Lyman-$\alpha$ flux is produced by the first stars. 

However, in scenarios where DM converts even a tiny fraction of its mass energy into Standard Model particles (e.g.~through annihilation or decay, or accretion or evaporation of black holes), those particles can also source a non-negligible Lyman-$\alpha$ flux. Public tools such as \texttt{DarkHistory}~\cite{Liu:2019bbm,Liu:2023nct,Liu:2023fgu}, \texttt{DM21cm}~\cite{Sun:2023acy}, and \texttt{exo21cmFAST}~\cite{Facchinetti:2023slb} have been written to compute how such particles deposit their energy, and how the resulting ionization, heating, and Lyman-$\alpha$ photons affect the 21-cm signal. Most of this previous work has not disentangled each of these contributions, but for DM with mass above the keV scale (or black hole processes producing primarily particles with keV+ energies), the effect on the global signal is generally similar to an exotic heating source.

In this \textit{Letter}, we aim to disentangle the effect of Lyman-$\alpha$ injected by decaying DM from the usual heating signal. We consider DM particles $\chi$ with mass $m_{\chi}$, decaying via $\chi \to \gamma \gamma$. Initially, we consider the case where the DM mass is in the range $20.4\textup{--}27.2$ eV, and the resulting photons are exactly in the Lyman-$\alpha$ band needed for efficient WF coupling, whereas the direct heating effect is negligible.

\begin{figure}[b]
		\includegraphics[width=\columnwidth]{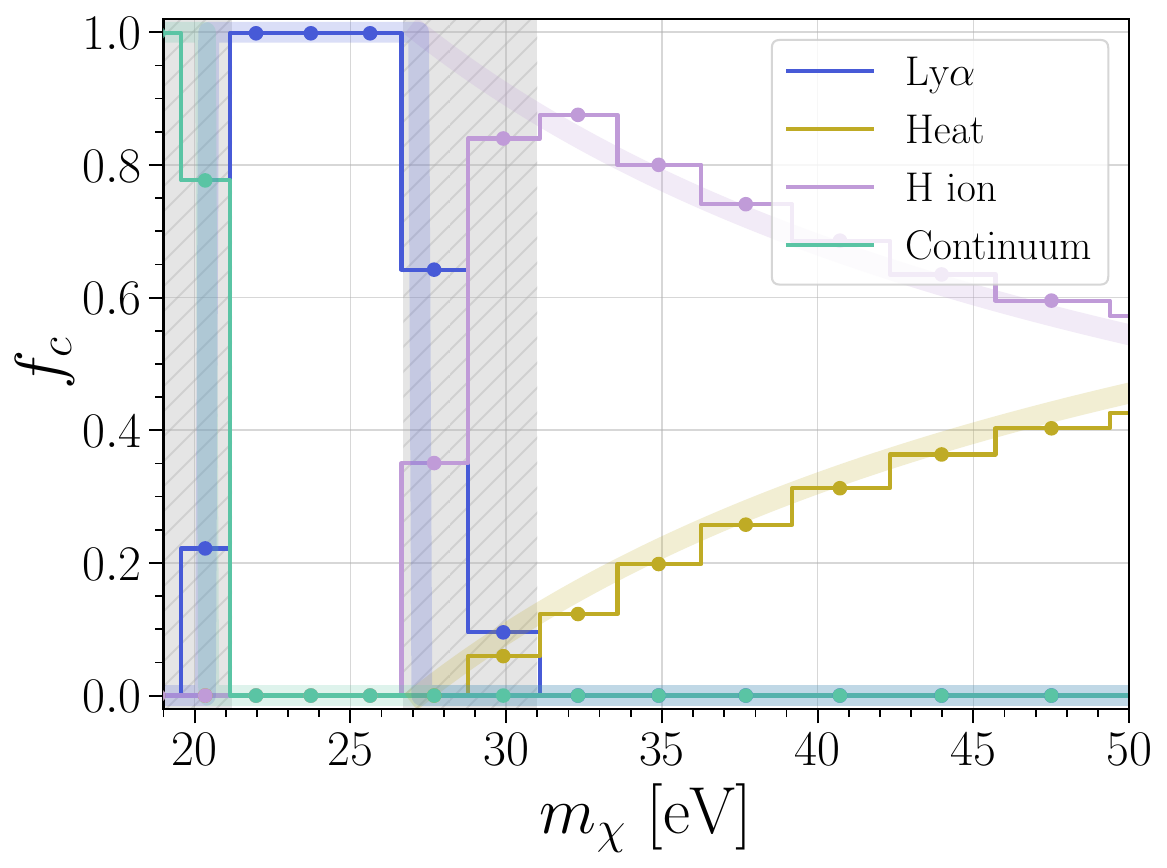}
		\caption{\label{fig:energy_deposition} \textbf{The deposition of energy by channel.} We show the fraction $f$ of energy deposited into the relevant channels $c$ for different masses $m_\chi$, at $z = 35$, computed using {\tt DarkHistory}~\cite{Liu:2019bbm}, with lifetime $\tau = 10^{28}$~s. The bin boundaries from {\tt DarkHistory} are shown as a step function, and bin centers marked with a dot. The shaded gray regions indicate where the {\tt DarkHistory} computation fails, due to binning issues, and we show with the colored bands the corrected behavior of each channel. }
	\end{figure}

We also consider injection of photons at modestly higher energies, with $m_{\chi} > 27.2$ eV, where an increasing fraction of the energy is diverted to ionization and (eventually) heating. The transition between these regimes is computed using {\tt DarkHistory} and shown in Fig.~\ref{fig:energy_deposition}.
 Any deviations from precise energy levels of the hydrogen atom as depicted in Fig.~\ref{fig:energy_deposition} should be interpreted as binning artifacts, where the explicit bin edges and centers are clearly labeled, and we have provided an outline of the corrected behavior of each channel (see supplementary material), that we have implemented in computing our bounds.

While we select this energy range to demonstrate the effect of physically distinct contributions to the 21-cm signal, it has recently provoked some independent interest in the literature, ignited by recent observations from James Webb Space Telescope~\cite{Gardner:2006ky,Gardner:2023} of high-redshift active galactic nuclei~\cite{bler_2023,Larson_2023,harikane2023jwstnirspec,Carnall_2023,Onoue_2023,kocevski2023hidden,fan2022quasarsintergalacticmediumcosmic,maiolino2023small}.  In particular, photon injection in this regime could dissociate molecular hydrogen in the early universe and hence facilitate the formation of  direct collapse supermassive black holes (DC-SMBHs), potentially explaining these observations \cite{Omukai:2000ic, Bromm:2002hb, Biermann:2006bu, Stasielak:2006br, Spolyar:2007qv, Dijkstra:2008jk, Cyr:2022urs, Friedlander:2022ovf, Lu:2023xoi, Lu:2024zwa, Aggarwal:2025pit}.

\textbf{21-cm Signature.}---
We simulate the 21-cm signal for DM decay in our mass range of interest using {\tt DM21cm}~\cite{Sun:2023acy}, a highly flexible code built upon the popular {\tt 21cmFAST}~\cite{mesinger_21cmFAST_2011,Murray2020}. The code evolves simulation boxes throughout the dark ages to the EoR, and self-consistently accounts for the impact of exotic injections. 

We caution that the treatment of Lyman-$\alpha$ in {\tt DM21cm} is somewhat crude in that by default it does not track the full spectrum within the Lyman-series band (10.2--13.6 eV), only the injected energy into this band, treating the total energy in such photons as a direct contribution to the Lyman-$\alpha$ flux $J_\alpha$. A more accurate approach would be to treat the Lyman-band photons from exotic energy injection in the same way that \texttt{21cmFAST} treats Lyman-band photons emitted by stars; as discussed in Ref.~\cite{mesinger_21cmFAST_2011}, these Lyman-band photons are treated as free-streaming until they redshift into one of the Lyman-series resonances, whereupon they excite a $np$ state of hydrogen. As pointed out in Refs.~\cite{Hirata:2005mz,Pritchard:2005an}, the subsequent decay of the excited $np$ state produces a Lyman-$\alpha$ photon with probability $f_\text{recycle}(n)$: for example, photons between $12.1\textup{--}12.8$ eV redshift until they excite the $3p$ state, which decays to the $2s$ state, which in turn decays through a two-photon transition to the ground state, without producing Lyman-$\alpha$ photons, so that $f_\text{recycle}(3)=0$. For higher excited states $f_\text{recycle}(n)$ is typically in the range $0.3\textup{--}0.4$; we adopt the values of $f_\text{recycle}(n)$ encoded in \texttt{21cmFAST}, reproduced from Ref.~\cite{Hirata:2005mz}.

To estimate the impact of these effects on our analysis, we note that the constraints from {\tt DM21cm} should be precisely correct for photons injected directly into the Lyman-$\alpha$ transition (i.e.~20.4 eV decaying DM). For 10.2--13.6 eV photons, we rescale these constraints to reflect (1) that photons injected at an energy $E_\gamma$ not overlapping with a line will lose energy by redshifting before being absorbed, weakening the constraints by a factor $E_n/E_\gamma$ where $E_n$ is the highest line energy below $E_\gamma$, and (2) the number of Lyman-$\alpha$ photons will be further reduced by a factor of $f_\text{recycle}(n)$. This approach does neglect the slight time-delay and spatial smearing of the Lyman-$\alpha$ flux from injected Lyman-band photons, due to the time/distance they travel before being absorbed. A full treatment of Lyman-$\alpha$ in the case of exotic energy injections, analogous to the contribution from stars, would require non-trivial modifications to \texttt{DM21cm}, which we leave for future work.

Furthermore, the behavior encoded in \texttt{DM21cm} for the photon energy bins around 10.2 eV and 13.6 eV also has numerical artifacts arising from the mismatch between the bin edges and the physical thresholds. At the level of the treatment in \texttt{DarkHistory v1.0}, which was used to build \texttt{DM21cm}, photons below 10.2 eV should be solely deposited into the ``continuum'' channel, while photons above 13.6 eV will each give rise to a photoionization and a secondary low-energy electron, which deposits its energy into heating (for electrons too low-energy to collisionally excite the gas). For energies in the ranges shown in the gray regions of Fig.~\ref{fig:energy_deposition}, \texttt{DM21cm} interpolates between these two regimes; for our limits, we instead calculate the constraints above and below these ambiguous regions and transition between the two at exactly the physical energy thresholds. In principle, accounting for the width of the relevant lines and the nonzero velocities of the primordial gas clouds could soften the step functions at the physical thresholds, and an update to \texttt{DM21cm} could incorporate a more in-depth treatment of these threshold regions; we leave this for future work.

\begin{figure}[ht]
	\includegraphics[width=\columnwidth]{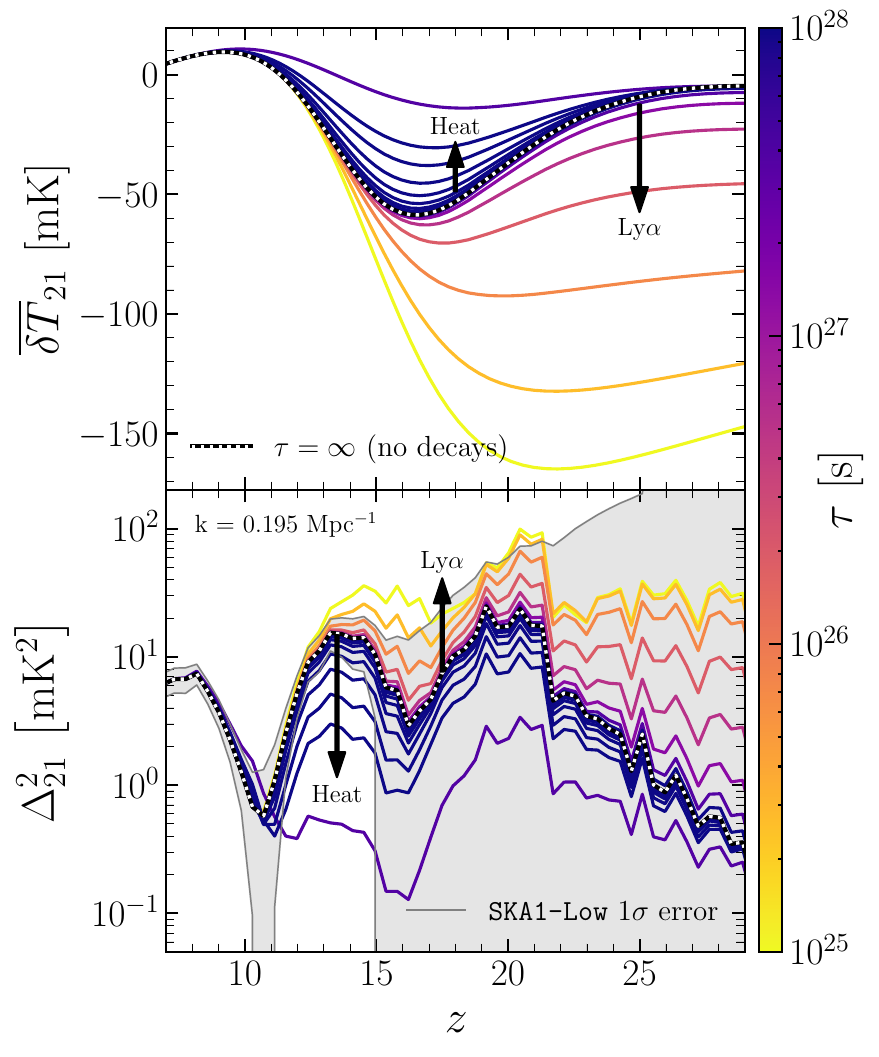}
	\caption{\label{fig:decaying_dm_global_PS} \textbf{The effect of decaying dark matter on the global signal and power spectrum.} We show the effect of decaying dark matter with various lifetimes $\tau$ and with masses $m_{\chi} = 20.4~{\rm eV}$ and  $m_{\chi} = 50~{\rm eV}$, on global signal (top) and power spectrum (bottom) at $k \simeq 0.2$ Mpc$^{-1}$. The $m_{\chi} = 20.4~{\rm eV}$ scenario, labeled by ``Ly$\alpha$'', enhances the signal in both figures, whereas the $m_{\chi} = 50~{\rm eV}$ scenario, labeled by ``Heat'', suppresses the signal in both figures. We include the 1$\sigma$ experimental thermal error computed for {\tt SKA1-Low} in the bottom panel. We use the same color bar for both scenarios, noting that they are separated by the dotted line denoting the decay-free scenario.}
\end{figure}

In Fig.~\ref{fig:decaying_dm_global_PS}, we show the impact of decaying DM with masses $m_{\chi} = 20.4 $~eV and $m_{\chi} = 50 $~eV, on both the global signal (top) and the power spectrum (bottom). 	

The signature from $m_\chi = 50$ eV can be understood as a typical DM heating signal, where the global absorption trough gets shallower with increasing energy injection, corresponding to a decreasing lifetime, $\tau$. This damping of the global signal corresponds to the suppression of the power spectrum shown in the bottom panel.  The new scenario we consider in this \textit{Letter}, where energy is maximally deposited into Lyman-$\alpha$, shows a very different phenomenology. In contrast, the global signal absorption trough deepens with decreasing lifetime, until the WF coupling saturates (for $\tau \sim 10^{25}$~s). In this regime, $T_S$ is driven to efficiently track the gas temperature $T_k$, causing the global signal to flatten at high redshifts. The 21-cm power spectrum (bottom panel) reflects this behavior, showing that the largest phenomenological impact occurs at these early times, well before stellar Lyman-$\alpha$ coupling begins. The intersection of the power spectrum  with the projected {\tt SKA1-Low} sensitivity indicates the approximate lifetime that will be probed by future observations. As shown in Fig.~\ref{fig:decaying_dm_global_PS}, heating signatures can be probed at much longer lifetimes for this fiducial astrophysical model choice. However, as recently pointed out~\cite{Agius:2025xbj}, the constraining power of such signatures is strongly tied to the fiducial model choice. On the contrary, by initiating an early Lyman-$\alpha$ flux independent of the first stars, strong bounds can be placed on this scenario  which are much less dependent on the choice of fiducial astrophysical model, compared to heating constraints. We expand on this claim in \textit{Appendix~B}. Furthermore, because the effect of the enhanced early Lyman-$\alpha$ flux is to increase the power spectrum (and the magnitude of the global signal), bounds can even be placed in the absence of any detection of the 21-cm power spectrum. 

In particular, {\tt HERA}  and {\tt NenuFAR} have already placed the strongest upper limits on the 21-cm power spectrum~\cite{HERA:2022wmy,Munshi:2025hgk}. Although neither experiment has yet reached the sensitivity  required to probe the scenarios considered here, future data releases from {\tt HERA}, {\tt NenuFAR} and the {\tt SKA}, are expected to be sensitive to this effect. Furthermore, any future lunar-based observatory targeting the  dark ages 21-cm signal, would provide improved  sensitivity compared to ground-based experiments~\cite{Jester:2009dw,Farside:2019,LuseeNight:2023,Liu:2019awk,Silk:2025znp}. 
We also point out that this physical effect shifts toward the anomalous {\tt EDGES} absorption feature~\cite{Bowman:2018yin}. However, as pointed out, e.g., in Refs.~\cite{Barkana:2018qrx,Barkana:2018lgd,Kovetz:2018zan}, an exotic contribution is needed to cool the gas further, making $T_S \to T_k$ alone not enough to be consistent with this signal. Finally, we note that  {\tt SARAS 3} reported that the best-fitting {\tt EDGES} profile is disfavored with 95.3\% confidence~\cite{Singh:2021mxo}.  

\textbf{Projected Constraints.}---
To quantify the constraining power of experiments {\tt HERA} and the {\tt SKA} to this signature, we perform a Fisher forecast for each observatory. Each experiment has sensitivity for the 21-cm power spectrum at comoving wavenumbers $k$ between $0.1\textup{--}1.0 \, {\rm Mpc}^{-1}$. Both {\tt HERA} and {\tt SKA1-Low} have lower frequency limits of $50$ MHz, and upper limits of 250 MHz and 350 MHz respectively. We use {\tt 21cmSense}~\cite{Pober:2012zz,Pober:2013jna,Murray:2024the} to compute the experimental sensitivities and assume the  fiducial astrophysical parameters of Ref.~\cite{Sun:2023acy}, corresponding to the ``best guess'' scenario of Ref.~\cite{Munoz:2021psm}. We add in quadrature both the cosmic variance and an additional 20\% modeling error, to the experimental sensitivity, to obtain our total sensitivity, with details on this procedure provided in \textit{Appendix~A}. 

We perform simulations using {\tt DM21cm} with a range of masses $m_\chi$ from $19\textup{--}50$~eV, using a box of comoving volume $256^3 \,{\rm Mpc}^3$, on a $128^3$ grid, following Ref.~\cite{Sun:2023acy}. 
	\begin{figure}[b]
		\includegraphics[width=\columnwidth]{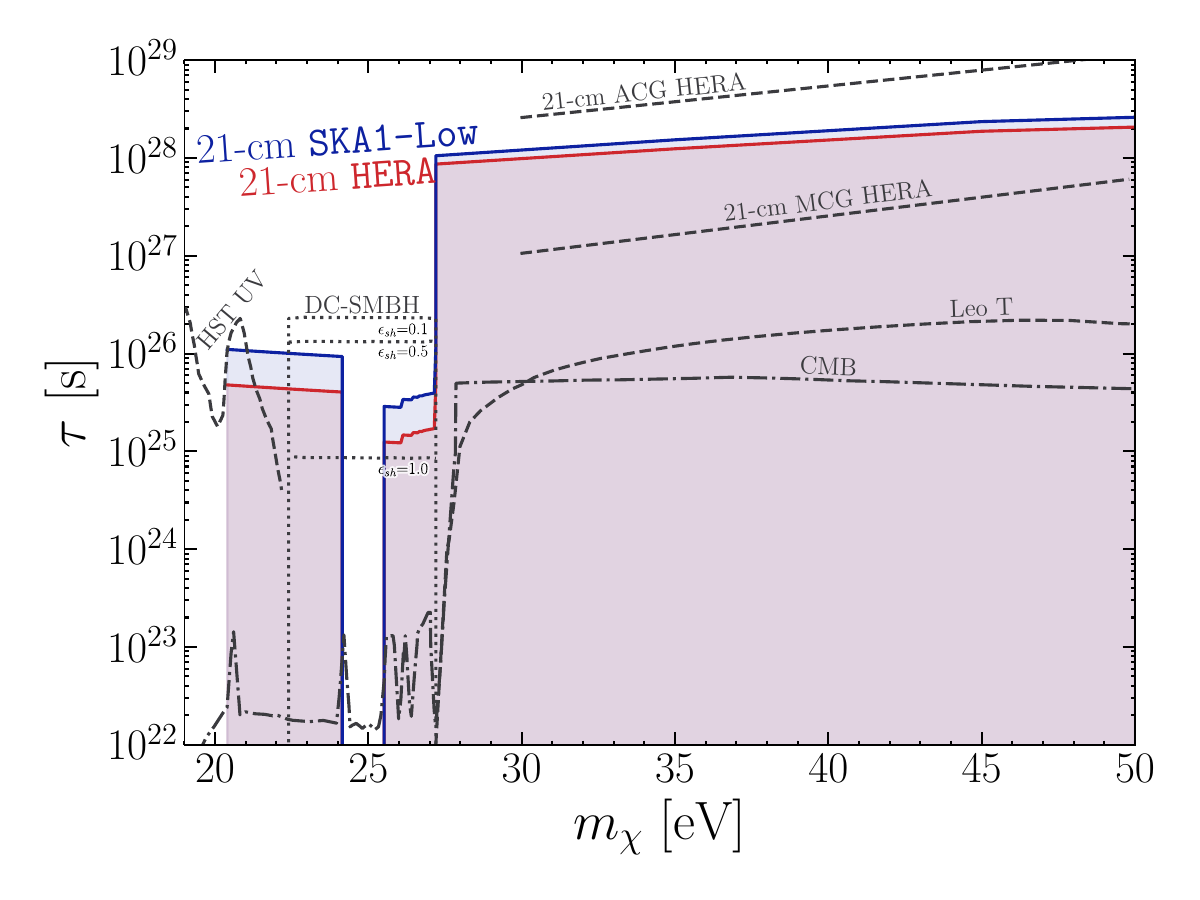}
		\caption{\textbf{Projected 95th percentile limits on monochromatic decays, with {\tt HERA} and {\tt SKA1-Low}.} We also show bounds computed using the 21-cm signal for molecularly cooled galaxies (21-cm MCG HERA) and atomically cooled galaxies (21-cm ACG HERA)~\cite{Facchinetti:2023slb}, CMB limits computed below and above the ionization thresholds (CMB)~\cite{xu_cmb_2024}, HST limits from galaxy UV searches (HST UV)~\cite{Todarello:2024qci}, limits from anomalous heating of the Leo T dwarf galaxy (Leo T)~\cite{Wadekar:2021qae},  and the region of interest identified for Lyman-Werner direct collapse black holes (DC-SMBH), with shielding parameters $\varepsilon_{sh} = 0.1, \, 0.5, \, 1.0$~\cite{Lu:2024zwa}.
} 
        \label{fig:bound_plot}
	\end{figure}
    
We show the resulting 95$^{\rm th}$ percentile projected constraints for both telescopes in Fig.~\ref{fig:bound_plot}, where we quote limits at $1.65 \sqrt{F_{ii}^{-1}}$, assuming the Cramér-Rao bound~\cite{Cramer:1946,Rao1992}. The shape of the bound can be understood through the energy deposition functions shown in Fig.~\ref{fig:energy_deposition}. Below $m_\chi = 20.4$ eV, the bound vanishes, since decays transfer energy into two photons of energy $\lesssim 10.2 $ eV, which are in the continuum band, where they do not interact with the gas (at least at the level of the approximations employed in {\tt DM21cm}). 

Between $ 20.4 \lesssim m_\chi  \lesssim 27.2$ eV, the resulting decay product photons are in the Lyman-$\alpha$ band, where they can contribute directly to an enhanced Wouthuysen--Field coupling, independent of stellar sources. The energy-dependence of the constraint between these masses is a consequence of the redshifting of photons to the nearest atomic line, and from the atomic effects captured in $f_{\rm recycle}$~\cite{Hirata:2005mz}; in particular, the gap in the constraints corresponds to the energy range where injected photons are initially absorbed into the $3p$ state (a more detailed treatment, capturing e.g.~the relevant linewidths and the effects of the gas velocity, might mitigate the loss of sensitivity in this range). We show the full covariance between astrophysical parameters and the decay rate, for $m_{\chi} = 20.4$~eV in Fig.~\ref{fig:Triangle_plot}. Then, for $m_\chi \gtrsim 27.2$ eV, the Lyman-$\alpha$ contribution becomes negligible, and heating and ionization dominate. This causes the steep feature in the bound at these masses; the additional ionization acts to heat the gas, by enhancing the Compton coupling between the gas temperature and the (higher) CMB temperature, and at even higher masses direct heating plays an increasing role. The scenario above the ionization threshold has analogous physics effects to previous studies in the literature for higher-mass DM.

In Fig.~\ref{fig:bound_plot}, we include previously computed 21-cm constraints by \cite{Facchinetti:2023slb}, showing both their results for a single population of atomic-cooling galaxies (ACGs), and both ACGs and molecular-cooling galaxies (MCGs), more similarly to our scenario (see also~\cite{Sun:2023acy}). We remark that their bounds were computed down to $m_{\chi} = 30$ eV, however, they did not discuss the atomic physics associated with this regime, and only computed two masses $m_\chi = 30$ eV and $ m_\chi = 50$ eV.~\footnote{We remark that~\cite{Lopez-Honorez:2024ant} also computed a point at $m_\chi = 25$ eV, but did not discuss that this bound arises from an efficient WF coupling.} It is the interpolation between these points that we show in the Fig.~\ref{fig:bound_plot}. We also include previously derived CMB constraints~\cite{xu_cmb_2024} (see also~\cite{Capozzi:2023xie} for CMB constraints above the ionization threshold), as well as Hubble Space Telescope (HST) limits from UV searches in dwarf spheroidal galaxies and galaxy clusters~\cite{Todarello:2024qci}, and limits from anomalous heating of the Leo T dwarf galaxy~\cite{Wadekar:2021qae}. For converting between axion limits on $g_{a\gamma \gamma}$, and the lifetime $\tau_a$, we use the conversion $
\tau_a =
\frac{64\pi}{g_{a\gamma \gamma}^2\, m_a^3 \hbar}
$. As shown in Fig.~\ref{fig:bound_plot}, the limits forecast in this work would provide competitive bounds, with the 95th percentile limit for $m_a = 20.4\,\textrm{eV}$ being $g_{a\gamma \gamma} < 4.1{\times}10^{-13}\, \textrm{GeV}^{-1}$, for \texttt{SKA1-Low}. 

Furthermore, we include the  parameter space identified by~\cite{Lu:2024zwa} as promising for DC-SMBH formation; we find that much of this region should  lead to an observable 21-cm signal from enhanced WF coupling, independent of any effect on compact objects.

\textbf{Conclusion and Future Perspectives.}---In this \textit{Letter} we have introduced a novel mechanism to constrain decaying dark matter using the 21-cm signal. By directly producing Lyman-$\alpha$ photons, DM with masses  $m_\chi \simeq 20.4\textup{--}27.2 $~eV can initiate early Wouthuysen--Field coupling, producing a distinct signature in both the global 21-cm signal and power spectrum. This approach provides powerful constraints that are more robust against the large uncertainties in early star formation astrophysics which affect traditional heating-based limits. Moreover, unlike typical heating constraints, in this case meaningful limits can be obtained even in the absence of 21-cm power spectrum detection, using upper limits alone. Our forecasts for {\tt HERA} and {\tt SKA} demonstrate the potential to probe currently unconstrained regions of parameter space for light decaying DM, improving existing bounds by orders of magnitude.

The mechanism explored here opens several new avenues for research, since the studied signature is not unique to diphoton decay and also arises in other models that deposit sufficient energy into the Lyman-band. In particular,  scenarios in which an $\mathcal{O}(1)$ fraction is deposited into the Lyman-band could produce similar effects, where the significance of the effect can be estimated based on the energy injected at lifetimes corresponding to our constraints at 20.4 eV. Such constraints may be particularly useful in the absence of any power spectrum detection, since they allow limits to be placed from upper bounds alone across a wide range of scenarios. For particle injections at higher energies, the fraction of energy deposited into the Lyman-band can be $\mathcal{O}(0.1)$ across a range of masses \cite{Liu:2019bbm}. The heating-based limits forecast by previous studies for heavier decaying DM (e.g.~\cite{Facchinetti:2023slb, Sun:2023acy}), corresponding to lifetimes around $10^{28}$ s, are thus not likely to be significantly affected by a more careful treatment of Lyman-$\alpha$ photons, but if the underlying astrophysics is such that these limits are markedly weakened \cite{Agius:2025xbj}, the exotic Lyman-band flux could become important.

Finally, since the effect is most pronounced at high redshifts, future lunar or space-based observatories targeting the 21-cm signal from the dark ages would offer even greater sensitivity.

\textit{Acknowledgments.}---We are grateful to Yitian Sun and Joshua Foster for their clarifications regarding the {\tt DM21cm} code. We are also grateful to Laura Lopez Honorez for clarification on her recent works related to decaying dark matter and the 21-cm signal. DA acknowledges support from the Generalitat Valenciana grant CIGRIS/2021/054, and grant PID2023-151418NB-I00, which is funded by MCIU/AEI/10.13039/501100011033/ FEDER, UE. DA also received support from the European Union’s Horizon 2020 research and innovation programme under the Marie Skłodowska-Curie grants H2020-MSCA-ITN-2019/860881-HIDDeN and the HORIZON-MSCA-2021-SE-01/101086085-ASYMMETRY Staff Exchange grant agreement. TRS' work is supported by the Simons Foundation (Grant Number 929255, T.R.S) and by the U.S. Department of Energy, Office of Science, Office of High Energy Physics of U.S. Department of Energy under grant contract Number DE-SC0012567. During the course of this work, T.R.S.~was supported in part by a Guggenheim Fellowship; the Edward, Frances, and Shirley B.~Daniels Fellowship of the Harvard Radcliffe Institute; and the Bershadsky Distinguished Fellowship of the Harvard Physics Department.  This work made use of resources provided by subMIT at MIT Physics~\cite{bendavid2025submitphysicsanalysisfacility}, as well as {\tt NumPy}~\cite{Harris:2020xlr}, {\tt SciPy}~\cite{Virtanen:2019joe}, {\tt matplotlib}~\cite{Hunter:2007}, and Webplotdigitizer~\cite{WebPlotDigitizer}. We made use of {\tt Zeus21}~\cite{Munoz:2023kkg} in the early stages of this work.  

	\bibliography{bibliography}

\begin{thebibliography}{118}%
\makeatletter
\providecommand \@ifxundefined [1]{%
 \@ifx{#1\undefined}
}%
\providecommand \@ifnum [1]{%
 \ifnum #1\expandafter \@firstoftwo
 \else \expandafter \@secondoftwo
 \fi
}%
\providecommand \@ifx [1]{%
 \ifx #1\expandafter \@firstoftwo
 \else \expandafter \@secondoftwo
 \fi
}%
\providecommand \natexlab [1]{#1}%
\providecommand \enquote  [1]{``#1''}%
\providecommand \bibnamefont  [1]{#1}%
\providecommand \bibfnamefont [1]{#1}%
\providecommand \citenamefont [1]{#1}%
\providecommand \href@noop [0]{\@secondoftwo}%
\providecommand \href [0]{\begingroup \@sanitize@url \@href}%
\providecommand \@href[1]{\@@startlink{#1}\@@href}%
\providecommand \@@href[1]{\endgroup#1\@@endlink}%
\providecommand \@sanitize@url [0]{\catcode `\\12\catcode `\$12\catcode `\&12\catcode `\#12\catcode `\^12\catcode `\_12\catcode `\%12\relax}%
\providecommand \@@startlink[1]{}%
\providecommand \@@endlink[0]{}%
\providecommand \url  [0]{\begingroup\@sanitize@url \@url }%
\providecommand \@url [1]{\endgroup\@href {#1}{\urlprefix }}%
\providecommand \urlprefix  [0]{URL }%
\providecommand \Eprint [0]{\href }%
\providecommand \doibase [0]{https://doi.org/}%
\providecommand \selectlanguage [0]{\@gobble}%
\providecommand \bibinfo  [0]{\@secondoftwo}%
\providecommand \bibfield  [0]{\@secondoftwo}%
\providecommand \translation [1]{[#1]}%
\providecommand \BibitemOpen [0]{}%
\providecommand \bibitemStop [0]{}%
\providecommand \bibitemNoStop [0]{.\EOS\space}%
\providecommand \EOS [0]{\spacefactor3000\relax}%
\providecommand \BibitemShut  [1]{\csname bibitem#1\endcsname}%
\let\auto@bib@innerbib\@empty
\bibitem [{\citenamefont {Barkana}\ and\ \citenamefont {Loeb}(2001)}]{Barkana:2000fd}%
  \BibitemOpen
  \bibfield  {author} {\bibinfo {author} {\bibfnamefont {R.}~\bibnamefont {Barkana}}\ and\ \bibinfo {author} {\bibfnamefont {A.}~\bibnamefont {Loeb}},\ }\bibfield  {title} {\bibinfo {title} {{In the beginning: The First sources of light and the reionization of the Universe}},\ }\href {https://doi.org/10.1016/S0370-1573(01)00019-9} {\bibfield  {journal} {\bibinfo  {journal} {Phys. Rept.}\ }\textbf {\bibinfo {volume} {349}},\ \bibinfo {pages} {125} (\bibinfo {year} {2001})},\ \Eprint {https://arxiv.org/abs/astro-ph/0010468} {arXiv:astro-ph/0010468} \BibitemShut {NoStop}%
\bibitem [{\citenamefont {Pritchard}\ and\ \citenamefont {Loeb}(2012)}]{Pritchard:2011xb}%
  \BibitemOpen
  \bibfield  {author} {\bibinfo {author} {\bibfnamefont {J.~R.}\ \bibnamefont {Pritchard}}\ and\ \bibinfo {author} {\bibfnamefont {A.}~\bibnamefont {Loeb}},\ }\bibfield  {title} {\bibinfo {title} {{21-cm cosmology}},\ }\href {https://doi.org/10.1088/0034-4885/75/8/086901} {\bibfield  {journal} {\bibinfo  {journal} {Rept. Prog. Phys.}\ }\textbf {\bibinfo {volume} {75}},\ \bibinfo {pages} {086901} (\bibinfo {year} {2012})},\ \Eprint {https://arxiv.org/abs/1109.6012} {arXiv:1109.6012 [astro-ph.CO]} \BibitemShut {NoStop}%
\bibitem [{\citenamefont {{Wouthuysen}}(1952)}]{Wouthuysen:1952}%
  \BibitemOpen
  \bibfield  {author} {\bibinfo {author} {\bibfnamefont {S.~A.}\ \bibnamefont {{Wouthuysen}}},\ }\bibfield  {title} {\bibinfo {title} {{On the excitation mechanism of the 21-cm (radio-frequency) interstellar hydrogen emission line.}},\ }\href {https://doi.org/10.1086/106661} {\bibfield  {journal} {\bibinfo  {journal} {Astrophys. J.}\ }\textbf {\bibinfo {volume} {57}},\ \bibinfo {pages} {31} (\bibinfo {year} {1952})}\BibitemShut {NoStop}%
\bibitem [{\citenamefont {{Field}}(1958)}]{Field:1958}%
  \BibitemOpen
  \bibfield  {author} {\bibinfo {author} {\bibfnamefont {G.~B.}\ \bibnamefont {{Field}}},\ }\bibfield  {title} {\bibinfo {title} {{Excitation of the hydrogen 21-cm line}},\ }\href {https://doi.org/10.1109/JRPROC.1958.286741} {\bibfield  {journal} {\bibinfo  {journal} {Proceedings of the IRE}\ }\textbf {\bibinfo {volume} {46}},\ \bibinfo {pages} {240} (\bibinfo {year} {1958})}\BibitemShut {NoStop}%
\bibitem [{\citenamefont {DeBoer}\ \emph {et~al.}(2017)\citenamefont {DeBoer} \emph {et~al.}}]{HERA:2016}%
  \BibitemOpen
  \bibfield  {author} {\bibinfo {author} {\bibfnamefont {D.~R.}\ \bibnamefont {DeBoer}} \emph {et~al.},\ }\bibfield  {title} {\bibinfo {title} {{Hydrogen Epoch of Reionization Array (HERA)}},\ }\href {https://doi.org/10.1088/1538-3873/129/974/045001} {\bibfield  {journal} {\bibinfo  {journal} {Publ. Astron. Soc. Pac.}\ }\textbf {\bibinfo {volume} {129}},\ \bibinfo {pages} {045001} (\bibinfo {year} {2017})},\ \Eprint {https://arxiv.org/abs/1606.07473} {arXiv:1606.07473 [astro-ph.IM]} \BibitemShut {NoStop}%
\bibitem [{\citenamefont {Mellema}\ \emph {et~al.}(2013)\citenamefont {Mellema}, \citenamefont {Koopmans}, \citenamefont {Abdalla}, \citenamefont {Bernardi}, \citenamefont {Ciardi}, \citenamefont {Daiboo}, \citenamefont {de~Bruyn}, \citenamefont {Datta}, \citenamefont {Falcke}, \citenamefont {Ferrara}, \citenamefont {Iliev}, \citenamefont {Iocco}, \citenamefont {Jelić}, \citenamefont {Jensen}, \citenamefont {Joseph}, \citenamefont {Labroupoulos}, \citenamefont {Meiksin}, \citenamefont {Mesinger}, \citenamefont {Offringa}, \citenamefont {Pandey}, \citenamefont {Pritchard}, \citenamefont {Santos}, \citenamefont {Schwarz}, \citenamefont {Semelin}, \citenamefont {Vedantham}, \citenamefont {Yatawatta},\ and\ \citenamefont {Zaroubi}}]{SKA:2012}%
  \BibitemOpen
  \bibfield  {author} {\bibinfo {author} {\bibfnamefont {G.}~\bibnamefont {Mellema}}, \bibinfo {author} {\bibfnamefont {L.~V.~E.}\ \bibnamefont {Koopmans}}, \bibinfo {author} {\bibfnamefont {F.~A.}\ \bibnamefont {Abdalla}}, \bibinfo {author} {\bibfnamefont {G.}~\bibnamefont {Bernardi}}, \bibinfo {author} {\bibfnamefont {B.}~\bibnamefont {Ciardi}}, \bibinfo {author} {\bibfnamefont {S.}~\bibnamefont {Daiboo}}, \bibinfo {author} {\bibfnamefont {A.~G.}\ \bibnamefont {de~Bruyn}}, \bibinfo {author} {\bibfnamefont {K.~K.}\ \bibnamefont {Datta}}, \bibinfo {author} {\bibfnamefont {H.}~\bibnamefont {Falcke}}, \bibinfo {author} {\bibfnamefont {A.}~\bibnamefont {Ferrara}}, \bibinfo {author} {\bibfnamefont {I.~T.}\ \bibnamefont {Iliev}}, \bibinfo {author} {\bibfnamefont {F.}~\bibnamefont {Iocco}}, \bibinfo {author} {\bibfnamefont {V.}~\bibnamefont {Jelić}}, \bibinfo {author} {\bibfnamefont {H.}~\bibnamefont {Jensen}}, \bibinfo {author} {\bibfnamefont {R.}~\bibnamefont {Joseph}}, \bibinfo {author} {\bibfnamefont
  {P.}~\bibnamefont {Labroupoulos}}, \bibinfo {author} {\bibfnamefont {A.}~\bibnamefont {Meiksin}}, \bibinfo {author} {\bibfnamefont {A.}~\bibnamefont {Mesinger}}, \bibinfo {author} {\bibfnamefont {A.~R.}\ \bibnamefont {Offringa}}, \bibinfo {author} {\bibfnamefont {V.~N.}\ \bibnamefont {Pandey}}, \bibinfo {author} {\bibfnamefont {J.~R.}\ \bibnamefont {Pritchard}}, \bibinfo {author} {\bibfnamefont {M.~G.}\ \bibnamefont {Santos}}, \bibinfo {author} {\bibfnamefont {D.~J.}\ \bibnamefont {Schwarz}}, \bibinfo {author} {\bibfnamefont {B.}~\bibnamefont {Semelin}}, \bibinfo {author} {\bibfnamefont {H.}~\bibnamefont {Vedantham}}, \bibinfo {author} {\bibfnamefont {S.}~\bibnamefont {Yatawatta}},\ and\ \bibinfo {author} {\bibfnamefont {S.}~\bibnamefont {Zaroubi}},\ }\bibfield  {title} {\bibinfo {title} {Reionization and the cosmic dawn with the square kilometre array},\ }\href {https://doi.org/10.1007/s10686-013-9334-5} {\bibfield  {journal} {\bibinfo  {journal} {Experimental Astronomy}\ }\textbf {\bibinfo {volume}
  {36}},\ \bibinfo {pages} {235–318} (\bibinfo {year} {2013})},\ \Eprint {https://arxiv.org/abs/1210.0197} {arXiv:1210.0197 [astro-ph.CO]} \BibitemShut {NoStop}%
\bibitem [{\citenamefont {Greig}\ and\ \citenamefont {Mesinger}(2015)}]{Greig:2015qca}%
  \BibitemOpen
  \bibfield  {author} {\bibinfo {author} {\bibfnamefont {B.}~\bibnamefont {Greig}}\ and\ \bibinfo {author} {\bibfnamefont {A.}~\bibnamefont {Mesinger}},\ }\bibfield  {title} {\bibinfo {title} {{21CMMC: an MCMC analysis tool enabling astrophysical parameter studies of the cosmic 21 cm signal}},\ }\href {https://doi.org/10.1093/mnras/stv571} {\bibfield  {journal} {\bibinfo  {journal} {Mon. Not. Roy. Astron. Soc.}\ }\textbf {\bibinfo {volume} {449}},\ \bibinfo {pages} {4246} (\bibinfo {year} {2015})},\ \Eprint {https://arxiv.org/abs/1501.06576} {arXiv:1501.06576 [astro-ph.CO]} \BibitemShut {NoStop}%
\bibitem [{\citenamefont {Greig}\ and\ \citenamefont {Mesinger}(2017)}]{Greig:2016wjs}%
  \BibitemOpen
  \bibfield  {author} {\bibinfo {author} {\bibfnamefont {B.}~\bibnamefont {Greig}}\ and\ \bibinfo {author} {\bibfnamefont {A.}~\bibnamefont {Mesinger}},\ }\bibfield  {title} {\bibinfo {title} {{The global history of reionization}},\ }\href {https://doi.org/10.1093/mnras/stw3026} {\bibfield  {journal} {\bibinfo  {journal} {Mon. Not. Roy. Astron. Soc.}\ }\textbf {\bibinfo {volume} {465}},\ \bibinfo {pages} {4838} (\bibinfo {year} {2017})},\ \Eprint {https://arxiv.org/abs/1605.05374} {arXiv:1605.05374 [astro-ph.CO]} \BibitemShut {NoStop}%
\bibitem [{\citenamefont {Park}\ \emph {et~al.}(2019)\citenamefont {Park}, \citenamefont {Mesinger}, \citenamefont {Greig},\ and\ \citenamefont {Gillet}}]{Park:2018ljd}%
  \BibitemOpen
  \bibfield  {author} {\bibinfo {author} {\bibfnamefont {J.}~\bibnamefont {Park}}, \bibinfo {author} {\bibfnamefont {A.}~\bibnamefont {Mesinger}}, \bibinfo {author} {\bibfnamefont {B.}~\bibnamefont {Greig}},\ and\ \bibinfo {author} {\bibfnamefont {N.}~\bibnamefont {Gillet}},\ }\bibfield  {title} {\bibinfo {title} {{Inferring the astrophysics of reionization and cosmic dawn from galaxy luminosity functions and the 21-cm signal}},\ }\href {https://doi.org/10.1093/mnras/stz032} {\bibfield  {journal} {\bibinfo  {journal} {Mon. Not. Roy. Astron. Soc.}\ }\textbf {\bibinfo {volume} {484}},\ \bibinfo {pages} {933} (\bibinfo {year} {2019})},\ \Eprint {https://arxiv.org/abs/1809.08995} {arXiv:1809.08995 [astro-ph.GA]} \BibitemShut {NoStop}%
\bibitem [{\citenamefont {Pochinda}\ \emph {et~al.}(2024)\citenamefont {Pochinda} \emph {et~al.}}]{Pochinda:2023uom}%
  \BibitemOpen
  \bibfield  {author} {\bibinfo {author} {\bibfnamefont {S.}~\bibnamefont {Pochinda}} \emph {et~al.},\ }\bibfield  {title} {\bibinfo {title} {{Constraining the properties of Population III galaxies with multiwavelength observations}},\ }\href {https://doi.org/10.1093/mnras/stae1185} {\bibfield  {journal} {\bibinfo  {journal} {Mon. Not. Roy. Astron. Soc.}\ }\textbf {\bibinfo {volume} {531}},\ \bibinfo {pages} {1113} (\bibinfo {year} {2024})},\ \Eprint {https://arxiv.org/abs/2312.08095} {arXiv:2312.08095 [astro-ph.CO]} \BibitemShut {NoStop}%
\bibitem [{\citenamefont {Katz}\ \emph {et~al.}(2024)\citenamefont {Katz}, \citenamefont {Outmezguine}, \citenamefont {Redigolo},\ and\ \citenamefont {Volansky}}]{Katz:2024ayw}%
  \BibitemOpen
  \bibfield  {author} {\bibinfo {author} {\bibfnamefont {O.~Z.}\ \bibnamefont {Katz}}, \bibinfo {author} {\bibfnamefont {N.}~\bibnamefont {Outmezguine}}, \bibinfo {author} {\bibfnamefont {D.}~\bibnamefont {Redigolo}},\ and\ \bibinfo {author} {\bibfnamefont {T.}~\bibnamefont {Volansky}},\ }\bibfield  {title} {\bibinfo {title} {{Probing new physics at cosmic dawn with 21-cm cosmology}},\ }\href {https://doi.org/10.1016/j.nuclphysb.2024.116502} {\bibfield  {journal} {\bibinfo  {journal} {Nucl. Phys. B}\ }\textbf {\bibinfo {volume} {1003}},\ \bibinfo {pages} {116502} (\bibinfo {year} {2024})},\ \Eprint {https://arxiv.org/abs/2401.10978} {arXiv:2401.10978 [hep-ph]} \BibitemShut {NoStop}%
\bibitem [{\citenamefont {Katz}\ \emph {et~al.}(2025)\citenamefont {Katz}, \citenamefont {Redigolo},\ and\ \citenamefont {Volansky}}]{Katz:2025sie}%
  \BibitemOpen
  \bibfield  {author} {\bibinfo {author} {\bibfnamefont {O.~Z.}\ \bibnamefont {Katz}}, \bibinfo {author} {\bibfnamefont {D.}~\bibnamefont {Redigolo}},\ and\ \bibinfo {author} {\bibfnamefont {T.}~\bibnamefont {Volansky}},\ }\href {https://arxiv.org/abs/2502.03525} {\bibinfo {title} {Closing in on pop-iii stars: Constraints and predictions across the spectrum}} (\bibinfo {year} {2025}),\ \Eprint {https://arxiv.org/abs/2502.03525} {arXiv:2502.03525 [astro-ph.CO]} \BibitemShut {NoStop}%
\bibitem [{\citenamefont {Dhandha}\ \emph {et~al.}(2025)\citenamefont {Dhandha}, \citenamefont {Fialkov}, \citenamefont {Gessey-Jones}, \citenamefont {Bevins}, \citenamefont {Tacchella}, \citenamefont {Pochinda}, \citenamefont {de~Lera~Acedo}, \citenamefont {Singh},\ and\ \citenamefont {Barkana}}]{Dhandha:2025dtn}%
  \BibitemOpen
  \bibfield  {author} {\bibinfo {author} {\bibfnamefont {J.}~\bibnamefont {Dhandha}}, \bibinfo {author} {\bibfnamefont {A.}~\bibnamefont {Fialkov}}, \bibinfo {author} {\bibfnamefont {T.}~\bibnamefont {Gessey-Jones}}, \bibinfo {author} {\bibfnamefont {H.~T.~J.}\ \bibnamefont {Bevins}}, \bibinfo {author} {\bibfnamefont {S.}~\bibnamefont {Tacchella}}, \bibinfo {author} {\bibfnamefont {S.}~\bibnamefont {Pochinda}}, \bibinfo {author} {\bibfnamefont {E.}~\bibnamefont {de~Lera~Acedo}}, \bibinfo {author} {\bibfnamefont {S.}~\bibnamefont {Singh}},\ and\ \bibinfo {author} {\bibfnamefont {R.}~\bibnamefont {Barkana}},\ }\href@noop {} {\bibinfo {title} {Narrowing the discovery space of the cosmological 21-cm signal using multi-wavelength constraints}} (\bibinfo {year} {2025}),\ \Eprint {https://arxiv.org/abs/2508.13761} {arXiv:2508.13761 [astro-ph.CO]} \BibitemShut {NoStop}%
\bibitem [{\citenamefont {Furlanetto}\ \emph {et~al.}(2006)\citenamefont {Furlanetto}, \citenamefont {Oh},\ and\ \citenamefont {Pierpaoli}}]{Furlanetto:2006wp}%
  \BibitemOpen
  \bibfield  {author} {\bibinfo {author} {\bibfnamefont {S.~R.}\ \bibnamefont {Furlanetto}}, \bibinfo {author} {\bibfnamefont {S.~P.}\ \bibnamefont {Oh}},\ and\ \bibinfo {author} {\bibfnamefont {E.}~\bibnamefont {Pierpaoli}},\ }\bibfield  {title} {\bibinfo {title} {{The effects of dark matter decay and annihilation on the high-redshift 21 cm background}},\ }\href {https://doi.org/10.1103/PhysRevD.74.103502} {\bibfield  {journal} {\bibinfo  {journal} {Phys. Rev. D}\ }\textbf {\bibinfo {volume} {74}},\ \bibinfo {pages} {103502} (\bibinfo {year} {2006})},\ \Eprint {https://arxiv.org/abs/astro-ph/0608385} {arXiv:astro-ph/0608385} \BibitemShut {NoStop}%
\bibitem [{\citenamefont {Valdes}\ \emph {et~al.}(2007)\citenamefont {Valdes}, \citenamefont {Ferrara}, \citenamefont {Mapelli},\ and\ \citenamefont {Ripamonti}}]{Valdes:2007cu}%
  \BibitemOpen
  \bibfield  {author} {\bibinfo {author} {\bibfnamefont {M.}~\bibnamefont {Valdes}}, \bibinfo {author} {\bibfnamefont {A.}~\bibnamefont {Ferrara}}, \bibinfo {author} {\bibfnamefont {M.}~\bibnamefont {Mapelli}},\ and\ \bibinfo {author} {\bibfnamefont {E.}~\bibnamefont {Ripamonti}},\ }\bibfield  {title} {\bibinfo {title} {{Constraining DM through 21 cm observations}},\ }\href {https://doi.org/10.1111/j.1365-2966.2007.11594.x} {\bibfield  {journal} {\bibinfo  {journal} {Mon. Not. Roy. Astron. Soc.}\ }\textbf {\bibinfo {volume} {377}},\ \bibinfo {pages} {245} (\bibinfo {year} {2007})},\ \Eprint {https://arxiv.org/abs/astro-ph/0701301} {arXiv:astro-ph/0701301} \BibitemShut {NoStop}%
\bibitem [{\citenamefont {Evoli}\ \emph {et~al.}(2014)\citenamefont {Evoli}, \citenamefont {Mesinger},\ and\ \citenamefont {Ferrara}}]{Evoli:2014pva}%
  \BibitemOpen
  \bibfield  {author} {\bibinfo {author} {\bibfnamefont {C.}~\bibnamefont {Evoli}}, \bibinfo {author} {\bibfnamefont {A.}~\bibnamefont {Mesinger}},\ and\ \bibinfo {author} {\bibfnamefont {A.}~\bibnamefont {Ferrara}},\ }\bibfield  {title} {\bibinfo {title} {Unveiling the nature of dark matter with high redshift 21 cm line experiments},\ }\href {https://doi.org/10.1088/1475-7516/2014/11/024} {\bibfield  {journal} {\bibinfo  {journal} {Journal of Cosmology and Astroparticle Physics}\ }\textbf {\bibinfo {volume} {2014}}\bibinfo  {number} { (11)},\ \bibinfo {pages} {024–024}}\BibitemShut {NoStop}%
\bibitem [{\citenamefont {Lopez-Honorez}\ \emph {et~al.}(2016)\citenamefont {Lopez-Honorez}, \citenamefont {Mena}, \citenamefont {Molin{\'e}}, \citenamefont {Palomares-Ruiz},\ and\ \citenamefont {Vincent}}]{Lopez-Honorez:2016sur}%
  \BibitemOpen
\bibfield  {number} {  }\bibfield  {author} {\bibinfo {author} {\bibfnamefont {L.}~\bibnamefont {Lopez-Honorez}}, \bibinfo {author} {\bibfnamefont {O.}~\bibnamefont {Mena}}, \bibinfo {author} {\bibfnamefont {{\'A}.}~\bibnamefont {Molin{\'e}}}, \bibinfo {author} {\bibfnamefont {S.}~\bibnamefont {Palomares-Ruiz}},\ and\ \bibinfo {author} {\bibfnamefont {A.~C.}\ \bibnamefont {Vincent}},\ }\bibfield  {title} {\bibinfo {title} {The 21 cm signal and the interplay between dark matter annihilations and astrophysical processes},\ }\href {https://doi.org/10.1088/1475-7516/2016/08/004} {\bibfield  {journal} {\bibinfo  {journal} {Journal of Cosmology and Astroparticle Physics}\ }\textbf {\bibinfo {volume} {2016}}\bibinfo  {number} { (08)},\ \bibinfo {pages} {004–004}}\BibitemShut {NoStop}%
\bibitem [{\citenamefont {Poulin}\ \emph {et~al.}(2017)\citenamefont {Poulin}, \citenamefont {Lesgourgues},\ and\ \citenamefont {Serpico}}]{Poulin:2016anj}%
  \BibitemOpen
\bibfield  {number} {  }\bibfield  {author} {\bibinfo {author} {\bibfnamefont {V.}~\bibnamefont {Poulin}}, \bibinfo {author} {\bibfnamefont {J.}~\bibnamefont {Lesgourgues}},\ and\ \bibinfo {author} {\bibfnamefont {P.~D.}\ \bibnamefont {Serpico}},\ }\bibfield  {title} {\bibinfo {title} {{Cosmological constraints on exotic injection of electromagnetic energy}},\ }\href {https://doi.org/10.1088/1475-7516/2017/03/043} {\bibfield  {journal} {\bibinfo  {journal} {JCAP}\ }\textbf {\bibinfo {volume} {03}},\ \bibinfo {pages} {043}},\ \Eprint {https://arxiv.org/abs/1610.10051} {arXiv:1610.10051 [astro-ph.CO]} \BibitemShut {NoStop}%
\bibitem [{\citenamefont {Clark}\ \emph {et~al.}(2018)\citenamefont {Clark}, \citenamefont {Dutta}, \citenamefont {Gao}, \citenamefont {Ma},\ and\ \citenamefont {Strigari}}]{Clark:2018ghm}%
  \BibitemOpen
  \bibfield  {author} {\bibinfo {author} {\bibfnamefont {S.}~\bibnamefont {Clark}}, \bibinfo {author} {\bibfnamefont {B.}~\bibnamefont {Dutta}}, \bibinfo {author} {\bibfnamefont {Y.}~\bibnamefont {Gao}}, \bibinfo {author} {\bibfnamefont {Y.-Z.}\ \bibnamefont {Ma}},\ and\ \bibinfo {author} {\bibfnamefont {L.~E.}\ \bibnamefont {Strigari}},\ }\bibfield  {title} {\bibinfo {title} {{21 cm limits on decaying dark matter and primordial black holes}},\ }\href {https://doi.org/10.1103/PhysRevD.98.043006} {\bibfield  {journal} {\bibinfo  {journal} {Phys. Rev. D}\ }\textbf {\bibinfo {volume} {98}},\ \bibinfo {pages} {043006} (\bibinfo {year} {2018})},\ \Eprint {https://arxiv.org/abs/1803.09390} {arXiv:1803.09390 [astro-ph.HE]} \BibitemShut {NoStop}%
\bibitem [{\citenamefont {D'Amico}\ \emph {et~al.}(2018)\citenamefont {D'Amico}, \citenamefont {Panci},\ and\ \citenamefont {Strumia}}]{DAmico:2018sxd}%
  \BibitemOpen
  \bibfield  {author} {\bibinfo {author} {\bibfnamefont {G.}~\bibnamefont {D'Amico}}, \bibinfo {author} {\bibfnamefont {P.}~\bibnamefont {Panci}},\ and\ \bibinfo {author} {\bibfnamefont {A.}~\bibnamefont {Strumia}},\ }\bibfield  {title} {\bibinfo {title} {{Bounds on Dark Matter annihilations from 21 cm data}},\ }\href {https://doi.org/10.1103/PhysRevLett.121.011103} {\bibfield  {journal} {\bibinfo  {journal} {Phys. Rev. Lett.}\ }\textbf {\bibinfo {volume} {121}},\ \bibinfo {pages} {011103} (\bibinfo {year} {2018})},\ \Eprint {https://arxiv.org/abs/1803.03629} {arXiv:1803.03629 [astro-ph.CO]} \BibitemShut {NoStop}%
\bibitem [{\citenamefont {Hektor}\ \emph {et~al.}(2018)\citenamefont {Hektor}, \citenamefont {H{\"u}tsi}, \citenamefont {Marzola}, \citenamefont {Raidal}, \citenamefont {Vaskonen},\ and\ \citenamefont {Veerm{\"a}e}}]{Hektor:2018qqw}%
  \BibitemOpen
  \bibfield  {author} {\bibinfo {author} {\bibfnamefont {A.}~\bibnamefont {Hektor}}, \bibinfo {author} {\bibfnamefont {G.}~\bibnamefont {H{\"u}tsi}}, \bibinfo {author} {\bibfnamefont {L.}~\bibnamefont {Marzola}}, \bibinfo {author} {\bibfnamefont {M.}~\bibnamefont {Raidal}}, \bibinfo {author} {\bibfnamefont {V.}~\bibnamefont {Vaskonen}},\ and\ \bibinfo {author} {\bibfnamefont {H.}~\bibnamefont {Veerm{\"a}e}},\ }\bibfield  {title} {\bibinfo {title} {{Constraining primordial black holes with the EDGES 21-cm absorption signal}},\ }\href {https://doi.org/10.1103/PhysRevD.98.023503} {\bibfield  {journal} {\bibinfo  {journal} {Phys. Rev. D}\ }\textbf {\bibinfo {volume} {98}},\ \bibinfo {pages} {023503} (\bibinfo {year} {2018})},\ \Eprint {https://arxiv.org/abs/1803.09697} {arXiv:1803.09697 [astro-ph.CO]} \BibitemShut {NoStop}%
\bibitem [{\citenamefont {Liu}\ and\ \citenamefont {Slatyer}(2018)}]{Liu:2018uzy}%
  \BibitemOpen
  \bibfield  {author} {\bibinfo {author} {\bibfnamefont {H.}~\bibnamefont {Liu}}\ and\ \bibinfo {author} {\bibfnamefont {T.~R.}\ \bibnamefont {Slatyer}},\ }\bibfield  {title} {\bibinfo {title} {{Implications of a 21-cm signal for dark matter annihilation and decay}},\ }\href {https://doi.org/10.1103/PhysRevD.98.023501} {\bibfield  {journal} {\bibinfo  {journal} {Phys. Rev. D}\ }\textbf {\bibinfo {volume} {98}},\ \bibinfo {pages} {023501} (\bibinfo {year} {2018})},\ \Eprint {https://arxiv.org/abs/1803.09739} {arXiv:1803.09739 [astro-ph.CO]} \BibitemShut {NoStop}%
\bibitem [{\citenamefont {Mitridate}\ and\ \citenamefont {Podo}(2018)}]{Mitridate:2018iag}%
  \BibitemOpen
  \bibfield  {author} {\bibinfo {author} {\bibfnamefont {A.}~\bibnamefont {Mitridate}}\ and\ \bibinfo {author} {\bibfnamefont {A.}~\bibnamefont {Podo}},\ }\bibfield  {title} {\bibinfo {title} {{Bounds on dark matter decay from 21 cm line}},\ }\href {https://doi.org/10.1088/1475-7516/2018/05/069} {\bibfield  {journal} {\bibinfo  {journal} {JCAP}\ }\textbf {\bibinfo {volume} {05}},\ \bibinfo {pages} {069}},\ \Eprint {https://arxiv.org/abs/1803.11169} {arXiv:1803.11169 [hep-ph]} \BibitemShut {NoStop}%
\bibitem [{\citenamefont {Mena}\ \emph {et~al.}(2019)\citenamefont {Mena}, \citenamefont {Palomares-Ruiz}, \citenamefont {Villanueva-Domingo},\ and\ \citenamefont {Witte}}]{Mena:2019nhm}%
  \BibitemOpen
  \bibfield  {author} {\bibinfo {author} {\bibfnamefont {O.}~\bibnamefont {Mena}}, \bibinfo {author} {\bibfnamefont {S.}~\bibnamefont {Palomares-Ruiz}}, \bibinfo {author} {\bibfnamefont {P.}~\bibnamefont {Villanueva-Domingo}},\ and\ \bibinfo {author} {\bibfnamefont {S.~J.}\ \bibnamefont {Witte}},\ }\bibfield  {title} {\bibinfo {title} {{Constraining the primordial black hole abundance with 21-cm cosmology}},\ }\href {https://doi.org/10.1103/PhysRevD.100.043540} {\bibfield  {journal} {\bibinfo  {journal} {Phys. Rev. D}\ }\textbf {\bibinfo {volume} {100}},\ \bibinfo {pages} {043540} (\bibinfo {year} {2019})},\ \Eprint {https://arxiv.org/abs/1906.07735} {arXiv:1906.07735 [astro-ph.CO]} \BibitemShut {NoStop}%
\bibitem [{\citenamefont {Yang}(2020)}]{Yang:2020egn}%
  \BibitemOpen
  \bibfield  {author} {\bibinfo {author} {\bibfnamefont {Y.}~\bibnamefont {Yang}},\ }\bibfield  {title} {\bibinfo {title} {{Constraints on primordial black holes and curvature perturbations from the global 21-cm signal}},\ }\href {https://doi.org/10.1103/PhysRevD.102.083538} {\bibfield  {journal} {\bibinfo  {journal} {Phys. Rev. D}\ }\textbf {\bibinfo {volume} {102}},\ \bibinfo {pages} {083538} (\bibinfo {year} {2020})},\ \Eprint {https://arxiv.org/abs/2009.11547} {arXiv:2009.11547 [astro-ph.CO]} \BibitemShut {NoStop}%
\bibitem [{\citenamefont {Cang}\ \emph {et~al.}(2022)\citenamefont {Cang}, \citenamefont {Gao},\ and\ \citenamefont {Ma}}]{Cang:2021owu}%
  \BibitemOpen
  \bibfield  {author} {\bibinfo {author} {\bibfnamefont {J.}~\bibnamefont {Cang}}, \bibinfo {author} {\bibfnamefont {Y.}~\bibnamefont {Gao}},\ and\ \bibinfo {author} {\bibfnamefont {Y.-Z.}\ \bibnamefont {Ma}},\ }\bibfield  {title} {\bibinfo {title} {{21-cm constraints on spinning primordial black holes}},\ }\href {https://doi.org/10.1088/1475-7516/2022/03/012} {\bibfield  {journal} {\bibinfo  {journal} {JCAP}\ }\textbf {\bibinfo {volume} {03}}\bibfield  {number} {\bibinfo  {number} { (03)},\ \bibinfo {pages} {012}},\ }\Eprint {https://arxiv.org/abs/2108.13256} {arXiv:2108.13256 [astro-ph.CO]} \BibitemShut {NoStop}%
\bibitem [{\citenamefont {Halder}\ and\ \citenamefont {Pandey}(2021)}]{Halder:2021jiv}%
  \BibitemOpen
  \bibfield  {author} {\bibinfo {author} {\bibfnamefont {A.}~\bibnamefont {Halder}}\ and\ \bibinfo {author} {\bibfnamefont {M.}~\bibnamefont {Pandey}},\ }\bibfield  {title} {\bibinfo {title} {{Probing the effects of primordial black holes on 21-cm EDGES signal along with interacting dark energy and dark matter{\textendash}baryon scattering}},\ }\href {https://doi.org/10.1093/mnras/stab2795} {\bibfield  {journal} {\bibinfo  {journal} {Mon. Not. Roy. Astron. Soc.}\ }\textbf {\bibinfo {volume} {508}},\ \bibinfo {pages} {3446} (\bibinfo {year} {2021})},\ \Eprint {https://arxiv.org/abs/2101.05228} {arXiv:2101.05228 [astro-ph.CO]} \BibitemShut {NoStop}%
\bibitem [{\citenamefont {Halder}\ and\ \citenamefont {Banerjee}(2021)}]{Halder:2021rbq}%
  \BibitemOpen
  \bibfield  {author} {\bibinfo {author} {\bibfnamefont {A.}~\bibnamefont {Halder}}\ and\ \bibinfo {author} {\bibfnamefont {S.}~\bibnamefont {Banerjee}},\ }\bibfield  {title} {\bibinfo {title} {{Bounds on abundance of primordial black hole and dark matter from EDGES 21-cm signal}},\ }\href {https://doi.org/10.1103/PhysRevD.103.063044} {\bibfield  {journal} {\bibinfo  {journal} {Phys. Rev. D}\ }\textbf {\bibinfo {volume} {103}},\ \bibinfo {pages} {063044} (\bibinfo {year} {2021})},\ \Eprint {https://arxiv.org/abs/2102.00959} {arXiv:2102.00959 [astro-ph.CO]} \BibitemShut {NoStop}%
\bibitem [{\citenamefont {Mittal}\ \emph {et~al.}(2022)\citenamefont {Mittal}, \citenamefont {Ray}, \citenamefont {Kulkarni},\ and\ \citenamefont {Dasgupta}}]{Mittal:2021egv}%
  \BibitemOpen
  \bibfield  {author} {\bibinfo {author} {\bibfnamefont {S.}~\bibnamefont {Mittal}}, \bibinfo {author} {\bibfnamefont {A.}~\bibnamefont {Ray}}, \bibinfo {author} {\bibfnamefont {G.}~\bibnamefont {Kulkarni}},\ and\ \bibinfo {author} {\bibfnamefont {B.}~\bibnamefont {Dasgupta}},\ }\bibfield  {title} {\bibinfo {title} {{Constraining primordial black holes as dark matter using the global 21-cm signal with X-ray heating and excess radio background}},\ }\href {https://doi.org/10.1088/1475-7516/2022/03/030} {\bibfield  {journal} {\bibinfo  {journal} {JCAP}\ }\textbf {\bibinfo {volume} {03}},\ \bibinfo {pages} {030}},\ \Eprint {https://arxiv.org/abs/2107.02190} {arXiv:2107.02190 [astro-ph.CO]} \BibitemShut {NoStop}%
\bibitem [{\citenamefont {Natwariya}\ \emph {et~al.}(2021)\citenamefont {Natwariya}, \citenamefont {Nayak},\ and\ \citenamefont {Srivastava}}]{Natwariya:2021xki}%
  \BibitemOpen
  \bibfield  {author} {\bibinfo {author} {\bibfnamefont {P.~K.}\ \bibnamefont {Natwariya}}, \bibinfo {author} {\bibfnamefont {A.~C.}\ \bibnamefont {Nayak}},\ and\ \bibinfo {author} {\bibfnamefont {T.}~\bibnamefont {Srivastava}},\ }\bibfield  {title} {\bibinfo {title} {{Constraining spinning primordial black holes with global 21-cm signal}},\ }\href {https://doi.org/10.1093/mnras/stab3754} {\bibfield  {journal} {\bibinfo  {journal} {Mon. Not. Roy. Astron. Soc.}\ }\textbf {\bibinfo {volume} {510}},\ \bibinfo {pages} {4236} (\bibinfo {year} {2021})},\ \Eprint {https://arxiv.org/abs/2107.12358} {arXiv:2107.12358 [astro-ph.CO]} \BibitemShut {NoStop}%
\bibitem [{\citenamefont {Saha}\ and\ \citenamefont {Laha}(2022)}]{Saha:2021pqf}%
  \BibitemOpen
  \bibfield  {author} {\bibinfo {author} {\bibfnamefont {A.~K.}\ \bibnamefont {Saha}}\ and\ \bibinfo {author} {\bibfnamefont {R.}~\bibnamefont {Laha}},\ }\bibfield  {title} {\bibinfo {title} {{Sensitivities on nonspinning and spinning primordial black hole dark matter with global 21-cm troughs}},\ }\href {https://doi.org/10.1103/PhysRevD.105.103026} {\bibfield  {journal} {\bibinfo  {journal} {Phys. Rev. D}\ }\textbf {\bibinfo {volume} {105}},\ \bibinfo {pages} {103026} (\bibinfo {year} {2022})},\ \Eprint {https://arxiv.org/abs/2112.10794} {arXiv:2112.10794 [astro-ph.CO]} \BibitemShut {NoStop}%
\bibitem [{\citenamefont {Yang}(2021)}]{Yang:2021idt}%
  \BibitemOpen
  \bibfield  {author} {\bibinfo {author} {\bibfnamefont {Y.}~\bibnamefont {Yang}},\ }\bibfield  {title} {\bibinfo {title} {{Constraints on accreting primordial black holes with the global 21-cm signal}},\ }\href {https://doi.org/10.1103/PhysRevD.104.063528} {\bibfield  {journal} {\bibinfo  {journal} {Phys. Rev. D}\ }\textbf {\bibinfo {volume} {104}},\ \bibinfo {pages} {063528} (\bibinfo {year} {2021})},\ \Eprint {https://arxiv.org/abs/2108.11130} {arXiv:2108.11130 [astro-ph.CO]} \BibitemShut {NoStop}%
\bibitem [{\citenamefont {Mukhopadhyay}\ \emph {et~al.}(2022)\citenamefont {Mukhopadhyay}, \citenamefont {Majumdar},\ and\ \citenamefont {Halder}}]{Mukhopadhyay:2022jqc}%
  \BibitemOpen
  \bibfield  {author} {\bibinfo {author} {\bibfnamefont {U.}~\bibnamefont {Mukhopadhyay}}, \bibinfo {author} {\bibfnamefont {D.}~\bibnamefont {Majumdar}},\ and\ \bibinfo {author} {\bibfnamefont {A.}~\bibnamefont {Halder}},\ }\bibfield  {title} {\bibinfo {title} {{Constraining PBH mass distributions from 21cm brightness temperature results and an analytical mapping between probability distribution of 21cm signal and PBH masses}},\ }\href {https://doi.org/10.1088/1475-7516/2022/10/099} {\bibfield  {journal} {\bibinfo  {journal} {JCAP}\ }\textbf {\bibinfo {volume} {10}},\ \bibinfo {pages} {099}},\ \Eprint {https://arxiv.org/abs/2203.13008} {arXiv:2203.13008 [astro-ph.CO]} \BibitemShut {NoStop}%
\bibitem [{\citenamefont {Yang}(2022)}]{Yang:2022puh}%
  \BibitemOpen
  \bibfield  {author} {\bibinfo {author} {\bibfnamefont {Y.}~\bibnamefont {Yang}},\ }\bibfield  {title} {\bibinfo {title} {{Impact of radiation from primordial black holes on the 21-cm angular-power spectrum in the dark ages}},\ }\href {https://doi.org/10.1103/PhysRevD.106.123508} {\bibfield  {journal} {\bibinfo  {journal} {Phys. Rev. D}\ }\textbf {\bibinfo {volume} {106}},\ \bibinfo {pages} {123508} (\bibinfo {year} {2022})},\ \Eprint {https://arxiv.org/abs/2209.00851} {arXiv:2209.00851 [astro-ph.CO]} \BibitemShut {NoStop}%
\bibitem [{\citenamefont {Facchinetti}\ \emph {et~al.}(2024)\citenamefont {Facchinetti}, \citenamefont {Lopez-Honorez}, \citenamefont {Qin},\ and\ \citenamefont {Mesinger}}]{Facchinetti:2023slb}%
  \BibitemOpen
  \bibfield  {author} {\bibinfo {author} {\bibfnamefont {G.}~\bibnamefont {Facchinetti}}, \bibinfo {author} {\bibfnamefont {L.}~\bibnamefont {Lopez-Honorez}}, \bibinfo {author} {\bibfnamefont {Y.}~\bibnamefont {Qin}},\ and\ \bibinfo {author} {\bibfnamefont {A.}~\bibnamefont {Mesinger}},\ }\bibfield  {title} {\bibinfo {title} {{21cm signal sensitivity to dark matter decay}},\ }\href {https://doi.org/10.1088/1475-7516/2024/01/005} {\bibfield  {journal} {\bibinfo  {journal} {JCAP}\ }\textbf {\bibinfo {volume} {01}},\ \bibinfo {pages} {005}},\ \Eprint {https://arxiv.org/abs/2308.16656} {arXiv:2308.16656 [astro-ph.CO]} \BibitemShut {NoStop}%
\bibitem [{\citenamefont {Qin}\ \emph {et~al.}(2024)\citenamefont {Qin}, \citenamefont {Munoz}, \citenamefont {Liu},\ and\ \citenamefont {Slatyer}}]{Qin:2023kkk}%
  \BibitemOpen
  \bibfield  {author} {\bibinfo {author} {\bibfnamefont {W.}~\bibnamefont {Qin}}, \bibinfo {author} {\bibfnamefont {J.~B.}\ \bibnamefont {Munoz}}, \bibinfo {author} {\bibfnamefont {H.}~\bibnamefont {Liu}},\ and\ \bibinfo {author} {\bibfnamefont {T.~R.}\ \bibnamefont {Slatyer}},\ }\bibfield  {title} {\bibinfo {title} {{Birth of the first stars amidst decaying and annihilating dark matter}},\ }\href {https://doi.org/10.1103/PhysRevD.109.103026} {\bibfield  {journal} {\bibinfo  {journal} {Phys. Rev. D}\ }\textbf {\bibinfo {volume} {109}},\ \bibinfo {pages} {103026} (\bibinfo {year} {2024})},\ \Eprint {https://arxiv.org/abs/2308.12992} {arXiv:2308.12992 [astro-ph.CO]} \BibitemShut {NoStop}%
\bibitem [{\citenamefont {Sun}\ \emph {et~al.}(2025{\natexlab{a}})\citenamefont {Sun}, \citenamefont {Foster}, \citenamefont {Liu}, \citenamefont {Mu{\~n}oz},\ and\ \citenamefont {Slatyer}}]{Sun:2023acy}%
  \BibitemOpen
  \bibfield  {author} {\bibinfo {author} {\bibfnamefont {Y.}~\bibnamefont {Sun}}, \bibinfo {author} {\bibfnamefont {J.~W.}\ \bibnamefont {Foster}}, \bibinfo {author} {\bibfnamefont {H.}~\bibnamefont {Liu}}, \bibinfo {author} {\bibfnamefont {J.~B.}\ \bibnamefont {Mu{\~n}oz}},\ and\ \bibinfo {author} {\bibfnamefont {T.~R.}\ \bibnamefont {Slatyer}},\ }\bibfield  {title} {\bibinfo {title} {{Inhomogeneous energy injection in the 21-cm power spectrum: Sensitivity to dark matter decay}},\ }\href {https://doi.org/10.1103/PhysRevD.111.043015} {\bibfield  {journal} {\bibinfo  {journal} {Phys. Rev. D}\ }\textbf {\bibinfo {volume} {111}},\ \bibinfo {pages} {043015} (\bibinfo {year} {2025}{\natexlab{a}})},\ \Eprint {https://arxiv.org/abs/2312.11608} {arXiv:2312.11608 [hep-ph]} \BibitemShut {NoStop}%
\bibitem [{\citenamefont {Novosyadlyj}\ \emph {et~al.}(2025)\citenamefont {Novosyadlyj}, \citenamefont {Kulinich},\ and\ \citenamefont {Koval}}]{Novosyadlyj:2024bie}%
  \BibitemOpen
  \bibfield  {author} {\bibinfo {author} {\bibfnamefont {B.}~\bibnamefont {Novosyadlyj}}, \bibinfo {author} {\bibfnamefont {Y.}~\bibnamefont {Kulinich}},\ and\ \bibinfo {author} {\bibfnamefont {D.}~\bibnamefont {Koval}},\ }\bibfield  {title} {\bibinfo {title} {{Global signal in the redshifted hydrogen 21-cm line from the dark ages and cosmic dawn: Dependence on the nature of dark matter and modeling of first light}},\ }\href {https://doi.org/10.1103/PhysRevD.111.083514} {\bibfield  {journal} {\bibinfo  {journal} {Phys. Rev. D}\ }\textbf {\bibinfo {volume} {111}},\ \bibinfo {pages} {083514} (\bibinfo {year} {2025})},\ \Eprint {https://arxiv.org/abs/2410.07380} {arXiv:2410.07380 [astro-ph.CO]} \BibitemShut {NoStop}%
\bibitem [{\citenamefont {Zhao}\ \emph {et~al.}(2025{\natexlab{a}})\citenamefont {Zhao}, \citenamefont {Wang},\ and\ \citenamefont {Zhang}}]{Zhao:2024jad}%
  \BibitemOpen
  \bibfield  {author} {\bibinfo {author} {\bibfnamefont {M.-L.}\ \bibnamefont {Zhao}}, \bibinfo {author} {\bibfnamefont {S.}~\bibnamefont {Wang}},\ and\ \bibinfo {author} {\bibfnamefont {X.}~\bibnamefont {Zhang}},\ }\bibfield  {title} {\bibinfo {title} {{Prospects for probing dark matter particles and primordial black holes with the Hongmeng mission using the 21 cm global spectrum at cosmic dawn}},\ }\href {https://doi.org/10.1088/1475-7516/2025/07/039} {\bibfield  {journal} {\bibinfo  {journal} {JCAP}\ }\textbf {\bibinfo {volume} {07}},\ \bibinfo {pages} {039}},\ \Eprint {https://arxiv.org/abs/2412.19257} {arXiv:2412.19257 [astro-ph.CO]} \BibitemShut {NoStop}%
\bibitem [{\citenamefont {Bae}\ \emph {et~al.}(2025)\citenamefont {Bae}, \citenamefont {Erickcek}, \citenamefont {Delos},\ and\ \citenamefont {Mu{\~n}oz}}]{Bae:2025uqa}%
  \BibitemOpen
  \bibfield  {author} {\bibinfo {author} {\bibfnamefont {H.}~\bibnamefont {Bae}}, \bibinfo {author} {\bibfnamefont {A.~L.}\ \bibnamefont {Erickcek}}, \bibinfo {author} {\bibfnamefont {M.~S.}\ \bibnamefont {Delos}},\ and\ \bibinfo {author} {\bibfnamefont {J.~B.}\ \bibnamefont {Mu{\~n}oz}},\ }\bibfield  {title} {\bibinfo {title} {{21-cm constraints on dark matter annihilation after an early matter-dominated era}},\ }\href {https://doi.org/10.1103/6rkn-tlrh} {\bibfield  {journal} {\bibinfo  {journal} {Phys. Rev. D}\ }\textbf {\bibinfo {volume} {112}},\ \bibinfo {pages} {083013} (\bibinfo {year} {2025})},\ \Eprint {https://arxiv.org/abs/2502.08719} {arXiv:2502.08719 [astro-ph.CO]} \BibitemShut {NoStop}%
\bibitem [{\citenamefont {Natwariya}\ \emph {et~al.}(2025)\citenamefont {Natwariya}, \citenamefont {Kadota},\ and\ \citenamefont {Nishizawa}}]{Natwariya:2025jlw}%
  \BibitemOpen
  \bibfield  {author} {\bibinfo {author} {\bibfnamefont {P.~K.}\ \bibnamefont {Natwariya}}, \bibinfo {author} {\bibfnamefont {K.}~\bibnamefont {Kadota}},\ and\ \bibinfo {author} {\bibfnamefont {A.~J.}\ \bibnamefont {Nishizawa}},\ }\href {https://arxiv.org/abs/2508.08251} {\bibinfo {title} {Sensitivity toward dark matter annihilation imprints on 21-cm signal with ska-low: A convolutional neural network approach}} (\bibinfo {year} {2025}),\ \Eprint {https://arxiv.org/abs/2508.08251} {arXiv:2508.08251 [astro-ph.CO]} \BibitemShut {NoStop}%
\bibitem [{\citenamefont {Sun}\ \emph {et~al.}(2025{\natexlab{b}})\citenamefont {Sun}, \citenamefont {Foster},\ and\ \citenamefont {Muñoz}}]{Sun:2025ksr}%
  \BibitemOpen
  \bibfield  {author} {\bibinfo {author} {\bibfnamefont {Y.}~\bibnamefont {Sun}}, \bibinfo {author} {\bibfnamefont {J.~W.}\ \bibnamefont {Foster}},\ and\ \bibinfo {author} {\bibfnamefont {J.~B.}\ \bibnamefont {Muñoz}},\ }\href {https://arxiv.org/abs/2509.22772} {\bibinfo {title} {Constraining inhomogeneous energy injection from annihilating dark matter and primordial black holes with 21-cm cosmology}} (\bibinfo {year} {2025}{\natexlab{b}}),\ \Eprint {https://arxiv.org/abs/2509.22772} {arXiv:2509.22772 [hep-ph]} \BibitemShut {NoStop}%
\bibitem [{\citenamefont {Zhao}\ \emph {et~al.}(2025{\natexlab{b}})\citenamefont {Zhao}, \citenamefont {Shao}, \citenamefont {Wang},\ and\ \citenamefont {Zhang}}]{Zhao:2025ddy}%
  \BibitemOpen
  \bibfield  {author} {\bibinfo {author} {\bibfnamefont {M.-L.}\ \bibnamefont {Zhao}}, \bibinfo {author} {\bibfnamefont {Y.}~\bibnamefont {Shao}}, \bibinfo {author} {\bibfnamefont {S.}~\bibnamefont {Wang}},\ and\ \bibinfo {author} {\bibfnamefont {X.}~\bibnamefont {Zhang}},\ }\href@noop {} {\bibinfo {title} {{Prospects for probing dark matter particles and primordial black holes with the Square Kilometre Array using the 21 cm power spectrum at cosmic dawn}}} (\bibinfo {year} {2025}{\natexlab{b}}),\ \Eprint {https://arxiv.org/abs/2507.02651} {arXiv:2507.02651 [astro-ph.CO]} \BibitemShut {NoStop}%
\bibitem [{\citenamefont {Agius}\ \emph {et~al.}(2025)\citenamefont {Agius}, \citenamefont {Essig}, \citenamefont {Gaggero}, \citenamefont {Palomares-Ruiz}, \citenamefont {Suczewski},\ and\ \citenamefont {Valli}}]{Agius:2025xbj}%
  \BibitemOpen
  \bibfield  {author} {\bibinfo {author} {\bibfnamefont {D.}~\bibnamefont {Agius}}, \bibinfo {author} {\bibfnamefont {R.}~\bibnamefont {Essig}}, \bibinfo {author} {\bibfnamefont {D.}~\bibnamefont {Gaggero}}, \bibinfo {author} {\bibfnamefont {S.}~\bibnamefont {Palomares-Ruiz}}, \bibinfo {author} {\bibfnamefont {G.}~\bibnamefont {Suczewski}},\ and\ \bibinfo {author} {\bibfnamefont {M.}~\bibnamefont {Valli}},\ }\href {https://arxiv.org/abs/2510.14877} {\bibinfo {title} {Astrophysical uncertainties challenge 21-cm forecasts: A primordial black hole case study}} (\bibinfo {year} {2025}),\ \Eprint {https://arxiv.org/abs/2510.14877} {arXiv:2510.14877 [astro-ph.CO]} \BibitemShut {NoStop}%
\bibitem [{\citenamefont {Fraser}\ \emph {et~al.}(2018)\citenamefont {Fraser} \emph {et~al.}}]{Fraser:2018acy}%
  \BibitemOpen
  \bibfield  {author} {\bibinfo {author} {\bibfnamefont {S.}~\bibnamefont {Fraser}} \emph {et~al.},\ }\bibfield  {title} {\bibinfo {title} {{The EDGES 21 cm Anomaly and Properties of Dark Matter}},\ }\href {https://doi.org/10.1016/j.physletb.2018.08.035} {\bibfield  {journal} {\bibinfo  {journal} {Phys. Lett. B}\ }\textbf {\bibinfo {volume} {785}},\ \bibinfo {pages} {159} (\bibinfo {year} {2018})},\ \Eprint {https://arxiv.org/abs/1803.03245} {arXiv:1803.03245 [hep-ph]} \BibitemShut {NoStop}%
\bibitem [{\citenamefont {Barkana}\ \emph {et~al.}(2018)\citenamefont {Barkana}, \citenamefont {Outmezguine}, \citenamefont {Redigolo},\ and\ \citenamefont {Volansky}}]{Barkana:2018qrx}%
  \BibitemOpen
  \bibfield  {author} {\bibinfo {author} {\bibfnamefont {R.}~\bibnamefont {Barkana}}, \bibinfo {author} {\bibfnamefont {N.~J.}\ \bibnamefont {Outmezguine}}, \bibinfo {author} {\bibfnamefont {D.}~\bibnamefont {Redigolo}},\ and\ \bibinfo {author} {\bibfnamefont {T.}~\bibnamefont {Volansky}},\ }\bibfield  {title} {\bibinfo {title} {{Strong constraints on light dark matter interpretation of the EDGES signal}},\ }\href {https://doi.org/10.1103/PhysRevD.98.103005} {\bibfield  {journal} {\bibinfo  {journal} {Phys. Rev. D}\ }\textbf {\bibinfo {volume} {98}},\ \bibinfo {pages} {103005} (\bibinfo {year} {2018})},\ \Eprint {https://arxiv.org/abs/1803.03091} {arXiv:1803.03091 [hep-ph]} \BibitemShut {NoStop}%
\bibitem [{\citenamefont {Fialkov}\ \emph {et~al.}(2018)\citenamefont {Fialkov}, \citenamefont {Barkana},\ and\ \citenamefont {Cohen}}]{Fialkov:2018xre}%
  \BibitemOpen
  \bibfield  {author} {\bibinfo {author} {\bibfnamefont {A.}~\bibnamefont {Fialkov}}, \bibinfo {author} {\bibfnamefont {R.}~\bibnamefont {Barkana}},\ and\ \bibinfo {author} {\bibfnamefont {A.}~\bibnamefont {Cohen}},\ }\bibfield  {title} {\bibinfo {title} {{Constraining Baryon--Dark Matter Scattering with the Cosmic Dawn 21-cm Signal}},\ }\href {https://doi.org/10.1103/PhysRevLett.121.011101} {\bibfield  {journal} {\bibinfo  {journal} {Phys. Rev. Lett.}\ }\textbf {\bibinfo {volume} {121}},\ \bibinfo {pages} {011101} (\bibinfo {year} {2018})},\ \Eprint {https://arxiv.org/abs/1802.10577} {arXiv:1802.10577 [astro-ph.CO]} \BibitemShut {NoStop}%
\bibitem [{\citenamefont {Berlin}\ \emph {et~al.}(2018)\citenamefont {Berlin}, \citenamefont {Hooper}, \citenamefont {Krnjaic},\ and\ \citenamefont {McDermott}}]{Berlin:2018sjs}%
  \BibitemOpen
  \bibfield  {author} {\bibinfo {author} {\bibfnamefont {A.}~\bibnamefont {Berlin}}, \bibinfo {author} {\bibfnamefont {D.}~\bibnamefont {Hooper}}, \bibinfo {author} {\bibfnamefont {G.}~\bibnamefont {Krnjaic}},\ and\ \bibinfo {author} {\bibfnamefont {S.~D.}\ \bibnamefont {McDermott}},\ }\bibfield  {title} {\bibinfo {title} {{Severely Constraining Dark Matter Interpretations of the 21-cm Anomaly}},\ }\href {https://doi.org/10.1103/PhysRevLett.121.011102} {\bibfield  {journal} {\bibinfo  {journal} {Phys. Rev. Lett.}\ }\textbf {\bibinfo {volume} {121}},\ \bibinfo {pages} {011102} (\bibinfo {year} {2018})},\ \Eprint {https://arxiv.org/abs/1803.02804} {arXiv:1803.02804 [hep-ph]} \BibitemShut {NoStop}%
\bibitem [{\citenamefont {Abdurashidova}\ \emph {et~al.}(2022)\citenamefont {Abdurashidova} \emph {et~al.}}]{HERA:2021noe}%
  \BibitemOpen
  \bibfield  {author} {\bibinfo {author} {\bibfnamefont {Z.}~\bibnamefont {Abdurashidova}} \emph {et~al.} (\bibinfo {collaboration} {HERA}),\ }\bibfield  {title} {\bibinfo {title} {{HERA Phase I Limits on the Cosmic 21 cm Signal: Constraints on Astrophysics and Cosmology during the Epoch of Reionization}},\ }\href {https://doi.org/10.3847/1538-4357/ac2ffc} {\bibfield  {journal} {\bibinfo  {journal} {Astrophys. J.}\ }\textbf {\bibinfo {volume} {924}},\ \bibinfo {pages} {51} (\bibinfo {year} {2022})},\ \Eprint {https://arxiv.org/abs/2108.07282} {arXiv:2108.07282 [astro-ph.CO]} \BibitemShut {NoStop}%
\bibitem [{\citenamefont {Abdurashidova}\ \emph {et~al.}(2023)\citenamefont {Abdurashidova} \emph {et~al.}}]{HERA:2022wmy}%
  \BibitemOpen
  \bibfield  {author} {\bibinfo {author} {\bibfnamefont {Z.}~\bibnamefont {Abdurashidova}} \emph {et~al.} (\bibinfo {collaboration} {HERA}),\ }\bibfield  {title} {\bibinfo {title} {{Improved Constraints on the 21 cm EoR Power Spectrum and the X-Ray Heating of the IGM with HERA Phase I Observations}},\ }\href {https://doi.org/10.3847/1538-4357/acaf50} {\bibfield  {journal} {\bibinfo  {journal} {Astrophys. J.}\ }\textbf {\bibinfo {volume} {945}},\ \bibinfo {pages} {124} (\bibinfo {year} {2023})},\ \Eprint {https://arxiv.org/abs/2210.04912} {arXiv:2210.04912 [astro-ph.CO]} \BibitemShut {NoStop}%
\bibitem [{\citenamefont {Hirata}(2006)}]{Hirata:2005mz}%
  \BibitemOpen
  \bibfield  {author} {\bibinfo {author} {\bibfnamefont {C.~M.}\ \bibnamefont {Hirata}},\ }\bibfield  {title} {\bibinfo {title} {{Wouthuysen-Field coupling strength and application to high-redshift 21 cm radiation}},\ }\href {https://doi.org/10.1111/j.1365-2966.2005.09949.x} {\bibfield  {journal} {\bibinfo  {journal} {Mon. Not. Roy. Astron. Soc.}\ }\textbf {\bibinfo {volume} {367}},\ \bibinfo {pages} {259} (\bibinfo {year} {2006})},\ \Eprint {https://arxiv.org/abs/astro-ph/0507102} {arXiv:astro-ph/0507102} \BibitemShut {NoStop}%
\bibitem [{\citenamefont {Liu}\ \emph {et~al.}(2020)\citenamefont {Liu}, \citenamefont {Ridgway},\ and\ \citenamefont {Slatyer}}]{Liu:2019bbm}%
  \BibitemOpen
  \bibfield  {author} {\bibinfo {author} {\bibfnamefont {H.}~\bibnamefont {Liu}}, \bibinfo {author} {\bibfnamefont {G.~W.}\ \bibnamefont {Ridgway}},\ and\ \bibinfo {author} {\bibfnamefont {T.~R.}\ \bibnamefont {Slatyer}},\ }\bibfield  {title} {\bibinfo {title} {{Code package for calculating modified cosmic ionization and thermal histories with dark matter and other exotic energy injections}},\ }\href {https://doi.org/10.1103/PhysRevD.101.023530} {\bibfield  {journal} {\bibinfo  {journal} {Phys. Rev. D}\ }\textbf {\bibinfo {volume} {101}},\ \bibinfo {pages} {023530} (\bibinfo {year} {2020})},\ \bibinfo {note} {arXiv:1904.09296 [astro-ph]},\ \Eprint {https://arxiv.org/abs/1904.09296} {arXiv:1904.09296 [astro-ph.CO]} \BibitemShut {NoStop}%
\bibitem [{\citenamefont {Liu}\ \emph {et~al.}(2023{\natexlab{a}})\citenamefont {Liu}, \citenamefont {Qin}, \citenamefont {Ridgway},\ and\ \citenamefont {Slatyer}}]{Liu:2023nct}%
  \BibitemOpen
  \bibfield  {author} {\bibinfo {author} {\bibfnamefont {H.}~\bibnamefont {Liu}}, \bibinfo {author} {\bibfnamefont {W.}~\bibnamefont {Qin}}, \bibinfo {author} {\bibfnamefont {G.~W.}\ \bibnamefont {Ridgway}},\ and\ \bibinfo {author} {\bibfnamefont {T.~R.}\ \bibnamefont {Slatyer}},\ }\bibfield  {title} {\bibinfo {title} {{Exotic energy injection in the early Universe. II. CMB spectral distortions and constraints on light dark matter}},\ }\href {https://doi.org/10.1103/PhysRevD.108.043531} {\bibfield  {journal} {\bibinfo  {journal} {Phys. Rev. D}\ }\textbf {\bibinfo {volume} {108}},\ \bibinfo {pages} {043531} (\bibinfo {year} {2023}{\natexlab{a}})},\ \Eprint {https://arxiv.org/abs/2303.07370} {arXiv:2303.07370 [astro-ph.CO]} \BibitemShut {NoStop}%
\bibitem [{\citenamefont {Liu}\ \emph {et~al.}(2023{\natexlab{b}})\citenamefont {Liu}, \citenamefont {Qin}, \citenamefont {Ridgway},\ and\ \citenamefont {Slatyer}}]{Liu:2023fgu}%
  \BibitemOpen
  \bibfield  {author} {\bibinfo {author} {\bibfnamefont {H.}~\bibnamefont {Liu}}, \bibinfo {author} {\bibfnamefont {W.}~\bibnamefont {Qin}}, \bibinfo {author} {\bibfnamefont {G.~W.}\ \bibnamefont {Ridgway}},\ and\ \bibinfo {author} {\bibfnamefont {T.~R.}\ \bibnamefont {Slatyer}},\ }\bibfield  {title} {\bibinfo {title} {{Exotic energy injection in the early Universe. I. A novel treatment for low-energy electrons and photons}},\ }\href {https://doi.org/10.1103/PhysRevD.108.043530} {\bibfield  {journal} {\bibinfo  {journal} {Phys. Rev. D}\ }\textbf {\bibinfo {volume} {108}},\ \bibinfo {pages} {043530} (\bibinfo {year} {2023}{\natexlab{b}})},\ \Eprint {https://arxiv.org/abs/2303.07366} {arXiv:2303.07366 [astro-ph.CO]} \BibitemShut {NoStop}%
\bibitem [{\citenamefont {Gardner}\ \emph {et~al.}(2006)\citenamefont {Gardner} \emph {et~al.}}]{Gardner:2006ky}%
  \BibitemOpen
  \bibfield  {author} {\bibinfo {author} {\bibfnamefont {J.~P.}\ \bibnamefont {Gardner}} \emph {et~al.},\ }\bibfield  {title} {\bibinfo {title} {{The James Webb Space Telescope}},\ }\href {https://doi.org/10.1007/s11214-006-8315-7} {\bibfield  {journal} {\bibinfo  {journal} {Space Sci. Rev.}\ }\textbf {\bibinfo {volume} {123}},\ \bibinfo {pages} {485} (\bibinfo {year} {2006})},\ \Eprint {https://arxiv.org/abs/astro-ph/0606175} {arXiv:astro-ph/0606175} \BibitemShut {NoStop}%
\bibitem [{\citenamefont {Gardner}\ \emph {et~al.}(2023)\citenamefont {Gardner} \emph {et~al.}}]{Gardner:2023}%
  \BibitemOpen
  \bibfield  {author} {\bibinfo {author} {\bibfnamefont {J.~P.}\ \bibnamefont {Gardner}} \emph {et~al.},\ }\bibfield  {title} {\bibinfo {title} {{The James Webb Space Telescope Mission}},\ }\href {https://doi.org/10.1088/1538-3873/acd1b5} {\bibfield  {journal} {\bibinfo  {journal} {Publ. Astron. Soc. Pac.}\ }\textbf {\bibinfo {volume} {135}},\ \bibinfo {pages} {068001} (\bibinfo {year} {2023})}\BibitemShut {NoStop}%
\bibitem [{\citenamefont {Übler}\ \emph {et~al.}(2023)\citenamefont {Übler}, \citenamefont {Maiolino}, \citenamefont {Curtis-Lake}, \citenamefont {Pérez-González}, \citenamefont {Curti}, \citenamefont {Perna}, \citenamefont {Arribas}, \citenamefont {Charlot}, \citenamefont {Marshall}, \citenamefont {D’Eugenio}, \citenamefont {Scholtz}, \citenamefont {Bunker}, \citenamefont {Carniani}, \citenamefont {Ferruit}, \citenamefont {Jakobsen}, \citenamefont {Rix}, \citenamefont {Rodríguez Del~Pino}, \citenamefont {Willott}, \citenamefont {Boeker}, \citenamefont {Cresci}, \citenamefont {Jones}, \citenamefont {Kumari},\ and\ \citenamefont {Rawle}}]{bler_2023}%
  \BibitemOpen
  \bibfield  {author} {\bibinfo {author} {\bibfnamefont {H.}~\bibnamefont {Übler}}, \bibinfo {author} {\bibfnamefont {R.}~\bibnamefont {Maiolino}}, \bibinfo {author} {\bibfnamefont {E.}~\bibnamefont {Curtis-Lake}}, \bibinfo {author} {\bibfnamefont {P.~G.}\ \bibnamefont {Pérez-González}}, \bibinfo {author} {\bibfnamefont {M.}~\bibnamefont {Curti}}, \bibinfo {author} {\bibfnamefont {M.}~\bibnamefont {Perna}}, \bibinfo {author} {\bibfnamefont {S.}~\bibnamefont {Arribas}}, \bibinfo {author} {\bibfnamefont {S.}~\bibnamefont {Charlot}}, \bibinfo {author} {\bibfnamefont {M.~A.}\ \bibnamefont {Marshall}}, \bibinfo {author} {\bibfnamefont {F.}~\bibnamefont {D’Eugenio}}, \bibinfo {author} {\bibfnamefont {J.}~\bibnamefont {Scholtz}}, \bibinfo {author} {\bibfnamefont {A.}~\bibnamefont {Bunker}}, \bibinfo {author} {\bibfnamefont {S.}~\bibnamefont {Carniani}}, \bibinfo {author} {\bibfnamefont {P.}~\bibnamefont {Ferruit}}, \bibinfo {author} {\bibfnamefont {P.}~\bibnamefont {Jakobsen}}, \bibinfo {author} {\bibfnamefont
  {H.-W.}\ \bibnamefont {Rix}}, \bibinfo {author} {\bibfnamefont {B.}~\bibnamefont {Rodríguez Del~Pino}}, \bibinfo {author} {\bibfnamefont {C.~J.}\ \bibnamefont {Willott}}, \bibinfo {author} {\bibfnamefont {T.}~\bibnamefont {Boeker}}, \bibinfo {author} {\bibfnamefont {G.}~\bibnamefont {Cresci}}, \bibinfo {author} {\bibfnamefont {G.~C.}\ \bibnamefont {Jones}}, \bibinfo {author} {\bibfnamefont {N.}~\bibnamefont {Kumari}},\ and\ \bibinfo {author} {\bibfnamefont {T.}~\bibnamefont {Rawle}},\ }\bibfield  {title} {\bibinfo {title} {Ga-nifs: A massive black hole in a low-metallicity agn at z $\sim$ 5.55 revealed by jwst/nirspec ifs},\ }\href {https://doi.org/10.1051/0004-6361/202346137} {\bibfield  {journal} {\bibinfo  {journal} {Astronomy \& Astrophysics}\ }\textbf {\bibinfo {volume} {677}},\ \bibinfo {pages} {A145} (\bibinfo {year} {2023})}\BibitemShut {NoStop}%
\bibitem [{\citenamefont {{Larson}}\ \emph {et~al.}(2023)\citenamefont {{Larson}}, \citenamefont {{Finkelstein}}, \citenamefont {{Kocevski}}, \citenamefont {{Hutchison}}, \citenamefont {{Trump}}, \citenamefont {{Arrabal Haro}}, \citenamefont {{Bromm}}, \citenamefont {{Cleri}}, \citenamefont {{Dickinson}}, \citenamefont {{Fujimoto}}, \citenamefont {{Kartaltepe}}, \citenamefont {{Koekemoer}}, \citenamefont {{Papovich}}, \citenamefont {{Pirzkal}}, \citenamefont {{Tacchella}}, \citenamefont {{Zavala}}, \citenamefont {{Bagley}}, \citenamefont {{Behroozi}}, \citenamefont {{Champagne}}, \citenamefont {{Cole}}, \citenamefont {{Jung}}, \citenamefont {{Morales}}, \citenamefont {{Yang}}, \citenamefont {{Zhang}}, \citenamefont {{Zitrin}}, \citenamefont {{Amor{\'\i}n}}, \citenamefont {{Burgarella}}, \citenamefont {{Casey}}, \citenamefont {{Ch{\'a}vez Ortiz}}, \citenamefont {{Cox}}, \citenamefont {{Chworowsky}}, \citenamefont {{Fontana}}, \citenamefont {{Gawiser}}, \citenamefont {{Grazian}}, \citenamefont {{Grogin}},
  \citenamefont {{Harish}}, \citenamefont {{Hathi}}, \citenamefont {{Hirschmann}}, \citenamefont {{Holwerda}}, \citenamefont {{Juneau}}, \citenamefont {{Leung}}, \citenamefont {{Lucas}}, \citenamefont {{McGrath}}, \citenamefont {{P{\'e}rez-Gonz{\'a}lez}}, \citenamefont {{Rigby}}, \citenamefont {{Seill{\'e}}}, \citenamefont {{Simons}}, \citenamefont {{de La Vega}}, \citenamefont {{Weiner}}, \citenamefont {{Wilkins}}, \citenamefont {{Yung}},\ and\ \citenamefont {{Ceers Team}}}]{Larson_2023}%
  \BibitemOpen
  \bibfield  {author} {\bibinfo {author} {\bibfnamefont {R.~L.}\ \bibnamefont {{Larson}}}, \bibinfo {author} {\bibfnamefont {S.~L.}\ \bibnamefont {{Finkelstein}}}, \bibinfo {author} {\bibfnamefont {D.~D.}\ \bibnamefont {{Kocevski}}}, \bibinfo {author} {\bibfnamefont {T.~A.}\ \bibnamefont {{Hutchison}}}, \bibinfo {author} {\bibfnamefont {J.~R.}\ \bibnamefont {{Trump}}}, \bibinfo {author} {\bibfnamefont {P.}~\bibnamefont {{Arrabal Haro}}}, \bibinfo {author} {\bibfnamefont {V.}~\bibnamefont {{Bromm}}}, \bibinfo {author} {\bibfnamefont {N.~J.}\ \bibnamefont {{Cleri}}}, \bibinfo {author} {\bibfnamefont {M.}~\bibnamefont {{Dickinson}}}, \bibinfo {author} {\bibfnamefont {S.}~\bibnamefont {{Fujimoto}}}, \bibinfo {author} {\bibfnamefont {J.~S.}\ \bibnamefont {{Kartaltepe}}}, \bibinfo {author} {\bibfnamefont {A.~M.}\ \bibnamefont {{Koekemoer}}}, \bibinfo {author} {\bibfnamefont {C.}~\bibnamefont {{Papovich}}}, \bibinfo {author} {\bibfnamefont {N.}~\bibnamefont {{Pirzkal}}}, \bibinfo {author} {\bibfnamefont
  {S.}~\bibnamefont {{Tacchella}}}, \bibinfo {author} {\bibfnamefont {J.~A.}\ \bibnamefont {{Zavala}}}, \bibinfo {author} {\bibfnamefont {M.}~\bibnamefont {{Bagley}}}, \bibinfo {author} {\bibfnamefont {P.}~\bibnamefont {{Behroozi}}}, \bibinfo {author} {\bibfnamefont {J.~B.}\ \bibnamefont {{Champagne}}}, \bibinfo {author} {\bibfnamefont {J.~W.}\ \bibnamefont {{Cole}}}, \bibinfo {author} {\bibfnamefont {I.}~\bibnamefont {{Jung}}}, \bibinfo {author} {\bibfnamefont {A.~M.}\ \bibnamefont {{Morales}}}, \bibinfo {author} {\bibfnamefont {G.}~\bibnamefont {{Yang}}}, \bibinfo {author} {\bibfnamefont {H.}~\bibnamefont {{Zhang}}}, \bibinfo {author} {\bibfnamefont {A.}~\bibnamefont {{Zitrin}}}, \bibinfo {author} {\bibfnamefont {R.~O.}\ \bibnamefont {{Amor{\'\i}n}}}, \bibinfo {author} {\bibfnamefont {D.}~\bibnamefont {{Burgarella}}}, \bibinfo {author} {\bibfnamefont {C.~M.}\ \bibnamefont {{Casey}}}, \bibinfo {author} {\bibfnamefont {{\'O}.~A.}\ \bibnamefont {{Ch{\'a}vez Ortiz}}}, \bibinfo {author} {\bibfnamefont {I.~G.}\
  \bibnamefont {{Cox}}}, \bibinfo {author} {\bibfnamefont {K.}~\bibnamefont {{Chworowsky}}}, \bibinfo {author} {\bibfnamefont {A.}~\bibnamefont {{Fontana}}}, \bibinfo {author} {\bibfnamefont {E.}~\bibnamefont {{Gawiser}}}, \bibinfo {author} {\bibfnamefont {A.}~\bibnamefont {{Grazian}}}, \bibinfo {author} {\bibfnamefont {N.~A.}\ \bibnamefont {{Grogin}}}, \bibinfo {author} {\bibfnamefont {S.}~\bibnamefont {{Harish}}}, \bibinfo {author} {\bibfnamefont {N.~P.}\ \bibnamefont {{Hathi}}}, \bibinfo {author} {\bibfnamefont {M.}~\bibnamefont {{Hirschmann}}}, \bibinfo {author} {\bibfnamefont {B.~W.}\ \bibnamefont {{Holwerda}}}, \bibinfo {author} {\bibfnamefont {S.}~\bibnamefont {{Juneau}}}, \bibinfo {author} {\bibfnamefont {G.~C.~K.}\ \bibnamefont {{Leung}}}, \bibinfo {author} {\bibfnamefont {R.~A.}\ \bibnamefont {{Lucas}}}, \bibinfo {author} {\bibfnamefont {E.~J.}\ \bibnamefont {{McGrath}}}, \bibinfo {author} {\bibfnamefont {P.~G.}\ \bibnamefont {{P{\'e}rez-Gonz{\'a}lez}}}, \bibinfo {author} {\bibfnamefont {J.~R.}\
  \bibnamefont {{Rigby}}}, \bibinfo {author} {\bibfnamefont {L.-M.}\ \bibnamefont {{Seill{\'e}}}}, \bibinfo {author} {\bibfnamefont {R.~C.}\ \bibnamefont {{Simons}}}, \bibinfo {author} {\bibfnamefont {A.}~\bibnamefont {{de La Vega}}}, \bibinfo {author} {\bibfnamefont {B.~J.}\ \bibnamefont {{Weiner}}}, \bibinfo {author} {\bibfnamefont {S.~M.}\ \bibnamefont {{Wilkins}}}, \bibinfo {author} {\bibfnamefont {L.~Y.~A.}\ \bibnamefont {{Yung}}},\ and\ \bibinfo {author} {\bibnamefont {{Ceers Team}}},\ }\bibfield  {title} {\bibinfo {title} {{A CEERS Discovery of an Accreting Supermassive Black Hole 570 Myr after the Big Bang: Identifying a Progenitor of Massive {z $>$ 6} Quasars}},\ }\href {https://doi.org/10.3847/2041-8213/ace619} {\bibfield  {journal} {\bibinfo  {journal} {\apjl}\ }\textbf {\bibinfo {volume} {953}},\ \bibinfo {eid} {L29} (\bibinfo {year} {2023})},\ \Eprint {https://arxiv.org/abs/2303.08918} {arXiv:2303.08918 [astro-ph.GA]} \BibitemShut {NoStop}%
\bibitem [{\citenamefont {{Harikane}}\ \emph {et~al.}(2023)\citenamefont {{Harikane}}, \citenamefont {{Zhang}}, \citenamefont {{Nakajima}}, \citenamefont {{Ouchi}}, \citenamefont {{Isobe}}, \citenamefont {{Ono}}, \citenamefont {{Hatano}}, \citenamefont {{Xu}},\ and\ \citenamefont {{Umeda}}}]{harikane2023jwstnirspec}%
  \BibitemOpen
  \bibfield  {author} {\bibinfo {author} {\bibfnamefont {Y.}~\bibnamefont {{Harikane}}}, \bibinfo {author} {\bibfnamefont {Y.}~\bibnamefont {{Zhang}}}, \bibinfo {author} {\bibfnamefont {K.}~\bibnamefont {{Nakajima}}}, \bibinfo {author} {\bibfnamefont {M.}~\bibnamefont {{Ouchi}}}, \bibinfo {author} {\bibfnamefont {Y.}~\bibnamefont {{Isobe}}}, \bibinfo {author} {\bibfnamefont {Y.}~\bibnamefont {{Ono}}}, \bibinfo {author} {\bibfnamefont {S.}~\bibnamefont {{Hatano}}}, \bibinfo {author} {\bibfnamefont {Y.}~\bibnamefont {{Xu}}},\ and\ \bibinfo {author} {\bibfnamefont {H.}~\bibnamefont {{Umeda}}},\ }\bibfield  {title} {\bibinfo {title} {{A JWST/NIRSpec First Census of Broad-line AGNs at z = 4-7: Detection of 10 Faint AGNs with M $_{BH}$ {}10$^{6}$-{}10$^{8}$ M $_{{\ensuremath{\odot}}}$ and Their Host Galaxy Properties}},\ }\href {https://doi.org/10.3847/1538-4357/ad029e} {\bibfield  {journal} {\bibinfo  {journal} {\apj}\ }\textbf {\bibinfo {volume} {959}},\ \bibinfo {eid} {39} (\bibinfo {year} {2023})},\ \Eprint
  {https://arxiv.org/abs/2303.11946} {arXiv:2303.11946 [astro-ph.GA]} \BibitemShut {NoStop}%
\bibitem [{\citenamefont {Carnall}\ \emph {et~al.}(2023)\citenamefont {Carnall}, \citenamefont {McLure}, \citenamefont {Dunlop}, \citenamefont {McLeod}, \citenamefont {Wild}, \citenamefont {Cullen}, \citenamefont {Magee}, \citenamefont {Begley}, \citenamefont {Cimatti}, \citenamefont {Donnan}, \citenamefont {Hamadouche}, \citenamefont {Jewell},\ and\ \citenamefont {Walker}}]{Carnall_2023}%
  \BibitemOpen
  \bibfield  {author} {\bibinfo {author} {\bibfnamefont {A.~C.}\ \bibnamefont {Carnall}}, \bibinfo {author} {\bibfnamefont {R.~J.}\ \bibnamefont {McLure}}, \bibinfo {author} {\bibfnamefont {J.~S.}\ \bibnamefont {Dunlop}}, \bibinfo {author} {\bibfnamefont {D.~J.}\ \bibnamefont {McLeod}}, \bibinfo {author} {\bibfnamefont {V.}~\bibnamefont {Wild}}, \bibinfo {author} {\bibfnamefont {F.}~\bibnamefont {Cullen}}, \bibinfo {author} {\bibfnamefont {D.}~\bibnamefont {Magee}}, \bibinfo {author} {\bibfnamefont {R.}~\bibnamefont {Begley}}, \bibinfo {author} {\bibfnamefont {A.}~\bibnamefont {Cimatti}}, \bibinfo {author} {\bibfnamefont {C.~T.}\ \bibnamefont {Donnan}}, \bibinfo {author} {\bibfnamefont {M.~L.}\ \bibnamefont {Hamadouche}}, \bibinfo {author} {\bibfnamefont {S.~M.}\ \bibnamefont {Jewell}},\ and\ \bibinfo {author} {\bibfnamefont {S.}~\bibnamefont {Walker}},\ }\bibfield  {title} {\bibinfo {title} {A massive quiescent galaxy at redshift 4.658},\ }\href {https://doi.org/10.1038/s41586-023-06158-6} {\bibfield
  {journal} {\bibinfo  {journal} {Nature}\ }\textbf {\bibinfo {volume} {619}},\ \bibinfo {pages} {716–719} (\bibinfo {year} {2023})}\BibitemShut {NoStop}%
\bibitem [{\citenamefont {Onoue}\ \emph {et~al.}(2023)\citenamefont {Onoue}, \citenamefont {Inayoshi}, \citenamefont {Ding}, \citenamefont {Li}, \citenamefont {Li}, \citenamefont {Molina}, \citenamefont {Inoue}, \citenamefont {Jiang},\ and\ \citenamefont {Ho}}]{Onoue_2023}%
  \BibitemOpen
  \bibfield  {author} {\bibinfo {author} {\bibfnamefont {M.}~\bibnamefont {Onoue}}, \bibinfo {author} {\bibfnamefont {K.}~\bibnamefont {Inayoshi}}, \bibinfo {author} {\bibfnamefont {X.}~\bibnamefont {Ding}}, \bibinfo {author} {\bibfnamefont {W.}~\bibnamefont {Li}}, \bibinfo {author} {\bibfnamefont {Z.}~\bibnamefont {Li}}, \bibinfo {author} {\bibfnamefont {J.}~\bibnamefont {Molina}}, \bibinfo {author} {\bibfnamefont {A.~K.}\ \bibnamefont {Inoue}}, \bibinfo {author} {\bibfnamefont {L.}~\bibnamefont {Jiang}},\ and\ \bibinfo {author} {\bibfnamefont {L.~C.}\ \bibnamefont {Ho}},\ }\bibfield  {title} {\bibinfo {title} {A candidate for the least-massive black hole in the first 1.1 billion years of the universe},\ }\href {https://doi.org/10.3847/2041-8213/aca9d3} {\bibfield  {journal} {\bibinfo  {journal} {The Astrophysical Journal Letters}\ }\textbf {\bibinfo {volume} {942}},\ \bibinfo {pages} {L17} (\bibinfo {year} {2023})}\BibitemShut {NoStop}%
\bibitem [{\citenamefont {Kocevski}\ \emph {et~al.}(2023)\citenamefont {Kocevski}, \citenamefont {Onoue}, \citenamefont {Inayoshi}, \citenamefont {Trump}, \citenamefont {Haro}, \citenamefont {Grazian}, \citenamefont {Dickinson}, \citenamefont {Finkelstein}, \citenamefont {Kartaltepe}, \citenamefont {Hirschmann}, \citenamefont {Fujimoto}, \citenamefont {Juneau}, \citenamefont {Amorin}, \citenamefont {Bagley}, \citenamefont {Barro}, \citenamefont {Bell}, \citenamefont {Bisigello}, \citenamefont {Calabro}, \citenamefont {Cleri}, \citenamefont {Cooper}, \citenamefont {Ding}, \citenamefont {Grogin}, \citenamefont {Ho}, \citenamefont {Inoue}, \citenamefont {Jiang}, \citenamefont {Jones}, \citenamefont {Koekemoer}, \citenamefont {Li}, \citenamefont {Li}, \citenamefont {McGrath}, \citenamefont {Molina}, \citenamefont {Papovich}, \citenamefont {Perez-Gonzalez}, \citenamefont {Pirzkal}, \citenamefont {Wilkins}, \citenamefont {Yang},\ and\ \citenamefont {Yung}}]{kocevski2023hidden}%
  \BibitemOpen
  \bibfield  {author} {\bibinfo {author} {\bibfnamefont {D.~D.}\ \bibnamefont {Kocevski}}, \bibinfo {author} {\bibfnamefont {M.}~\bibnamefont {Onoue}}, \bibinfo {author} {\bibfnamefont {K.}~\bibnamefont {Inayoshi}}, \bibinfo {author} {\bibfnamefont {J.~R.}\ \bibnamefont {Trump}}, \bibinfo {author} {\bibfnamefont {P.~A.}\ \bibnamefont {Haro}}, \bibinfo {author} {\bibfnamefont {A.}~\bibnamefont {Grazian}}, \bibinfo {author} {\bibfnamefont {M.}~\bibnamefont {Dickinson}}, \bibinfo {author} {\bibfnamefont {S.~L.}\ \bibnamefont {Finkelstein}}, \bibinfo {author} {\bibfnamefont {J.~S.}\ \bibnamefont {Kartaltepe}}, \bibinfo {author} {\bibfnamefont {M.}~\bibnamefont {Hirschmann}}, \bibinfo {author} {\bibfnamefont {S.}~\bibnamefont {Fujimoto}}, \bibinfo {author} {\bibfnamefont {S.}~\bibnamefont {Juneau}}, \bibinfo {author} {\bibfnamefont {R.~O.}\ \bibnamefont {Amorin}}, \bibinfo {author} {\bibfnamefont {M.~B.}\ \bibnamefont {Bagley}}, \bibinfo {author} {\bibfnamefont {G.}~\bibnamefont {Barro}}, \bibinfo {author}
  {\bibfnamefont {E.~F.}\ \bibnamefont {Bell}}, \bibinfo {author} {\bibfnamefont {L.}~\bibnamefont {Bisigello}}, \bibinfo {author} {\bibfnamefont {A.}~\bibnamefont {Calabro}}, \bibinfo {author} {\bibfnamefont {N.~J.}\ \bibnamefont {Cleri}}, \bibinfo {author} {\bibfnamefont {M.~C.}\ \bibnamefont {Cooper}}, \bibinfo {author} {\bibfnamefont {X.}~\bibnamefont {Ding}}, \bibinfo {author} {\bibfnamefont {N.~A.}\ \bibnamefont {Grogin}}, \bibinfo {author} {\bibfnamefont {L.~C.}\ \bibnamefont {Ho}}, \bibinfo {author} {\bibfnamefont {A.~K.}\ \bibnamefont {Inoue}}, \bibinfo {author} {\bibfnamefont {L.}~\bibnamefont {Jiang}}, \bibinfo {author} {\bibfnamefont {B.}~\bibnamefont {Jones}}, \bibinfo {author} {\bibfnamefont {A.~M.}\ \bibnamefont {Koekemoer}}, \bibinfo {author} {\bibfnamefont {W.}~\bibnamefont {Li}}, \bibinfo {author} {\bibfnamefont {Z.}~\bibnamefont {Li}}, \bibinfo {author} {\bibfnamefont {E.~J.}\ \bibnamefont {McGrath}}, \bibinfo {author} {\bibfnamefont {J.}~\bibnamefont {Molina}}, \bibinfo {author}
  {\bibfnamefont {C.}~\bibnamefont {Papovich}}, \bibinfo {author} {\bibfnamefont {P.~G.}\ \bibnamefont {Perez-Gonzalez}}, \bibinfo {author} {\bibfnamefont {N.}~\bibnamefont {Pirzkal}}, \bibinfo {author} {\bibfnamefont {S.~M.}\ \bibnamefont {Wilkins}}, \bibinfo {author} {\bibfnamefont {G.}~\bibnamefont {Yang}},\ and\ \bibinfo {author} {\bibfnamefont {L.~Y.~A.}\ \bibnamefont {Yung}},\ }\href@noop {} {\bibinfo {title} {Hidden little monsters: Spectroscopic identification of low-mass, broad-line agn at $z>5$ with ceers}} (\bibinfo {year} {2023}),\ \Eprint {https://arxiv.org/abs/2302.00012} {arXiv:2302.00012 [astro-ph.GA]} \BibitemShut {NoStop}%
\bibitem [{\citenamefont {Fan}\ \emph {et~al.}(2022)\citenamefont {Fan}, \citenamefont {Banados},\ and\ \citenamefont {Simcoe}}]{fan2022quasarsintergalacticmediumcosmic}%
  \BibitemOpen
  \bibfield  {author} {\bibinfo {author} {\bibfnamefont {X.}~\bibnamefont {Fan}}, \bibinfo {author} {\bibfnamefont {E.}~\bibnamefont {Banados}},\ and\ \bibinfo {author} {\bibfnamefont {R.~A.}\ \bibnamefont {Simcoe}},\ }\href {https://arxiv.org/abs/2212.06907} {\bibinfo {title} {Quasars and the intergalactic medium at cosmic dawn}} (\bibinfo {year} {2022}),\ \Eprint {https://arxiv.org/abs/2212.06907} {arXiv:2212.06907 [astro-ph.GA]} \BibitemShut {NoStop}%
\bibitem [{\citenamefont {Maiolino}\ \emph {et~al.}(2023)\citenamefont {Maiolino}, \citenamefont {Scholtz}, \citenamefont {Witstok}, \citenamefont {Carniani}, \citenamefont {D'Eugenio}, \citenamefont {de~Graaff}, \citenamefont {Uebler}, \citenamefont {Tacchella}, \citenamefont {Curtis-Lake}, \citenamefont {Arribas}, \citenamefont {Bunker}, \citenamefont {Charlot}, \citenamefont {Chevallard}, \citenamefont {Curti}, \citenamefont {Looser}, \citenamefont {Maseda}, \citenamefont {Rawle}, \citenamefont {Pino}, \citenamefont {Willott}, \citenamefont {Egami}, \citenamefont {Eisenstein}, \citenamefont {Hainline}, \citenamefont {Robertson}, \citenamefont {Williams}, \citenamefont {Willmer}, \citenamefont {Baker}, \citenamefont {Boyett}, \citenamefont {DeCoursey}, \citenamefont {Fabian}, \citenamefont {Helton}, \citenamefont {Ji}, \citenamefont {Jones}, \citenamefont {Kumari}, \citenamefont {Laporte}, \citenamefont {Nelson}, \citenamefont {Perna}, \citenamefont {Sandles}, \citenamefont {Shivaei},\ and\ \citenamefont
  {Sun}}]{maiolino2023small}%
  \BibitemOpen
  \bibfield  {author} {\bibinfo {author} {\bibfnamefont {R.}~\bibnamefont {Maiolino}}, \bibinfo {author} {\bibfnamefont {J.}~\bibnamefont {Scholtz}}, \bibinfo {author} {\bibfnamefont {J.}~\bibnamefont {Witstok}}, \bibinfo {author} {\bibfnamefont {S.}~\bibnamefont {Carniani}}, \bibinfo {author} {\bibfnamefont {F.}~\bibnamefont {D'Eugenio}}, \bibinfo {author} {\bibfnamefont {A.}~\bibnamefont {de~Graaff}}, \bibinfo {author} {\bibfnamefont {H.}~\bibnamefont {Uebler}}, \bibinfo {author} {\bibfnamefont {S.}~\bibnamefont {Tacchella}}, \bibinfo {author} {\bibfnamefont {E.}~\bibnamefont {Curtis-Lake}}, \bibinfo {author} {\bibfnamefont {S.}~\bibnamefont {Arribas}}, \bibinfo {author} {\bibfnamefont {A.}~\bibnamefont {Bunker}}, \bibinfo {author} {\bibfnamefont {S.}~\bibnamefont {Charlot}}, \bibinfo {author} {\bibfnamefont {J.}~\bibnamefont {Chevallard}}, \bibinfo {author} {\bibfnamefont {M.}~\bibnamefont {Curti}}, \bibinfo {author} {\bibfnamefont {T.~J.}\ \bibnamefont {Looser}}, \bibinfo {author} {\bibfnamefont {M.~V.}\
  \bibnamefont {Maseda}}, \bibinfo {author} {\bibfnamefont {T.}~\bibnamefont {Rawle}}, \bibinfo {author} {\bibfnamefont {B.~R.~D.}\ \bibnamefont {Pino}}, \bibinfo {author} {\bibfnamefont {C.~J.}\ \bibnamefont {Willott}}, \bibinfo {author} {\bibfnamefont {E.}~\bibnamefont {Egami}}, \bibinfo {author} {\bibfnamefont {D.}~\bibnamefont {Eisenstein}}, \bibinfo {author} {\bibfnamefont {K.}~\bibnamefont {Hainline}}, \bibinfo {author} {\bibfnamefont {B.}~\bibnamefont {Robertson}}, \bibinfo {author} {\bibfnamefont {C.~C.}\ \bibnamefont {Williams}}, \bibinfo {author} {\bibfnamefont {C.~N.~A.}\ \bibnamefont {Willmer}}, \bibinfo {author} {\bibfnamefont {W.~M.}\ \bibnamefont {Baker}}, \bibinfo {author} {\bibfnamefont {K.}~\bibnamefont {Boyett}}, \bibinfo {author} {\bibfnamefont {C.}~\bibnamefont {DeCoursey}}, \bibinfo {author} {\bibfnamefont {A.~C.}\ \bibnamefont {Fabian}}, \bibinfo {author} {\bibfnamefont {J.~M.}\ \bibnamefont {Helton}}, \bibinfo {author} {\bibfnamefont {Z.}~\bibnamefont {Ji}}, \bibinfo {author}
  {\bibfnamefont {G.~C.}\ \bibnamefont {Jones}}, \bibinfo {author} {\bibfnamefont {N.}~\bibnamefont {Kumari}}, \bibinfo {author} {\bibfnamefont {N.}~\bibnamefont {Laporte}}, \bibinfo {author} {\bibfnamefont {E.}~\bibnamefont {Nelson}}, \bibinfo {author} {\bibfnamefont {M.}~\bibnamefont {Perna}}, \bibinfo {author} {\bibfnamefont {L.}~\bibnamefont {Sandles}}, \bibinfo {author} {\bibfnamefont {I.}~\bibnamefont {Shivaei}},\ and\ \bibinfo {author} {\bibfnamefont {F.}~\bibnamefont {Sun}},\ }\href@noop {} {\bibinfo {title} {A small and vigorous black hole in the early universe}} (\bibinfo {year} {2023}),\ \Eprint {https://arxiv.org/abs/2305.12492} {arXiv:2305.12492 [astro-ph.GA]} \BibitemShut {NoStop}%
\bibitem [{\citenamefont {Omukai}(2001)}]{Omukai:2000ic}%
  \BibitemOpen
  \bibfield  {author} {\bibinfo {author} {\bibfnamefont {K.}~\bibnamefont {Omukai}},\ }\bibfield  {title} {\bibinfo {title} {{Primordial star formation under far-ultraviolet radiation}},\ }\href {https://doi.org/10.1086/318296} {\bibfield  {journal} {\bibinfo  {journal} {Astrophys. J.}\ }\textbf {\bibinfo {volume} {546}},\ \bibinfo {pages} {635} (\bibinfo {year} {2001})},\ \Eprint {https://arxiv.org/abs/astro-ph/0011446} {arXiv:astro-ph/0011446} \BibitemShut {NoStop}%
\bibitem [{\citenamefont {Bromm}\ and\ \citenamefont {Loeb}(2003)}]{Bromm:2002hb}%
  \BibitemOpen
  \bibfield  {author} {\bibinfo {author} {\bibfnamefont {V.}~\bibnamefont {Bromm}}\ and\ \bibinfo {author} {\bibfnamefont {A.}~\bibnamefont {Loeb}},\ }\bibfield  {title} {\bibinfo {title} {{Formation of the first supermassive black holes}},\ }\href {https://doi.org/10.1086/377529} {\bibfield  {journal} {\bibinfo  {journal} {Astrophys. J.}\ }\textbf {\bibinfo {volume} {596}},\ \bibinfo {pages} {34} (\bibinfo {year} {2003})},\ \Eprint {https://arxiv.org/abs/astro-ph/0212400} {arXiv:astro-ph/0212400} \BibitemShut {NoStop}%
\bibitem [{\citenamefont {Biermann}\ and\ \citenamefont {Kusenko}(2006)}]{Biermann:2006bu}%
  \BibitemOpen
  \bibfield  {author} {\bibinfo {author} {\bibfnamefont {P.~L.}\ \bibnamefont {Biermann}}\ and\ \bibinfo {author} {\bibfnamefont {A.}~\bibnamefont {Kusenko}},\ }\bibfield  {title} {\bibinfo {title} {{Relic keV sterile neutrinos and reionization}},\ }\href {https://doi.org/10.1103/PhysRevLett.96.091301} {\bibfield  {journal} {\bibinfo  {journal} {Phys. Rev. Lett.}\ }\textbf {\bibinfo {volume} {96}},\ \bibinfo {pages} {091301} (\bibinfo {year} {2006})},\ \Eprint {https://arxiv.org/abs/astro-ph/0601004} {arXiv:astro-ph/0601004} \BibitemShut {NoStop}%
\bibitem [{\citenamefont {Stasielak}\ \emph {et~al.}(2007)\citenamefont {Stasielak}, \citenamefont {Biermann},\ and\ \citenamefont {Kusenko}}]{Stasielak:2006br}%
  \BibitemOpen
  \bibfield  {author} {\bibinfo {author} {\bibfnamefont {J.}~\bibnamefont {Stasielak}}, \bibinfo {author} {\bibfnamefont {P.~L.}\ \bibnamefont {Biermann}},\ and\ \bibinfo {author} {\bibfnamefont {A.}~\bibnamefont {Kusenko}},\ }\bibfield  {title} {\bibinfo {title} {{Thermal evolution of the primordial clouds in warm dark matter models with keV sterile neutrinos}},\ }\href {https://doi.org/10.1086/509066} {\bibfield  {journal} {\bibinfo  {journal} {Astrophys. J.}\ }\textbf {\bibinfo {volume} {654}},\ \bibinfo {pages} {290} (\bibinfo {year} {2007})},\ \Eprint {https://arxiv.org/abs/astro-ph/0606435} {arXiv:astro-ph/0606435} \BibitemShut {NoStop}%
\bibitem [{\citenamefont {Spolyar}\ \emph {et~al.}(2008)\citenamefont {Spolyar}, \citenamefont {Freese},\ and\ \citenamefont {Gondolo}}]{Spolyar:2007qv}%
  \BibitemOpen
  \bibfield  {author} {\bibinfo {author} {\bibfnamefont {D.}~\bibnamefont {Spolyar}}, \bibinfo {author} {\bibfnamefont {K.}~\bibnamefont {Freese}},\ and\ \bibinfo {author} {\bibfnamefont {P.}~\bibnamefont {Gondolo}},\ }\bibfield  {title} {\bibinfo {title} {{Dark matter and the first stars: a new phase of stellar evolution}},\ }\href {https://doi.org/10.1103/PhysRevLett.100.051101} {\bibfield  {journal} {\bibinfo  {journal} {Phys. Rev. Lett.}\ }\textbf {\bibinfo {volume} {100}},\ \bibinfo {pages} {051101} (\bibinfo {year} {2008})},\ \Eprint {https://arxiv.org/abs/0705.0521} {arXiv:0705.0521 [astro-ph]} \BibitemShut {NoStop}%
\bibitem [{\citenamefont {Dijkstra}\ \emph {et~al.}(2008)\citenamefont {Dijkstra}, \citenamefont {Haiman}, \citenamefont {Mesinger},\ and\ \citenamefont {Wyithe}}]{Dijkstra:2008jk}%
  \BibitemOpen
  \bibfield  {author} {\bibinfo {author} {\bibfnamefont {M.}~\bibnamefont {Dijkstra}}, \bibinfo {author} {\bibfnamefont {Z.}~\bibnamefont {Haiman}}, \bibinfo {author} {\bibfnamefont {A.}~\bibnamefont {Mesinger}},\ and\ \bibinfo {author} {\bibfnamefont {S.}~\bibnamefont {Wyithe}},\ }\bibfield  {title} {\bibinfo {title} {{Fluctuations in the High-Redshift Lyman-Werner Background: Close Halo Pairs as the Origin of Supermassive Black Holes}},\ }\href {https://doi.org/10.1111/j.1365-2966.2008.14031.x} {\bibfield  {journal} {\bibinfo  {journal} {Mon. Not. Roy. Astron. Soc.}\ }\textbf {\bibinfo {volume} {391}},\ \bibinfo {pages} {1961} (\bibinfo {year} {2008})},\ \Eprint {https://arxiv.org/abs/0810.0014} {arXiv:0810.0014 [astro-ph]} \BibitemShut {NoStop}%
\bibitem [{\citenamefont {Cyr}\ \emph {et~al.}(2022)\citenamefont {Cyr}, \citenamefont {Jiao},\ and\ \citenamefont {Brandenberger}}]{Cyr:2022urs}%
  \BibitemOpen
  \bibfield  {author} {\bibinfo {author} {\bibfnamefont {B.}~\bibnamefont {Cyr}}, \bibinfo {author} {\bibfnamefont {H.}~\bibnamefont {Jiao}},\ and\ \bibinfo {author} {\bibfnamefont {R.}~\bibnamefont {Brandenberger}},\ }\bibfield  {title} {\bibinfo {title} {{Massive black holes at high redshifts from superconducting cosmic strings}},\ }\href {https://doi.org/10.1093/mnras/stac1939} {\bibfield  {journal} {\bibinfo  {journal} {Mon. Not. Roy. Astron. Soc.}\ }\textbf {\bibinfo {volume} {517}},\ \bibinfo {pages} {2221} (\bibinfo {year} {2022})},\ \Eprint {https://arxiv.org/abs/2202.01799} {arXiv:2202.01799 [astro-ph.CO]} \BibitemShut {NoStop}%
\bibitem [{\citenamefont {Friedlander}\ \emph {et~al.}(2023)\citenamefont {Friedlander}, \citenamefont {Schon},\ and\ \citenamefont {Vincent}}]{Friedlander:2022ovf}%
  \BibitemOpen
  \bibfield  {author} {\bibinfo {author} {\bibfnamefont {A.}~\bibnamefont {Friedlander}}, \bibinfo {author} {\bibfnamefont {S.}~\bibnamefont {Schon}},\ and\ \bibinfo {author} {\bibfnamefont {A.~C.}\ \bibnamefont {Vincent}},\ }\bibfield  {title} {\bibinfo {title} {{Supermassive black hole seeds from sub-keV dark matter}},\ }\href {https://doi.org/10.1088/1475-7516/2023/06/033} {\bibfield  {journal} {\bibinfo  {journal} {JCAP}\ }\textbf {\bibinfo {volume} {06}},\ \bibinfo {pages} {033}},\ \Eprint {https://arxiv.org/abs/2212.11100} {arXiv:2212.11100 [hep-ph]} \BibitemShut {NoStop}%
\bibitem [{\citenamefont {Lu}\ \emph {et~al.}(2024{\natexlab{a}})\citenamefont {Lu}, \citenamefont {Picker},\ and\ \citenamefont {Kusenko}}]{Lu:2023xoi}%
  \BibitemOpen
  \bibfield  {author} {\bibinfo {author} {\bibfnamefont {Y.}~\bibnamefont {Lu}}, \bibinfo {author} {\bibfnamefont {Z.~S.~C.}\ \bibnamefont {Picker}},\ and\ \bibinfo {author} {\bibfnamefont {A.}~\bibnamefont {Kusenko}},\ }\bibfield  {title} {\bibinfo {title} {{High-redshift supermassive black holes from tiny black hole explosions}},\ }\href {https://doi.org/10.1103/PhysRevD.109.123016} {\bibfield  {journal} {\bibinfo  {journal} {Phys. Rev. D}\ }\textbf {\bibinfo {volume} {109}},\ \bibinfo {pages} {123016} (\bibinfo {year} {2024}{\natexlab{a}})},\ \Eprint {https://arxiv.org/abs/2312.15062} {arXiv:2312.15062 [astro-ph.GA]} \BibitemShut {NoStop}%
\bibitem [{\citenamefont {Lu}\ \emph {et~al.}(2024{\natexlab{b}})\citenamefont {Lu}, \citenamefont {Picker},\ and\ \citenamefont {Kusenko}}]{Lu:2024zwa}%
  \BibitemOpen
  \bibfield  {author} {\bibinfo {author} {\bibfnamefont {Y.}~\bibnamefont {Lu}}, \bibinfo {author} {\bibfnamefont {Z.~S.~C.}\ \bibnamefont {Picker}},\ and\ \bibinfo {author} {\bibfnamefont {A.}~\bibnamefont {Kusenko}},\ }\bibfield  {title} {\bibinfo {title} {{Direct Collapse Supermassive Black Holes from Relic Particle Decay}},\ }\href {https://doi.org/10.1103/PhysRevLett.133.091001} {\bibfield  {journal} {\bibinfo  {journal} {Phys. Rev. Lett.}\ }\textbf {\bibinfo {volume} {133}},\ \bibinfo {pages} {091001} (\bibinfo {year} {2024}{\natexlab{b}})},\ \Eprint {https://arxiv.org/abs/2404.03909} {arXiv:2404.03909 [astro-ph.GA]} \BibitemShut {NoStop}%
\bibitem [{\citenamefont {Aggarwal}\ \emph {et~al.}(2025)\citenamefont {Aggarwal}, \citenamefont {Dent}, \citenamefont {Tanedo},\ and\ \citenamefont {Xu}}]{Aggarwal:2025pit}%
  \BibitemOpen
  \bibfield  {author} {\bibinfo {author} {\bibfnamefont {Y.}~\bibnamefont {Aggarwal}}, \bibinfo {author} {\bibfnamefont {J.~B.}\ \bibnamefont {Dent}}, \bibinfo {author} {\bibfnamefont {P.}~\bibnamefont {Tanedo}},\ and\ \bibinfo {author} {\bibfnamefont {T.}~\bibnamefont {Xu}},\ }\href@noop {} {\bibinfo {title} {{Direct Collapse Black Hole Candidates from Decaying Dark Matter}}} (\bibinfo {year} {2025}),\ \Eprint {https://arxiv.org/abs/2509.25325} {arXiv:2509.25325 [hep-ph]} \BibitemShut {NoStop}%
\bibitem [{\citenamefont {Mesinger}\ \emph {et~al.}(2011)\citenamefont {Mesinger}, \citenamefont {Furlanetto},\ and\ \citenamefont {Cen}}]{mesinger_21cmFAST_2011}%
  \BibitemOpen
  \bibfield  {author} {\bibinfo {author} {\bibfnamefont {A.}~\bibnamefont {Mesinger}}, \bibinfo {author} {\bibfnamefont {S.}~\bibnamefont {Furlanetto}},\ and\ \bibinfo {author} {\bibfnamefont {R.}~\bibnamefont {Cen}},\ }\bibfield  {title} {\bibinfo {title} {{21cmFAST}: {A} {Fast}, {Semi}-{Numerical} {Simulation} of the {High}-{Redshift} 21-cm {Signal}},\ }\href {https://doi.org/10.1111/j.1365-2966.2010.17731.x} {\bibfield  {journal} {\bibinfo  {journal} {Monthly Notices of the Royal Astronomical Society}\ }\textbf {\bibinfo {volume} {411}},\ \bibinfo {pages} {955} (\bibinfo {year} {2011})},\ \bibinfo {note} {arXiv:1003.3878 [astro-ph]},\ \Eprint {https://arxiv.org/abs/1003.3878} {arXiv:1003.3878 [astro-ph.CO]} \BibitemShut {NoStop}%
\bibitem [{\citenamefont {Murray}\ \emph {et~al.}(2020)\citenamefont {Murray}, \citenamefont {Greig}, \citenamefont {Mesinger}, \citenamefont {Muñoz}, \citenamefont {Qin}, \citenamefont {Park},\ and\ \citenamefont {Watkinson}}]{Murray2020}%
  \BibitemOpen
  \bibfield  {author} {\bibinfo {author} {\bibfnamefont {S.~G.}\ \bibnamefont {Murray}}, \bibinfo {author} {\bibfnamefont {B.}~\bibnamefont {Greig}}, \bibinfo {author} {\bibfnamefont {A.}~\bibnamefont {Mesinger}}, \bibinfo {author} {\bibfnamefont {J.~B.}\ \bibnamefont {Muñoz}}, \bibinfo {author} {\bibfnamefont {Y.}~\bibnamefont {Qin}}, \bibinfo {author} {\bibfnamefont {J.}~\bibnamefont {Park}},\ and\ \bibinfo {author} {\bibfnamefont {C.~A.}\ \bibnamefont {Watkinson}},\ }\bibfield  {title} {\bibinfo {title} {21cmfast v3: A python-integrated c code for generating 3d realizations of the cosmic 21cm signal.},\ }\href {https://doi.org/10.21105/joss.02582} {\bibfield  {journal} {\bibinfo  {journal} {Journal of Open Source Software}\ }\textbf {\bibinfo {volume} {5}},\ \bibinfo {pages} {2582} (\bibinfo {year} {2020})},\ \Eprint {https://arxiv.org/abs/2010.15121} {arXiv:2010.15121 [astro-ph.IM]} \BibitemShut {NoStop}%
\bibitem [{\citenamefont {Pritchard}\ and\ \citenamefont {Furlanetto}(2006)}]{Pritchard:2005an}%
  \BibitemOpen
  \bibfield  {author} {\bibinfo {author} {\bibfnamefont {J.~R.}\ \bibnamefont {Pritchard}}\ and\ \bibinfo {author} {\bibfnamefont {S.~R.}\ \bibnamefont {Furlanetto}},\ }\bibfield  {title} {\bibinfo {title} {{Descending from on high: lyman series cascades and spin-kinetic temperature coupling in the 21 cm line}},\ }\href {https://doi.org/10.1111/j.1365-2966.2006.10028.x} {\bibfield  {journal} {\bibinfo  {journal} {Mon. Not. Roy. Astron. Soc.}\ }\textbf {\bibinfo {volume} {367}},\ \bibinfo {pages} {1057} (\bibinfo {year} {2006})},\ \Eprint {https://arxiv.org/abs/astro-ph/0508381} {arXiv:astro-ph/0508381} \BibitemShut {NoStop}%
\bibitem [{\citenamefont {Munshi}\ \emph {et~al.}(2025)\citenamefont {Munshi} \emph {et~al.}}]{Munshi:2025hgk}%
  \BibitemOpen
  \bibfield  {author} {\bibinfo {author} {\bibfnamefont {S.}~\bibnamefont {Munshi}} \emph {et~al.},\ }\bibfield  {title} {\bibinfo {title} {{Improved upper limits on the 21-cm signal power spectrum at $z=17.0$ and $z=20.3$ from an optimal field observed with NenuFAR}},\ }\href {https://doi.org/10.1093/mnras/staf1386} {\bibfield  {journal} {\bibinfo  {journal} {Mon. Not. Roy. Astron. Soc.}\ }\textbf {\bibinfo {volume} {2785}},\ \bibinfo {pages} {2807} (\bibinfo {year} {2025})},\ \Eprint {https://arxiv.org/abs/2507.10533} {arXiv:2507.10533 [astro-ph.CO]} \BibitemShut {NoStop}%
\bibitem [{\citenamefont {Jester}\ and\ \citenamefont {Falcke}(2009)}]{Jester:2009dw}%
  \BibitemOpen
  \bibfield  {author} {\bibinfo {author} {\bibfnamefont {S.}~\bibnamefont {Jester}}\ and\ \bibinfo {author} {\bibfnamefont {H.}~\bibnamefont {Falcke}},\ }\bibfield  {title} {\bibinfo {title} {{Science with a lunar low-frequency array: from the dark ages of the Universe to nearby exoplanets}},\ }\href {https://doi.org/10.1016/j.newar.2009.02.001} {\bibfield  {journal} {\bibinfo  {journal} {New Astron. Rev.}\ }\textbf {\bibinfo {volume} {53}},\ \bibinfo {pages} {1} (\bibinfo {year} {2009})},\ \Eprint {https://arxiv.org/abs/0902.0493} {arXiv:0902.0493 [astro-ph.CO]} \BibitemShut {NoStop}%
\bibitem [{\citenamefont {{Burns}}\ \emph {et~al.}(2019)\citenamefont {{Burns}} \emph {et~al.}}]{Farside:2019}%
  \BibitemOpen
  \bibfield  {author} {\bibinfo {author} {\bibfnamefont {J.}~\bibnamefont {{Burns}}} \emph {et~al.},\ }\bibfield  {title} {\bibinfo {title} {{FARSIDE: A low radio frequency interferometric array on the lunar farside}},\ }in\ \href {https://doi.org/10.48550/arXiv.1907.05407} {\emph {\bibinfo {booktitle} {Bulletin of the American Astronomical Society}}},\ Vol.~\bibinfo {volume} {51}\ (\bibinfo {year} {2019})\ p.\ \bibinfo {pages} {178},\ \Eprint {https://arxiv.org/abs/1907.05407} {arXiv:1907.05407 [astro-ph.IM]} \BibitemShut {NoStop}%
\bibitem [{\citenamefont {{Bale}}\ \emph {et~al.}(2023)\citenamefont {{Bale}}, \citenamefont {{Bassett}}, \citenamefont {{Burns}}, \citenamefont {{Dorigo Jones}}, \citenamefont {{Goetz}}, \citenamefont {{Hellum-Bye}}, \citenamefont {{Hermann}}, \citenamefont {{Hibbard}}, \citenamefont {{Maksimovic}}, \citenamefont {{McLean}}, \citenamefont {{Monsalve}}, \citenamefont {{O'Connor}}, \citenamefont {{Parsons}}, \citenamefont {{Pulupa}}, \citenamefont {{Pund}}, \citenamefont {{Rapetti}}, \citenamefont {{Rotermund}}, \citenamefont {{Saliwanchik}}, \citenamefont {{Slosar}}, \citenamefont {{Sundkvist}},\ and\ \citenamefont {{Suzuki}}}]{LuseeNight:2023}%
  \BibitemOpen
  \bibfield  {author} {\bibinfo {author} {\bibfnamefont {S.~D.}\ \bibnamefont {{Bale}}}, \bibinfo {author} {\bibfnamefont {N.}~\bibnamefont {{Bassett}}}, \bibinfo {author} {\bibfnamefont {J.~O.}\ \bibnamefont {{Burns}}}, \bibinfo {author} {\bibfnamefont {J.}~\bibnamefont {{Dorigo Jones}}}, \bibinfo {author} {\bibfnamefont {K.}~\bibnamefont {{Goetz}}}, \bibinfo {author} {\bibfnamefont {C.}~\bibnamefont {{Hellum-Bye}}}, \bibinfo {author} {\bibfnamefont {S.}~\bibnamefont {{Hermann}}}, \bibinfo {author} {\bibfnamefont {J.}~\bibnamefont {{Hibbard}}}, \bibinfo {author} {\bibfnamefont {M.}~\bibnamefont {{Maksimovic}}}, \bibinfo {author} {\bibfnamefont {R.}~\bibnamefont {{McLean}}}, \bibinfo {author} {\bibfnamefont {R.}~\bibnamefont {{Monsalve}}}, \bibinfo {author} {\bibfnamefont {P.}~\bibnamefont {{O'Connor}}}, \bibinfo {author} {\bibfnamefont {A.}~\bibnamefont {{Parsons}}}, \bibinfo {author} {\bibfnamefont {M.}~\bibnamefont {{Pulupa}}}, \bibinfo {author} {\bibfnamefont {R.}~\bibnamefont {{Pund}}}, \bibinfo {author}
  {\bibfnamefont {D.}~\bibnamefont {{Rapetti}}}, \bibinfo {author} {\bibfnamefont {K.~M.}\ \bibnamefont {{Rotermund}}}, \bibinfo {author} {\bibfnamefont {B.}~\bibnamefont {{Saliwanchik}}}, \bibinfo {author} {\bibfnamefont {A.}~\bibnamefont {{Slosar}}}, \bibinfo {author} {\bibfnamefont {D.}~\bibnamefont {{Sundkvist}}},\ and\ \bibinfo {author} {\bibfnamefont {A.}~\bibnamefont {{Suzuki}}},\ }\bibfield  {title} {\bibinfo {title} {{LuSEE 'Night': The Lunar Surface Electromagnetics Experiment}},\ }\href {https://doi.org/10.48550/arXiv.2301.10345} {\bibfield  {journal} {\bibinfo  {journal} {arXiv e-prints}\ ,\ \bibinfo {eid} {arXiv:2301.10345}} (\bibinfo {year} {2023})},\ \Eprint {https://arxiv.org/abs/2301.10345} {arXiv:2301.10345 [astro-ph.IM]} \BibitemShut {NoStop}%
\bibitem [{\citenamefont {Liu}\ and\ \citenamefont {Shaw}(2020)}]{Liu:2019awk}%
  \BibitemOpen
  \bibfield  {author} {\bibinfo {author} {\bibfnamefont {A.}~\bibnamefont {Liu}}\ and\ \bibinfo {author} {\bibfnamefont {J.~R.}\ \bibnamefont {Shaw}},\ }\bibfield  {title} {\bibinfo {title} {{Data analysis for precision 21 cm cosmology}},\ }\href {https://doi.org/10.1088/1538-3873/ab5bfd} {\bibfield  {journal} {\bibinfo  {journal} {Publ. Astron. Soc. Pac.}\ }\textbf {\bibinfo {volume} {132}},\ \bibinfo {pages} {062001} (\bibinfo {year} {2020})},\ \Eprint {https://arxiv.org/abs/1907.08211} {arXiv:1907.08211 [astro-ph.IM]} \BibitemShut {NoStop}%
\bibitem [{\citenamefont {Silk}(2025)}]{Silk:2025znp}%
  \BibitemOpen
  \bibfield  {author} {\bibinfo {author} {\bibfnamefont {J.}~\bibnamefont {Silk}},\ }\bibfield  {title} {\bibinfo {title} {{The limits of cosmology}},\ }\href@noop {} {\bibfield  {journal} {\bibinfo  {journal} {Gen. Rel. Grav.}\ }\textbf {\bibinfo {volume} {57}},\ \bibinfo {pages} {127} (\bibinfo {year} {2025})},\ \Eprint {https://arxiv.org/abs/2509.08066} {arXiv:2509.08066 [astro-ph.CO]} \BibitemShut {NoStop}%
\bibitem [{\citenamefont {Bowman}\ \emph {et~al.}(2018)\citenamefont {Bowman}, \citenamefont {Rogers}, \citenamefont {Monsalve}, \citenamefont {Mozdzen},\ and\ \citenamefont {Mahesh}}]{Bowman:2018yin}%
  \BibitemOpen
  \bibfield  {author} {\bibinfo {author} {\bibfnamefont {J.~D.}\ \bibnamefont {Bowman}}, \bibinfo {author} {\bibfnamefont {A.~E.~E.}\ \bibnamefont {Rogers}}, \bibinfo {author} {\bibfnamefont {R.~A.}\ \bibnamefont {Monsalve}}, \bibinfo {author} {\bibfnamefont {T.~J.}\ \bibnamefont {Mozdzen}},\ and\ \bibinfo {author} {\bibfnamefont {N.}~\bibnamefont {Mahesh}},\ }\bibfield  {title} {\bibinfo {title} {{An absorption profile centred at 78 megahertz in the sky-averaged spectrum}},\ }\href {https://doi.org/10.1038/nature25792} {\bibfield  {journal} {\bibinfo  {journal} {Nature}\ }\textbf {\bibinfo {volume} {555}},\ \bibinfo {pages} {67} (\bibinfo {year} {2018})},\ \Eprint {https://arxiv.org/abs/1810.05912} {arXiv:1810.05912 [astro-ph.CO]} \BibitemShut {NoStop}%
\bibitem [{\citenamefont {Barkana}(2018)}]{Barkana:2018lgd}%
  \BibitemOpen
  \bibfield  {author} {\bibinfo {author} {\bibfnamefont {R.}~\bibnamefont {Barkana}},\ }\bibfield  {title} {\bibinfo {title} {{Possible interaction between baryons and dark-matter particles revealed by the first stars}},\ }\href {https://doi.org/10.1038/nature25791} {\bibfield  {journal} {\bibinfo  {journal} {Nature}\ }\textbf {\bibinfo {volume} {555}},\ \bibinfo {pages} {71} (\bibinfo {year} {2018})},\ \Eprint {https://arxiv.org/abs/1803.06698} {arXiv:1803.06698 [astro-ph.CO]} \BibitemShut {NoStop}%
\bibitem [{\citenamefont {Kovetz}\ \emph {et~al.}(2018)\citenamefont {Kovetz}, \citenamefont {Poulin}, \citenamefont {Gluscevic}, \citenamefont {Boddy}, \citenamefont {Barkana},\ and\ \citenamefont {Kamionkowski}}]{Kovetz:2018zan}%
  \BibitemOpen
  \bibfield  {author} {\bibinfo {author} {\bibfnamefont {E.~D.}\ \bibnamefont {Kovetz}}, \bibinfo {author} {\bibfnamefont {V.}~\bibnamefont {Poulin}}, \bibinfo {author} {\bibfnamefont {V.}~\bibnamefont {Gluscevic}}, \bibinfo {author} {\bibfnamefont {K.~K.}\ \bibnamefont {Boddy}}, \bibinfo {author} {\bibfnamefont {R.}~\bibnamefont {Barkana}},\ and\ \bibinfo {author} {\bibfnamefont {M.}~\bibnamefont {Kamionkowski}},\ }\bibfield  {title} {\bibinfo {title} {{Tighter limits on dark matter explanations of the anomalous EDGES 21 cm signal}},\ }\href {https://doi.org/10.1103/PhysRevD.98.103529} {\bibfield  {journal} {\bibinfo  {journal} {Phys. Rev. D}\ }\textbf {\bibinfo {volume} {98}},\ \bibinfo {pages} {103529} (\bibinfo {year} {2018})},\ \Eprint {https://arxiv.org/abs/1807.11482} {arXiv:1807.11482 [astro-ph.CO]} \BibitemShut {NoStop}%
\bibitem [{\citenamefont {Singh}\ \emph {et~al.}(2022)\citenamefont {Singh}, \citenamefont {Nambissan~T.}, \citenamefont {Subrahmanyan}, \citenamefont {Udaya~Shankar}, \citenamefont {Girish}, \citenamefont {Raghunathan}, \citenamefont {Somashekar}, \citenamefont {Srivani},\ and\ \citenamefont {Sathyanarayana~Rao}}]{Singh:2021mxo}%
  \BibitemOpen
  \bibfield  {author} {\bibinfo {author} {\bibfnamefont {S.}~\bibnamefont {Singh}}, \bibinfo {author} {\bibfnamefont {J.}~\bibnamefont {Nambissan~T.}}, \bibinfo {author} {\bibfnamefont {R.}~\bibnamefont {Subrahmanyan}}, \bibinfo {author} {\bibfnamefont {N.}~\bibnamefont {Udaya~Shankar}}, \bibinfo {author} {\bibfnamefont {B.~S.}\ \bibnamefont {Girish}}, \bibinfo {author} {\bibfnamefont {A.}~\bibnamefont {Raghunathan}}, \bibinfo {author} {\bibfnamefont {R.}~\bibnamefont {Somashekar}}, \bibinfo {author} {\bibfnamefont {K.~S.}\ \bibnamefont {Srivani}},\ and\ \bibinfo {author} {\bibfnamefont {M.}~\bibnamefont {Sathyanarayana~Rao}},\ }\bibfield  {title} {\bibinfo {title} {{On the detection of a cosmic dawn signal in the radio background}},\ }\href {https://doi.org/10.1038/s41550-022-01610-5} {\bibfield  {journal} {\bibinfo  {journal} {Nature Astron.}\ }\textbf {\bibinfo {volume} {6}},\ \bibinfo {pages} {607} (\bibinfo {year} {2022})},\ \Eprint {https://arxiv.org/abs/2112.06778} {arXiv:2112.06778 [astro-ph.CO]}
  \BibitemShut {NoStop}%
\bibitem [{\citenamefont {Pober}\ \emph {et~al.}(2013)\citenamefont {Pober}, \citenamefont {Parsons}, \citenamefont {DeBoer}, \citenamefont {McDonald}, \citenamefont {McQuinn}, \citenamefont {Aguirre}, \citenamefont {Ali}, \citenamefont {Bradley}, \citenamefont {Chang},\ and\ \citenamefont {Morales}}]{Pober:2012zz}%
  \BibitemOpen
  \bibfield  {author} {\bibinfo {author} {\bibfnamefont {J.~C.}\ \bibnamefont {Pober}}, \bibinfo {author} {\bibfnamefont {A.~R.}\ \bibnamefont {Parsons}}, \bibinfo {author} {\bibfnamefont {D.~R.}\ \bibnamefont {DeBoer}}, \bibinfo {author} {\bibfnamefont {P.}~\bibnamefont {McDonald}}, \bibinfo {author} {\bibfnamefont {M.}~\bibnamefont {McQuinn}}, \bibinfo {author} {\bibfnamefont {J.~E.}\ \bibnamefont {Aguirre}}, \bibinfo {author} {\bibfnamefont {Z.}~\bibnamefont {Ali}}, \bibinfo {author} {\bibfnamefont {R.~F.}\ \bibnamefont {Bradley}}, \bibinfo {author} {\bibfnamefont {T.-C.}\ \bibnamefont {Chang}},\ and\ \bibinfo {author} {\bibfnamefont {M.~F.}\ \bibnamefont {Morales}},\ }\bibfield  {title} {\bibinfo {title} {{The Baryon Acoustic Oscillation Broadband and Broad-beam Array: Design Overview and Sensitivity Forecasts}},\ }\href {https://doi.org/10.1088/0004-6256/145/3/65} {\bibfield  {journal} {\bibinfo  {journal} {Astron. J.}\ }\textbf {\bibinfo {volume} {145}},\ \bibinfo {pages} {65} (\bibinfo {year}
  {2013})},\ \Eprint {https://arxiv.org/abs/1210.2413} {arXiv:1210.2413 [astro-ph.CO]} \BibitemShut {NoStop}%
\bibitem [{\citenamefont {Pober}\ \emph {et~al.}(2014)\citenamefont {Pober} \emph {et~al.}}]{Pober:2013jna}%
  \BibitemOpen
  \bibfield  {author} {\bibinfo {author} {\bibfnamefont {J.~C.}\ \bibnamefont {Pober}} \emph {et~al.},\ }\bibfield  {title} {\bibinfo {title} {{What Next-Generation 21 cm Power Spectrum Measurements Can Teach Us About the Epoch of Reionization}},\ }\href {https://doi.org/10.1088/0004-637X/782/2/66} {\bibfield  {journal} {\bibinfo  {journal} {Astrophys. J.}\ }\textbf {\bibinfo {volume} {782}},\ \bibinfo {pages} {66} (\bibinfo {year} {2014})},\ \Eprint {https://arxiv.org/abs/1310.7031} {arXiv:1310.7031 [astro-ph.CO]} \BibitemShut {NoStop}%
\bibitem [{\citenamefont {Murray}\ \emph {et~al.}(2024)\citenamefont {Murray}, \citenamefont {Pober},\ and\ \citenamefont {Kolopanis}}]{Murray:2024the}%
  \BibitemOpen
  \bibfield  {author} {\bibinfo {author} {\bibfnamefont {S.~G.}\ \bibnamefont {Murray}}, \bibinfo {author} {\bibfnamefont {J.}~\bibnamefont {Pober}},\ and\ \bibinfo {author} {\bibfnamefont {M.}~\bibnamefont {Kolopanis}},\ }\bibfield  {title} {\bibinfo {title} {{21cmSense v2: a modular, open-source 21cm sensitivity calculator}},\ }\href {https://doi.org/10.21105/joss.06501} {\bibfield  {journal} {\bibinfo  {journal} {J. Open Source Softw.}\ }\textbf {\bibinfo {volume} {9}},\ \bibinfo {pages} {6501} (\bibinfo {year} {2024})},\ \Eprint {https://arxiv.org/abs/2406.02415} {arXiv:2406.02415 [astro-ph.CO]} \BibitemShut {NoStop}%
\bibitem [{\citenamefont {Mu{\~n}oz}\ \emph {et~al.}(2022)\citenamefont {Mu{\~n}oz}, \citenamefont {Qin}, \citenamefont {Mesinger}, \citenamefont {Murray}, \citenamefont {Greig},\ and\ \citenamefont {Mason}}]{Munoz:2021psm}%
  \BibitemOpen
  \bibfield  {author} {\bibinfo {author} {\bibfnamefont {J.~B.}\ \bibnamefont {Mu{\~n}oz}}, \bibinfo {author} {\bibfnamefont {Y.}~\bibnamefont {Qin}}, \bibinfo {author} {\bibfnamefont {A.}~\bibnamefont {Mesinger}}, \bibinfo {author} {\bibfnamefont {S.~G.}\ \bibnamefont {Murray}}, \bibinfo {author} {\bibfnamefont {B.}~\bibnamefont {Greig}},\ and\ \bibinfo {author} {\bibfnamefont {C.}~\bibnamefont {Mason}},\ }\bibfield  {title} {\bibinfo {title} {{The impact of the first galaxies on cosmic dawn and reionization}},\ }\href {https://doi.org/10.1093/mnras/stac185} {\bibfield  {journal} {\bibinfo  {journal} {Mon. Not. Roy. Astron. Soc.}\ }\textbf {\bibinfo {volume} {511}},\ \bibinfo {pages} {3657} (\bibinfo {year} {2022})},\ \Eprint {https://arxiv.org/abs/2110.13919} {arXiv:2110.13919 [astro-ph.CO]} \BibitemShut {NoStop}%
\bibitem [{\citenamefont {Xu}\ \emph {et~al.}(2024)\citenamefont {Xu}, \citenamefont {Qin},\ and\ \citenamefont {Slatyer}}]{xu_cmb_2024}%
  \BibitemOpen
  \bibfield  {author} {\bibinfo {author} {\bibfnamefont {C.}~\bibnamefont {Xu}}, \bibinfo {author} {\bibfnamefont {W.}~\bibnamefont {Qin}},\ and\ \bibinfo {author} {\bibfnamefont {T.~R.}\ \bibnamefont {Slatyer}},\ }\bibfield  {title} {\bibinfo {title} {{CMB} limits on decaying dark matter beyond the ionization threshold},\ }\href {https://doi.org/10.1103/PhysRevD.110.123529} {\bibfield  {journal} {\bibinfo  {journal} {Physical Review D}\ }\textbf {\bibinfo {volume} {110}},\ \bibinfo {pages} {123529} (\bibinfo {year} {2024})}\BibitemShut {NoStop}%
\bibitem [{\citenamefont {Todarello}\ and\ \citenamefont {Regis}(2025)}]{Todarello:2024qci}%
  \BibitemOpen
  \bibfield  {author} {\bibinfo {author} {\bibfnamefont {E.}~\bibnamefont {Todarello}}\ and\ \bibinfo {author} {\bibfnamefont {M.}~\bibnamefont {Regis}},\ }\bibfield  {title} {\bibinfo {title} {{Bounds on axions-like particles shining in the ultra-violet}},\ }\href {https://doi.org/10.1088/1475-7516/2025/05/070} {\bibfield  {journal} {\bibinfo  {journal} {JCAP}\ }\textbf {\bibinfo {volume} {05}},\ \bibinfo {pages} {070}},\ \Eprint {https://arxiv.org/abs/2412.02543} {arXiv:2412.02543 [hep-ph]} \BibitemShut {NoStop}%
\bibitem [{\citenamefont {Wadekar}\ and\ \citenamefont {Wang}(2022)}]{Wadekar:2021qae}%
  \BibitemOpen
  \bibfield  {author} {\bibinfo {author} {\bibfnamefont {D.}~\bibnamefont {Wadekar}}\ and\ \bibinfo {author} {\bibfnamefont {Z.}~\bibnamefont {Wang}},\ }\bibfield  {title} {\bibinfo {title} {{Strong constraints on decay and annihilation of dark matter from heating of gas-rich dwarf galaxies}},\ }\href {https://doi.org/10.1103/PhysRevD.106.075007} {\bibfield  {journal} {\bibinfo  {journal} {Phys. Rev. D}\ }\textbf {\bibinfo {volume} {106}},\ \bibinfo {pages} {075007} (\bibinfo {year} {2022})},\ \Eprint {https://arxiv.org/abs/2111.08025} {arXiv:2111.08025 [hep-ph]} \BibitemShut {NoStop}%
\bibitem [{\citenamefont {Cram{\'e}r}(1946)}]{Cramer:1946}%
  \BibitemOpen
  \bibfield  {author} {\bibinfo {author} {\bibfnamefont {H.}~\bibnamefont {Cram{\'e}r}},\ }\href {https://books.google.nl/books?id=_db1jwEACAAJ} {\emph {\bibinfo {title} {Mathematical Methods of Statistics}}},\ Goldstine Printed Materials\ (\bibinfo  {publisher} {Princeton University Press},\ \bibinfo {year} {1946})\BibitemShut {NoStop}%
\bibitem [{\citenamefont {Rao}(1992)}]{Rao1992}%
  \BibitemOpen
  \bibfield  {author} {\bibinfo {author} {\bibfnamefont {C.~R.}\ \bibnamefont {Rao}},\ }\bibinfo {title} {Information and the accuracy attainable in the estimation of statistical parameters},\ in\ \href {https://doi.org/10.1007/978-1-4612-0919-5_16} {\emph {\bibinfo {booktitle} {Breakthroughs in Statistics: Foundations and Basic Theory}}},\ \bibinfo {editor} {edited by\ \bibinfo {editor} {\bibfnamefont {S.}~\bibnamefont {Kotz}}\ and\ \bibinfo {editor} {\bibfnamefont {N.~L.}\ \bibnamefont {Johnson}}}\ (\bibinfo  {publisher} {Springer New York},\ \bibinfo {address} {New York, NY},\ \bibinfo {year} {1992})\ pp.\ \bibinfo {pages} {235--247}\BibitemShut {NoStop}%
\bibitem [{\citenamefont {Lopez-Honorez}(2024)}]{Lopez-Honorez:2024ant}%
  \BibitemOpen
  \bibfield  {author} {\bibinfo {author} {\bibfnamefont {L.}~\bibnamefont {Lopez-Honorez}},\ }\bibfield  {title} {\bibinfo {title} {{Future 21cm constraints on dark matter energy injection: Application to ALPs}},\ }\href {https://doi.org/10.22323/1.463.0111} {\bibfield  {journal} {\bibinfo  {journal} {PoS}\ }\textbf {\bibinfo {volume} {CORFU2023}},\ \bibinfo {pages} {111} (\bibinfo {year} {2024})},\ \Eprint {https://arxiv.org/abs/2406.15378} {arXiv:2406.15378 [astro-ph.CO]} \BibitemShut {NoStop}%
\bibitem [{\citenamefont {Capozzi}\ \emph {et~al.}(2023)\citenamefont {Capozzi}, \citenamefont {Ferreira}, \citenamefont {Lopez-Honorez},\ and\ \citenamefont {Mena}}]{Capozzi:2023xie}%
  \BibitemOpen
  \bibfield  {author} {\bibinfo {author} {\bibfnamefont {F.}~\bibnamefont {Capozzi}}, \bibinfo {author} {\bibfnamefont {R.~Z.}\ \bibnamefont {Ferreira}}, \bibinfo {author} {\bibfnamefont {L.}~\bibnamefont {Lopez-Honorez}},\ and\ \bibinfo {author} {\bibfnamefont {O.}~\bibnamefont {Mena}},\ }\bibfield  {title} {\bibinfo {title} {{CMB and Lyman-{\ensuremath{\alpha}} constraints on dark matter decays to photons}},\ }\href {https://doi.org/10.1088/1475-7516/2023/06/060} {\bibfield  {journal} {\bibinfo  {journal} {JCAP}\ }\textbf {\bibinfo {volume} {06}},\ \bibinfo {pages} {060}},\ \Eprint {https://arxiv.org/abs/2303.07426} {arXiv:2303.07426 [astro-ph.CO]} \BibitemShut {NoStop}%
\bibitem [{\citenamefont {Bendavid}\ \emph {et~al.}(2025)\citenamefont {Bendavid}, \citenamefont {D'Alfonso}, \citenamefont {Eysermans}, \citenamefont {Freer}, \citenamefont {Goncharov}, \citenamefont {Heine}, \citenamefont {Lavezzo}, \citenamefont {Moore}, \citenamefont {Paus}, \citenamefont {Shen}, \citenamefont {Walter},\ and\ \citenamefont {Wang}}]{bendavid2025submitphysicsanalysisfacility}%
  \BibitemOpen
  \bibfield  {author} {\bibinfo {author} {\bibfnamefont {J.}~\bibnamefont {Bendavid}}, \bibinfo {author} {\bibfnamefont {M.}~\bibnamefont {D'Alfonso}}, \bibinfo {author} {\bibfnamefont {J.}~\bibnamefont {Eysermans}}, \bibinfo {author} {\bibfnamefont {C.}~\bibnamefont {Freer}}, \bibinfo {author} {\bibfnamefont {M.}~\bibnamefont {Goncharov}}, \bibinfo {author} {\bibfnamefont {M.}~\bibnamefont {Heine}}, \bibinfo {author} {\bibfnamefont {L.}~\bibnamefont {Lavezzo}}, \bibinfo {author} {\bibfnamefont {M.}~\bibnamefont {Moore}}, \bibinfo {author} {\bibfnamefont {C.}~\bibnamefont {Paus}}, \bibinfo {author} {\bibfnamefont {X.}~\bibnamefont {Shen}}, \bibinfo {author} {\bibfnamefont {D.}~\bibnamefont {Walter}},\ and\ \bibinfo {author} {\bibfnamefont {Z.}~\bibnamefont {Wang}},\ }\href {https://arxiv.org/abs/2506.01958} {\bibinfo {title} {Submit: A physics analysis facility at mit}} (\bibinfo {year} {2025}),\ \Eprint {https://arxiv.org/abs/2506.01958} {arXiv:2506.01958 [cs.DC]} \BibitemShut {NoStop}%
\bibitem [{\citenamefont {Harris}\ \emph {et~al.}(2020)\citenamefont {Harris} \emph {et~al.}}]{Harris:2020xlr}%
  \BibitemOpen
  \bibfield  {author} {\bibinfo {author} {\bibfnamefont {C.~R.}\ \bibnamefont {Harris}} \emph {et~al.},\ }\bibfield  {title} {\bibinfo {title} {{Array programming with NumPy}},\ }\href {https://doi.org/10.1038/s41586-020-2649-2} {\bibfield  {journal} {\bibinfo  {journal} {Nature}\ }\textbf {\bibinfo {volume} {585}},\ \bibinfo {pages} {357} (\bibinfo {year} {2020})},\ \Eprint {https://arxiv.org/abs/2006.10256} {arXiv:2006.10256 [cs.MS]} \BibitemShut {NoStop}%
\bibitem [{\citenamefont {Virtanen}\ \emph {et~al.}(2020)\citenamefont {Virtanen} \emph {et~al.}}]{Virtanen:2019joe}%
  \BibitemOpen
  \bibfield  {author} {\bibinfo {author} {\bibfnamefont {P.}~\bibnamefont {Virtanen}} \emph {et~al.},\ }\bibfield  {title} {\bibinfo {title} {{SciPy 1.0--Fundamental algorithms for scientific computing in Python}},\ }\href {https://doi.org/10.1038/s41592-019-0686-2} {\bibfield  {journal} {\bibinfo  {journal} {Nature Meth.}\ }\textbf {\bibinfo {volume} {17}},\ \bibinfo {pages} {261} (\bibinfo {year} {2020})},\ \Eprint {https://arxiv.org/abs/1907.10121} {arXiv:1907.10121 [cs.MS]} \BibitemShut {NoStop}%
\bibitem [{\citenamefont {Hunter}(2007)}]{Hunter:2007}%
  \BibitemOpen
  \bibfield  {author} {\bibinfo {author} {\bibfnamefont {J.~D.}\ \bibnamefont {Hunter}},\ }\bibfield  {title} {\bibinfo {title} {Matplotlib: A 2d graphics environment},\ }\href {https://doi.org/10.1109/MCSE.2007.55} {\bibfield  {journal} {\bibinfo  {journal} {Computing in Science \& Engineering}\ }\textbf {\bibinfo {volume} {9}},\ \bibinfo {pages} {90} (\bibinfo {year} {2007})}\BibitemShut {NoStop}%
\bibitem [{\citenamefont {Rohatgi}()}]{WebPlotDigitizer}%
  \BibitemOpen
  \bibfield  {author} {\bibinfo {author} {\bibfnamefont {A.}~\bibnamefont {Rohatgi}},\ }\href {https://automeris.io} {\bibinfo {title} {Webplotdigitizer}}\BibitemShut {NoStop}%
\bibitem [{\citenamefont {Mu{\~n}oz}(2023)}]{Munoz:2023kkg}%
  \BibitemOpen
  \bibfield  {author} {\bibinfo {author} {\bibfnamefont {J.~B.}\ \bibnamefont {Mu{\~n}oz}},\ }\bibfield  {title} {\bibinfo {title} {{An effective model for the cosmic-dawn 21-cm signal}},\ }\href {https://doi.org/10.1093/mnras/stad1512} {\bibfield  {journal} {\bibinfo  {journal} {Mon. Not. Roy. Astron. Soc.}\ }\textbf {\bibinfo {volume} {523}},\ \bibinfo {pages} {2587} (\bibinfo {year} {2023})},\ \Eprint {https://arxiv.org/abs/2302.08506} {arXiv:2302.08506 [astro-ph.CO]} \BibitemShut {NoStop}%
\bibitem [{\citenamefont {Mason}\ \emph {et~al.}(2023)\citenamefont {Mason}, \citenamefont {Mu{\~n}oz}, \citenamefont {Greig}, \citenamefont {Mesinger},\ and\ \citenamefont {Park}}]{Mason:2022obt}%
  \BibitemOpen
  \bibfield  {author} {\bibinfo {author} {\bibfnamefont {C.~A.}\ \bibnamefont {Mason}}, \bibinfo {author} {\bibfnamefont {J.~B.}\ \bibnamefont {Mu{\~n}oz}}, \bibinfo {author} {\bibfnamefont {B.}~\bibnamefont {Greig}}, \bibinfo {author} {\bibfnamefont {A.}~\bibnamefont {Mesinger}},\ and\ \bibinfo {author} {\bibfnamefont {J.}~\bibnamefont {Park}},\ }\bibfield  {title} {\bibinfo {title} {{21cmfish: Fisher-matrix framework for fast parameter forecasts from the cosmic 21-cm signal}},\ }\href {https://doi.org/10.1093/mnras/stad2145} {\bibfield  {journal} {\bibinfo  {journal} {Mon. Not. Roy. Astron. Soc.}\ }\textbf {\bibinfo {volume} {524}},\ \bibinfo {pages} {4711} (\bibinfo {year} {2023})},\ \Eprint {https://arxiv.org/abs/2212.09797} {arXiv:2212.09797 [astro-ph.CO]} \BibitemShut {NoStop}%
\bibitem [{\citenamefont {Seethapuram~Sridhar}\ \emph {et~al.}(2025)\citenamefont {Seethapuram~Sridhar}, \citenamefont {Williams}, \citenamefont {Price}, \citenamefont {Breen},\ and\ \citenamefont {Ball}}]{SKA_MEMO:2025}%
  \BibitemOpen
  \bibfield  {author} {\bibinfo {author} {\bibfnamefont {S.}~\bibnamefont {Seethapuram~Sridhar}}, \bibinfo {author} {\bibfnamefont {W.}~\bibnamefont {Williams}}, \bibinfo {author} {\bibfnamefont {D.}~\bibnamefont {Price}}, \bibinfo {author} {\bibfnamefont {s.}~\bibnamefont {Breen}},\ and\ \bibinfo {author} {\bibfnamefont {L.}~\bibnamefont {Ball}},\ }\href {https://doi.org/10.5281/zenodo.16951088} {\bibinfo {title} {Ska low and mid subarray templates}},\ \bibinfo {howpublished} {SKAO} (\bibinfo {year} {2025})\BibitemShut {NoStop}%
\bibitem [{\citenamefont {{Mozdzen}}\ \emph {et~al.}(2017)\citenamefont {{Mozdzen}}, \citenamefont {{Bowman}}, \citenamefont {{Monsalve}},\ and\ \citenamefont {{Rogers}}}]{Mozdzen:2017}%
  \BibitemOpen
  \bibfield  {author} {\bibinfo {author} {\bibfnamefont {T.~J.}\ \bibnamefont {{Mozdzen}}}, \bibinfo {author} {\bibfnamefont {J.~D.}\ \bibnamefont {{Bowman}}}, \bibinfo {author} {\bibfnamefont {R.~A.}\ \bibnamefont {{Monsalve}}},\ and\ \bibinfo {author} {\bibfnamefont {A.~E.~E.}\ \bibnamefont {{Rogers}}},\ }\bibfield  {title} {\bibinfo {title} {{Improved measurement of the spectral index of the diffuse radio background between 90 and 190 MHz}},\ }\href {https://doi.org/10.1093/mnras/stw2696} {\bibfield  {journal} {\bibinfo  {journal} {Mon. Not. Roy. Astron. Soc.}\ }\textbf {\bibinfo {volume} {464}},\ \bibinfo {pages} {4995} (\bibinfo {year} {2017})},\ \Eprint {https://arxiv.org/abs/1609.08705} {arXiv:1609.08705 [astro-ph.IM]} \BibitemShut {NoStop}%
\bibitem [{\citenamefont {Fialkov}\ \emph {et~al.}(2017)\citenamefont {Fialkov}, \citenamefont {Cohen}, \citenamefont {Barkana},\ and\ \citenamefont {Silk}}]{Fialkov:2016zyq}%
  \BibitemOpen
  \bibfield  {author} {\bibinfo {author} {\bibfnamefont {A.}~\bibnamefont {Fialkov}}, \bibinfo {author} {\bibfnamefont {A.}~\bibnamefont {Cohen}}, \bibinfo {author} {\bibfnamefont {R.}~\bibnamefont {Barkana}},\ and\ \bibinfo {author} {\bibfnamefont {J.}~\bibnamefont {Silk}},\ }\bibfield  {title} {\bibinfo {title} {{Constraining the redshifted 21-cm signal with the unresolved soft X-ray background}},\ }\href {https://doi.org/10.1093/mnras/stw2540} {\bibfield  {journal} {\bibinfo  {journal} {Mon. Not. Roy. Astron. Soc.}\ }\textbf {\bibinfo {volume} {464}},\ \bibinfo {pages} {3498} (\bibinfo {year} {2017})},\ \Eprint {https://arxiv.org/abs/1602.07322} {arXiv:1602.07322 [astro-ph.CO]} \BibitemShut {NoStop}%
\bibitem [{\citenamefont {Cohen}\ \emph {et~al.}(2017)\citenamefont {Cohen}, \citenamefont {Fialkov}, \citenamefont {Barkana},\ and\ \citenamefont {Lotem}}]{Cohen:2016jbh}%
  \BibitemOpen
  \bibfield  {author} {\bibinfo {author} {\bibfnamefont {A.}~\bibnamefont {Cohen}}, \bibinfo {author} {\bibfnamefont {A.}~\bibnamefont {Fialkov}}, \bibinfo {author} {\bibfnamefont {R.}~\bibnamefont {Barkana}},\ and\ \bibinfo {author} {\bibfnamefont {M.}~\bibnamefont {Lotem}},\ }\bibfield  {title} {\bibinfo {title} {{Charting the parameter space of the global 21-cm signal}},\ }\href {https://doi.org/10.1093/mnras/stx2065} {\bibfield  {journal} {\bibinfo  {journal} {Mon. Not. Roy. Astron. Soc.}\ }\textbf {\bibinfo {volume} {472}},\ \bibinfo {pages} {1915} (\bibinfo {year} {2017})},\ \Eprint {https://arxiv.org/abs/1609.02312} {arXiv:1609.02312 [astro-ph.CO]} \BibitemShut {NoStop}%
\bibitem [{\citenamefont {Cohen}\ \emph {et~al.}(2018)\citenamefont {Cohen}, \citenamefont {Fialkov},\ and\ \citenamefont {Barkana}}]{Cohen:2017xpx}%
  \BibitemOpen
  \bibfield  {author} {\bibinfo {author} {\bibfnamefont {A.}~\bibnamefont {Cohen}}, \bibinfo {author} {\bibfnamefont {A.}~\bibnamefont {Fialkov}},\ and\ \bibinfo {author} {\bibfnamefont {R.}~\bibnamefont {Barkana}},\ }\bibfield  {title} {\bibinfo {title} {{Charting the Parameter Space of the 21-cm Power Spectrum}},\ }\href {https://doi.org/10.1093/mnras/sty1094} {\bibfield  {journal} {\bibinfo  {journal} {Mon. Not. Roy. Astron. Soc.}\ }\textbf {\bibinfo {volume} {478}},\ \bibinfo {pages} {2193} (\bibinfo {year} {2018})},\ \Eprint {https://arxiv.org/abs/1709.02122} {arXiv:1709.02122 [astro-ph.CO]} \BibitemShut {NoStop}%
\bibitem [{\citenamefont {Sims}\ \emph {et~al.}(2025)\citenamefont {Sims}, \citenamefont {Bevins}, \citenamefont {Fialkov}, \citenamefont {Anstey}, \citenamefont {Handley}, \citenamefont {Heimersheim}, \citenamefont {de~Lera~Acedo}, \citenamefont {Mondal},\ and\ \citenamefont {Barkana}}]{Sims:2025hfm}%
  \BibitemOpen
  \bibfield  {author} {\bibinfo {author} {\bibfnamefont {P.~H.}\ \bibnamefont {Sims}}, \bibinfo {author} {\bibfnamefont {H.~T.~J.}\ \bibnamefont {Bevins}}, \bibinfo {author} {\bibfnamefont {A.}~\bibnamefont {Fialkov}}, \bibinfo {author} {\bibfnamefont {D.}~\bibnamefont {Anstey}}, \bibinfo {author} {\bibfnamefont {W.~J.}\ \bibnamefont {Handley}}, \bibinfo {author} {\bibfnamefont {S.}~\bibnamefont {Heimersheim}}, \bibinfo {author} {\bibfnamefont {E.}~\bibnamefont {de~Lera~Acedo}}, \bibinfo {author} {\bibfnamefont {R.}~\bibnamefont {Mondal}},\ and\ \bibinfo {author} {\bibfnamefont {R.}~\bibnamefont {Barkana}},\ }\href {https://arxiv.org/abs/2504.09725} {\bibinfo {title} {Rapid and late cosmic reionization driven by massive galaxies: a joint analysis of constraints from 21-cm, lyman line \& cmb data sets}} (\bibinfo {year} {2025}),\ \Eprint {https://arxiv.org/abs/2504.09725} {arXiv:2504.09725 [astro-ph.CO]} \BibitemShut {NoStop}%
\bibitem [{\citenamefont {Decant}\ \emph {et~al.}(2025)\citenamefont {Decant}, \citenamefont {Dimitriou}, \citenamefont {Honorez},\ and\ \citenamefont {Zaldivar}}]{Decant:2024bpg}%
  \BibitemOpen
  \bibfield  {author} {\bibinfo {author} {\bibfnamefont {Q.}~\bibnamefont {Decant}}, \bibinfo {author} {\bibfnamefont {A.}~\bibnamefont {Dimitriou}}, \bibinfo {author} {\bibfnamefont {L.~L.}\ \bibnamefont {Honorez}},\ and\ \bibinfo {author} {\bibfnamefont {B.}~\bibnamefont {Zaldivar}},\ }\bibfield  {title} {\bibinfo {title} {{Simulation-based inference on warm dark matter from HERA forecasts}},\ }\href {https://doi.org/10.1088/1475-7516/2025/07/004} {\bibfield  {journal} {\bibinfo  {journal} {JCAP}\ }\textbf {\bibinfo {volume} {07}},\ \bibinfo {pages} {004}},\ \Eprint {https://arxiv.org/abs/2412.10310} {arXiv:2412.10310 [astro-ph.CO]} \BibitemShut {NoStop}%
\bibitem [{\citenamefont {{Mesinger}}\ \emph {et~al.}(2013)\citenamefont {{Mesinger}}, \citenamefont {{Ferrara}},\ and\ \citenamefont {{Spiegel}}}]{Mesinger:2013_Xray}%
  \BibitemOpen
  \bibfield  {author} {\bibinfo {author} {\bibfnamefont {A.}~\bibnamefont {{Mesinger}}}, \bibinfo {author} {\bibfnamefont {A.}~\bibnamefont {{Ferrara}}},\ and\ \bibinfo {author} {\bibfnamefont {D.~S.}\ \bibnamefont {{Spiegel}}},\ }\bibfield  {title} {\bibinfo {title} {{Signatures of X-rays in the early Universe}},\ }\href {https://doi.org/10.1093/mnras/stt198} {\bibfield  {journal} {\bibinfo  {journal} {\mnras}\ }\textbf {\bibinfo {volume} {431}},\ \bibinfo {pages} {621} (\bibinfo {year} {2013})},\ \Eprint {https://arxiv.org/abs/1210.7319} {arXiv:1210.7319 [astro-ph.CO]} \BibitemShut {NoStop}%
\bibitem [{\citenamefont {Qin}\ \emph {et~al.}(2020)\citenamefont {Qin}, \citenamefont {Mesinger}, \citenamefont {Park}, \citenamefont {Greig},\ and\ \citenamefont {Mu{\~n}oz}}]{Qin:2020xyh}%
  \BibitemOpen
  \bibfield  {author} {\bibinfo {author} {\bibfnamefont {Y.}~\bibnamefont {Qin}}, \bibinfo {author} {\bibfnamefont {A.}~\bibnamefont {Mesinger}}, \bibinfo {author} {\bibfnamefont {J.}~\bibnamefont {Park}}, \bibinfo {author} {\bibfnamefont {B.}~\bibnamefont {Greig}},\ and\ \bibinfo {author} {\bibfnamefont {J.~B.}\ \bibnamefont {Mu{\~n}oz}},\ }\bibfield  {title} {\bibinfo {title} {{A tale of two sites {\textendash} I. Inferring the properties of minihalo-hosted galaxies from current observations}},\ }\href {https://doi.org/10.1093/mnras/staa1131} {\bibfield  {journal} {\bibinfo  {journal} {Mon. Not. Roy. Astron. Soc.}\ }\textbf {\bibinfo {volume} {495}},\ \bibinfo {pages} {123} (\bibinfo {year} {2020})},\ \Eprint {https://arxiv.org/abs/2003.04442} {arXiv:2003.04442 [astro-ph.CO]} \BibitemShut {NoStop}%
\bibitem [{\citenamefont {Leitherer}\ \emph {et~al.}(1999)\citenamefont {Leitherer}, \citenamefont {Schaerer}, \citenamefont {Goldader}, \citenamefont {Gonzalez~Delgado}, \citenamefont {Robert}, \citenamefont {Kune}, \citenamefont {Mello}, \citenamefont {Devost},\ and\ \citenamefont {Heckman}}]{Leitherer:1999rq}%
  \BibitemOpen
  \bibfield  {author} {\bibinfo {author} {\bibfnamefont {C.}~\bibnamefont {Leitherer}}, \bibinfo {author} {\bibfnamefont {D.}~\bibnamefont {Schaerer}}, \bibinfo {author} {\bibfnamefont {J.~D.}\ \bibnamefont {Goldader}}, \bibinfo {author} {\bibfnamefont {R.~M.}\ \bibnamefont {Gonzalez~Delgado}}, \bibinfo {author} {\bibfnamefont {C.}~\bibnamefont {Robert}}, \bibinfo {author} {\bibfnamefont {D.~F.}\ \bibnamefont {Kune}}, \bibinfo {author} {\bibfnamefont {D.~F.~d.}\ \bibnamefont {Mello}}, \bibinfo {author} {\bibfnamefont {D.}~\bibnamefont {Devost}},\ and\ \bibinfo {author} {\bibfnamefont {T.~M.}\ \bibnamefont {Heckman}},\ }\bibfield  {title} {\bibinfo {title} {{Starburst99: synthesis models for galaxies with active star formation}},\ }\href {https://doi.org/10.1086/313233} {\bibfield  {journal} {\bibinfo  {journal} {Astrophys. J. Suppl.}\ }\textbf {\bibinfo {volume} {123}},\ \bibinfo {pages} {3} (\bibinfo {year} {1999})},\ \Eprint {https://arxiv.org/abs/astro-ph/9902334} {arXiv:astro-ph/9902334} \BibitemShut
  {NoStop}%
\bibitem [{\citenamefont {Bromm}\ \emph {et~al.}(2001)\citenamefont {Bromm}, \citenamefont {Kudritzki},\ and\ \citenamefont {Loeb}}]{Bromm:2000nn}%
  \BibitemOpen
  \bibfield  {author} {\bibinfo {author} {\bibfnamefont {V.}~\bibnamefont {Bromm}}, \bibinfo {author} {\bibfnamefont {R.~P.}\ \bibnamefont {Kudritzki}},\ and\ \bibinfo {author} {\bibfnamefont {A.}~\bibnamefont {Loeb}},\ }\bibfield  {title} {\bibinfo {title} {{Generic spectrum and ionization efficiency of a heavy initial mass function for the first stars}},\ }\href {https://doi.org/10.1086/320549} {\bibfield  {journal} {\bibinfo  {journal} {Astrophys. J.}\ }\textbf {\bibinfo {volume} {552}},\ \bibinfo {pages} {464} (\bibinfo {year} {2001})},\ \Eprint {https://arxiv.org/abs/astro-ph/0007248} {arXiv:astro-ph/0007248} \BibitemShut {NoStop}%
\bibitem [{\citenamefont {Tumlinson}\ and\ \citenamefont {Shull}(2000)}]{Tumlinson:1999iu}%
  \BibitemOpen
  \bibfield  {author} {\bibinfo {author} {\bibfnamefont {J.}~\bibnamefont {Tumlinson}}\ and\ \bibinfo {author} {\bibfnamefont {J.~M.}\ \bibnamefont {Shull}},\ }\bibfield  {title} {\bibinfo {title} {{Zero-metallicity stars and the effects of the first stars on reionization}},\ }\href {https://doi.org/10.1086/312432} {\bibfield  {journal} {\bibinfo  {journal} {Astrophys. J. Lett.}\ }\textbf {\bibinfo {volume} {528}},\ \bibinfo {pages} {L65} (\bibinfo {year} {2000})},\ \Eprint {https://arxiv.org/abs/astro-ph/9911339} {arXiv:astro-ph/9911339} \BibitemShut {NoStop}%
\end{thebibliography}%

\vspace{0.5cm}
\textit{Appendix A: 21-cm analysis details}---Fluctuations in the 21-cm signal defined in Eq.~\eqref{eqn:T21} are characterized to first order with the 21-cm power spectrum $P_{21}(\mathbf{k}, z)$, given by
\begin{equation}
    \langle \delta T_{21}(\mathbf{k}, z) \delta T_{21}^*(\mathbf{k}', z) \rangle = (2\pi)^3 \delta_D(\mathbf{k} - \mathbf{k}') P_{21}(\mathbf{k}, z),
\end{equation}
where $\delta T_{21}(\mathbf{k}, z)$ corresponds to the Fourier transform of $T_{21}(\mathbf{x}, z) - \overline{T}_{21}(\mathbf{x}, z)$. Throughout this work we use the reduced power spectrum $ \Delta_{21}^2(\mathbf{k}, z) = {k^3 P_{21}(\mathbf{k}, z)}/{2\pi^2} \, $, with dimension $\mathrm{mK}^2$.

We use this power spectrum as our observable to provide 21-cm forecasts via a Fisher forecast.  We follow the methodology outlined in Ref.~\cite{Facchinetti:2023slb}, with elements of the Fisher Matrix given by 
		\begin{eqnarray}\label{eq:Fisher_elements}
			F_{ij} = \sum_{i_k, \, i_z}^{N_k, \, N_z} \frac{1}{\sigma^2(k_{i_k}, z_{i_z})}   \frac{\Delta_{21}^2(k_{i_k}, z_{i_z} )}{\partial \theta_i} \frac{\Delta_{21}^2(k_{i_k}, z_{i_z} )}{\partial \theta_j},
		\end{eqnarray}
		where $\sigma^2$ is the one-sigma error due to measurements in the power spectrum. The sums over $i_k$ and $i_z$ run between $k \simeq 0.1\textup{--}1 \, {\rm Mpc}^{-1}$, and redshifts $z \simeq  5\textup{--}28$, with binning consistent with that used in~\cite{Mason:2022obt}. We remark that the redshift binning corresponds to the frequency range $50 - 234 $~MHz, with binning using an 8 MHz bandwidth, where we cut off at a frequency corresponding to $z \simeq 5$. We compute the partial derivatives of the power spectrum with respect to each parameter numerically, using a central difference derivative for our astrophysical parameters, following exactly the variations provided in Ref.~\cite{Mason:2022obt}, and a forward difference derivative for the decay rate. For each mass $m_\chi$ we run several simulation boxes with different decay rates $\Gamma = \tau^{-1}$, computing the power spectrum and the forward difference derivative with respect to $\Gamma$ for each, and choose a value of $\Gamma$ for which the power spectrum derivative is numerically stable.
   \begin{figure*}[t]
    \includegraphics[width=2\columnwidth]{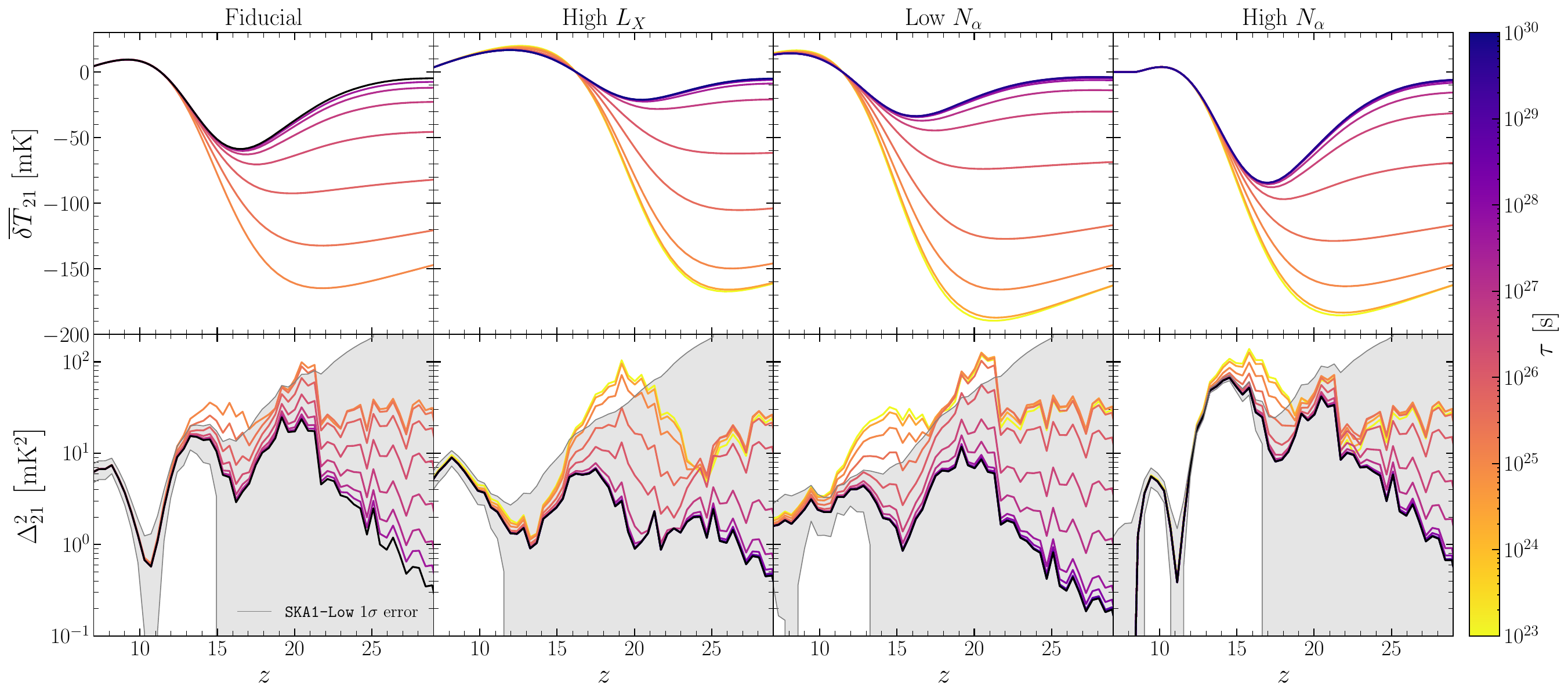}
    \caption{ \textbf{Comparison of the impact of exotic Lyman-$\boldsymbol{\alpha}$ on the 21-cm observables for different astrophysics scenarios.} We show the Lyman-$\alpha$--induced shift to the power spectrum for a decaying dark matter particle with mass $m_\chi = 20.4$~eV, at wavenumber $k \simeq 0.2$~Mpc$^{-1}$, as in Fig.~\ref{fig:decaying_dm_global_PS}. The different astrophysical scenarios are defined in Table~\ref{tab:fiducial_scenarios}.}
    \label{fig:panel_plot}
\end{figure*}	     
        For $\sigma^2$, we account for three contributions 
		\begin{equation}\label{eq:sigmas}
			\sigma^2 = \sigma_{\rm exp}^2 + \sigma_{\rm c.v.}^2 + 0.2 \Delta_{21}^2 ,
		\end{equation}
		where the first term is the experimental thermal error, computed for the {\tt SKA} using {\tt 21cmSense}~\cite{Pober:2012zz,Pober:2013jna,Murray:2024the}, for each fiducial astrophysical model (see \textit{Appendix B}). The second term is the cosmic variance, and the third term includes a 20\% theory uncertainty following~\cite{Park:2018ljd}. For our fiducial astrophysical model, we use the parameters from~\cite{Sun:2023acy}. We compute experimental sensitivities following the specifications for the {\tt SKA1-Low}~\cite{SKA_MEMO:2025}  and {\tt HERA}~\cite{HERA:2016} telescopes, and assuming the moderate foreground scenario considered in~\cite{Mason:2022obt}. We follow  explicitly the implementation in the {\tt SKA\_forecast} notebook provided in {\tt 21cmSense}~\cite{Pober:2012zz,Pober:2013jna,Murray:2024the}. For  the {\tt SKA1-Low} sensitivity, we use a deep survey, where we set the number of tracking hours to be the same as the number of observation hours per day, whereas for {\tt HERA}, we perform a drift-scan where the number of tracking hours is equal to the beam crossing time. For both experiments, we assume a fixed bandwidth of 8 MHz, and an observation time of 1080 hrs (6 hrs per day for 180 days). We model the system temperature equivalently to~\cite{Agius:2025xbj,Facchinetti:2023slb}, accounting for a receiver temperature of $100 {\ \rm K} $, and a sky temperature consistent with the measured diffuse radio background at $\sim 100$~MHz~\cite{Mozdzen:2017}. 

\textit{Appendix B: 21-cm constraining power with different astrophysical scenarios}---The goal of this appendix is to test the robustness of our bounds by varying the astrophysical assumptions that we expect to most strongly affect our derived constraints. We therefore study scenarios with high or low Lyman-$\alpha$ fluxes and enhanced X-ray heating (see Table~\ref{tab:fiducial_scenarios}).

The 21-cm signal during the cosmic dawn depends sensitively on uncertain astrophysical parameters~\cite{Fialkov:2016zyq, Cohen:2016jbh, Cohen:2017xpx, Katz:2024ayw, Katz:2025sie, Sims:2025hfm, Dhandha:2025dtn}. These uncertainties propagate into forecasts for exotic energy injection, particularly when the constraint is derived from additional heating and ionization of the IGM from dark matter~\cite{Lopez-Honorez:2016sur,Decant:2024bpg,Agius:2025xbj}. Fig.~\ref{fig:decaying_dm_global_PS} shows that heating and ionization suppresses the signal with respect to the fiducial case, while additional Lyman-$\alpha$ coupling enhances it. We further expand on this effect in Fig.~\ref{fig:panel_plot}, where we show the impact of the decaying dark matter of mass $m_{\chi} = 20.4$~eV on the global signal and power spectrum for different astrophysics scenarios chosen to illustrate the strongest expected effects.       
Two of these different scenarios were chosen to  correspond to the scenarios that were pointed out in Ref.~\cite{Agius:2025xbj} to be particularly unconstraining for heating from exotic energy injection, and the third to provide the highest expected degeneracy with Lyman-$\alpha$ photons from DM. A typically less-constraining scenario would be one with low Lyman-$\alpha$ and high X-ray fluxes from the first stars, since WF coupling would be inefficient and additional heating from the DM would be less distinguishable from astrophysical heating. It was shown that in such a scenario, the constraining power for accreting primordial black holes (whose main imprint on the 21-cm signal is through additional heating) essentially vanishes \cite{Agius:2025xbj}. The third scenario we choose is one with a high Lyman-$\alpha$ flux, which is typically highly constraining for heating scenarios, since the spin temperature closely tracks the gas temperature. We omit a low $L_X$ scenario, but note that this case would further strengthen our bounds. Furthermore, we remark that our high $L_X$ scenario is close to the limit of $L_{X} / \dot{M}_* < 10^{42}~\mathrm{erg~s^{-1}~M_\odot^{-1}~yr},
$ set by early X-ray induced reionization~\cite{HERA:2022wmy,Mesinger:2013_Xray}.

We characterize these different models explicitly by modifying in each case the number of ionizing photons emitted
per stellar baryon, $N_{\rm ion}$, or the X-ray luminosity, $L_X$, normalized to the star formation rate, $\dot{M}_{*}$. Details on the implementation of these parameters in \texttt{21cmFAST} can be found in~\cite{Park:2018ljd,Qin:2020xyh}. Importantly, any change to $N_{\rm ion}$ alters the normalization of the stellar ultraviolet spectrum, thus self-consistently changing the number of Lyman-$\alpha$ photons per baryon, $N_\alpha$.  We consider two stellar populations of atomically cooling galaxies (ACGs) and molecularly cooled galaxies (MCGs), where we take different values of $N_\alpha$ for each, due to the different metallicities, recent star forming histories and initial mass functions~\cite{Leitherer:1999rq,Bromm:2000nn,Tumlinson:1999iu}, but keep $L_X$ fixed to the same value for both populations, following~\cite{Munoz:2021psm}. The values of these parameters are given in Table~\ref{tab:fiducial_scenarios}, where we explicitly write $N_{\rm ion}$, the input parameter for {\tt 21cmFAST}, and the computed value of $N_\alpha$ from the resulting re-normalized spectrum. For the purpose of this study, $L_X / \dot{M}_*$ can be understood as the normalization regulating the X-ray heating, and $N_\alpha$ that governing the WF coupling from the first stars. 

\begin{table}[h]
\centering
\renewcommand{\arraystretch}{1.2}
\setlength{\tabcolsep}{4.2pt}
\begin{tabular}{lccccc}
\hline
Scenario & $L_X / \dot{M}_*$ & $N_{\rm ion}^{\mathrm{ACG}}$ & $N_{\rm Ly\alpha}^{\mathrm{ACG}}$ & $N_{\rm ion}^{\mathrm{MCG}}$ & $N_{\rm Ly\alpha}^{\mathrm{MCG}}$ \\
\hline
Fiducial        & $10^{40.5}$ & $5{\times}10^{3}$  & $1.1{\times}10^{4}$ & $4.4{\times}10^{4}$ & $4.8{\times}10^{3}$ \\
High $L_X$      & $10^{41.5}$ & $5{\times}10^{3}$  & $1.1{\times}10^{4}$ & $4.4{\times}10^{4}$ & $4.8{\times}10^{3}$ \\
Low $N_\alpha$  & $10^{40.5}$ & $1{\times}10^{3}$  & $2.2{\times}10^{3}$ & $2{\times}10^{4}$   & $2.1{\times}10^{3}$ \\
High $N_\alpha$ & $10^{40.5}$ & $2{\times}10^{4}$  & $4.4{\times}10^{4}$ & $8{\times}10^{4}$   & $8.7{\times}10^{3}$ \\
\hline
\end{tabular}
\caption{\textbf{Astrophysical scenarios and associated parameters.}
The X-ray luminosity per star formation rate, $L_X / \dot{M}_*$, is given in
erg\,s$^{-1}$\,M$_\odot^{-1}$\,yr$^{-1}$.
$N_{\rm ion}$ and $N_{\rm Ly\alpha}$ denote the number of ionizing and Lyman-$\alpha$ photons per baryon
for atomic-cooling (ACG) and molecular-cooling (MCG) galaxies, respectively.}
\label{tab:fiducial_scenarios}
\end{table}

As shown in Fig.~\ref{fig:panel_plot}, in all three  astrophysical scenarios, the effect Lyman-$\alpha$ photon injection is distinguishable from the baseline model. This is particularly relevant for both the high $L_X$ and low $N_\alpha$ scenarios, where the suppression from a typical heating signal is difficult to distinguish from the baseline, whose power spectrum is already suppressed. Moreover, in the third scenario, we show that despite extreme Lyman-$\alpha$ brightness originating from stellar sources, the DM injection is also distinguishable from the baseline, particularly at higher redshifts. 

We quantify the impact of different fiducial astrophysical scenarios, by performing a Fisher analysis for each scenario, and computing the 95th percentile limits on the lifetime for a DM particle of mass $m_{\chi} = 20.4$~eV. The resulting limits are shown in Table~\ref{tab:new_limits}, where it is apparent that none of the extreme astrophysical scenarios considered alter the constraint by more than a factor of $\sim 2$. 
In particular, the constraint from a low $N_\alpha$ scenario is remarkably robust compared to the fiducial, and a high $L_X$ scenario only weakens the constraint by a factor of $\sim 2$. The strongest constraint arises from the high $N_\alpha$ scenario, where this stems from this scenario having the largest baseline power spectrum, so experiments are more sensitive to measuring deviations from this baseline. This is in contrast to the orders of magnitude differences in constraining power reported for heating constraints under varying astrophysical models~\cite{Agius:2025xbj}.

\begin{table}[h!]
\centering
\renewcommand{\arraystretch}{1.2}
\setlength{\tabcolsep}{10pt}
\begin{tabular}{lc}
\hline
Scenario & $\tau_{95}~[\mathrm{s}]$ \\
\hline
Fiducial          & $1.14{\times}10^{26}$ \\
High $L_X$       & $5.15{\times}10^{25}$ \\
High $N_{\alpha}$& $2.17{\times}10^{26}$ \\
Low $N_{\alpha}$ & $1.07{\times}10^{26}$ \\
\hline
\end{tabular}
\caption{Derived 95th percentile limits on the lifetime $\tau_{95}$ for the four astrophysical scenarios in Table~\ref{tab:fiducial_scenarios}.}
\label{tab:new_limits}
\end{table}

\clearpage
\onecolumngrid
\begin{center}
\textbf{\large Boosting the cosmic 21-cm signal with exotic Lyman-$\alpha$ from dark matter} \\
\textit{\large Supplementary Material} \\ 
Dominic Agius, Tracy R. Slatyer
\end{center}
\setcounter{section}{0}
\setcounter{figure}{0}
\setcounter{table}{0}
\renewcommand{\thefigure}{S\arabic{figure}}
\renewcommand{\thetable}{S\arabic{table}}
\renewcommand{\thesection}{S\arabic{section}}
\renewcommand{\theHfigure}{S\arabic{figure}} 
\makeatother
\renewcommand{\theHtable}{S\arabic{table}}   
\makeatother

\begin{figure*}[h]
    \includegraphics[width=0.9\columnwidth]{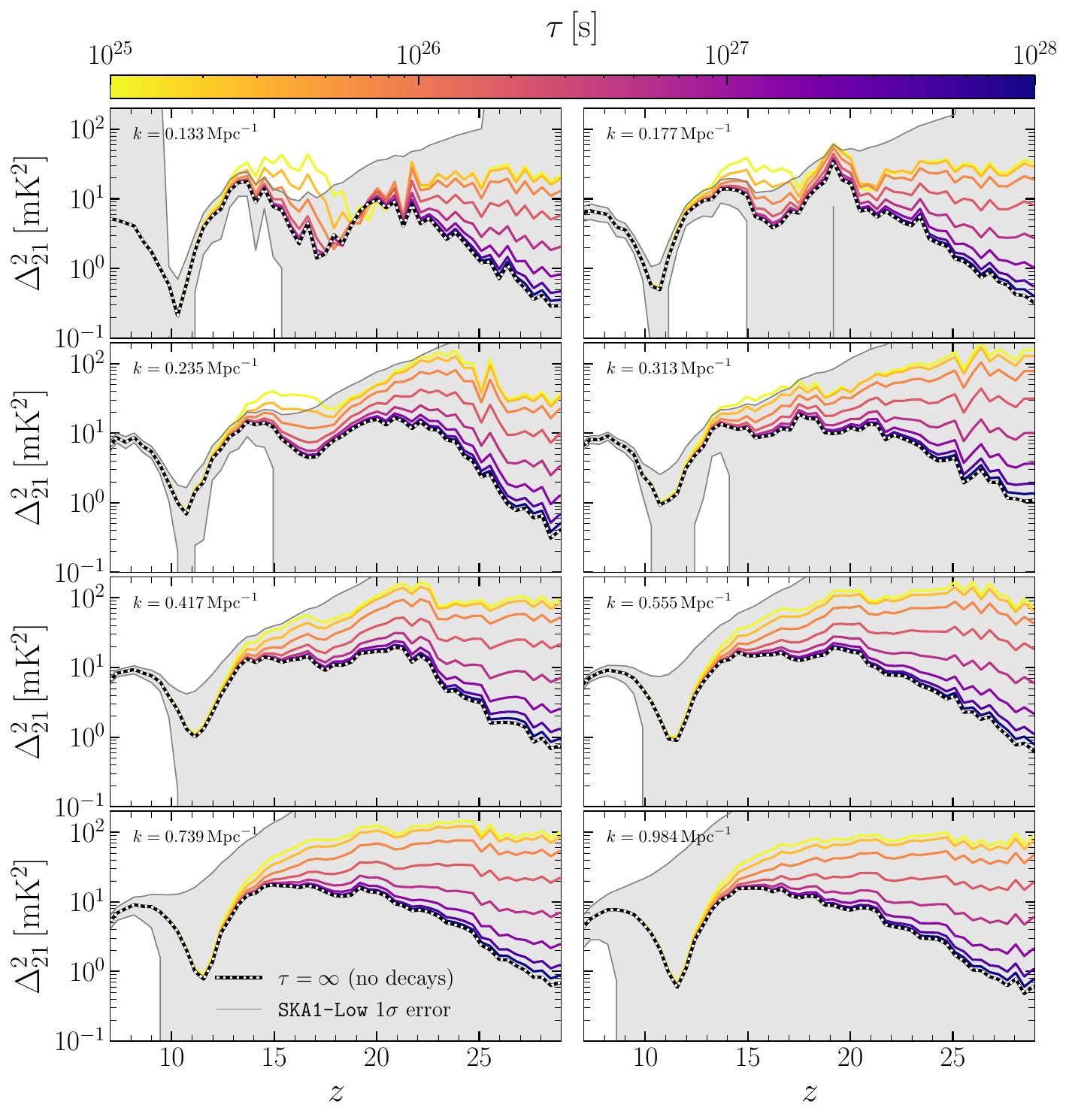}
    \caption{ \textbf{Comparison of the 21-cm power spectrum at different scales.} We show the Lyman-$\alpha$--induced shift to the power spectrum for a decaying dark matter particle with mass $m_\chi = 20.4$~eV, as in Fig~\ref{fig:decaying_dm_global_PS}, but for different wavenumbers, $k$. All panels include the 1$\sigma$ experimental thermal error computed for {\tt SKA1-Low}, and the fiducial scenario without decays. }
    \label{fig:panel_plot_different_k}
\end{figure*}

\begin{figure*}[b]
    \includegraphics[width=1\columnwidth]{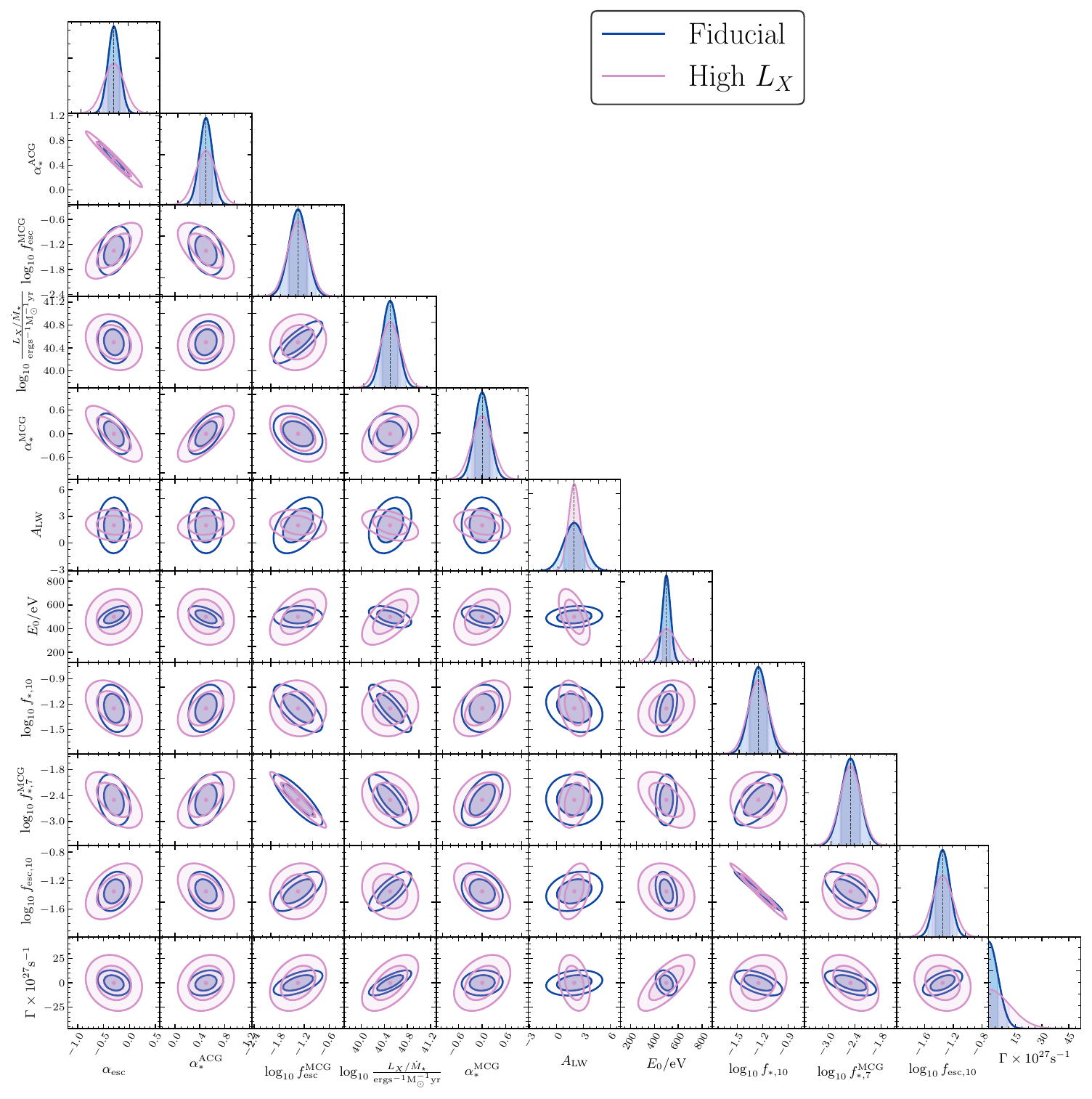}
    \caption{\textbf{Triangle plot comparing the covariance between the decay rate and the astrophysical parameters for two different X-ray heating scenarios.}  Results are shown for a decaying DM particle of mass $m_\chi = 20.4~{\rm eV}$, decaying via $\chi \to \gamma \gamma$, using the projected sensitivity of {\tt SKA1-Low}. We compare the fiducial scenario to a scenario with enhanced X-ray heating from the first stars (see Table~\ref{tab:fiducial_scenarios}).}
    \label{fig:Triangle_L_X}
\end{figure*}

\begin{figure*}[b]
    \includegraphics[width=1\columnwidth]{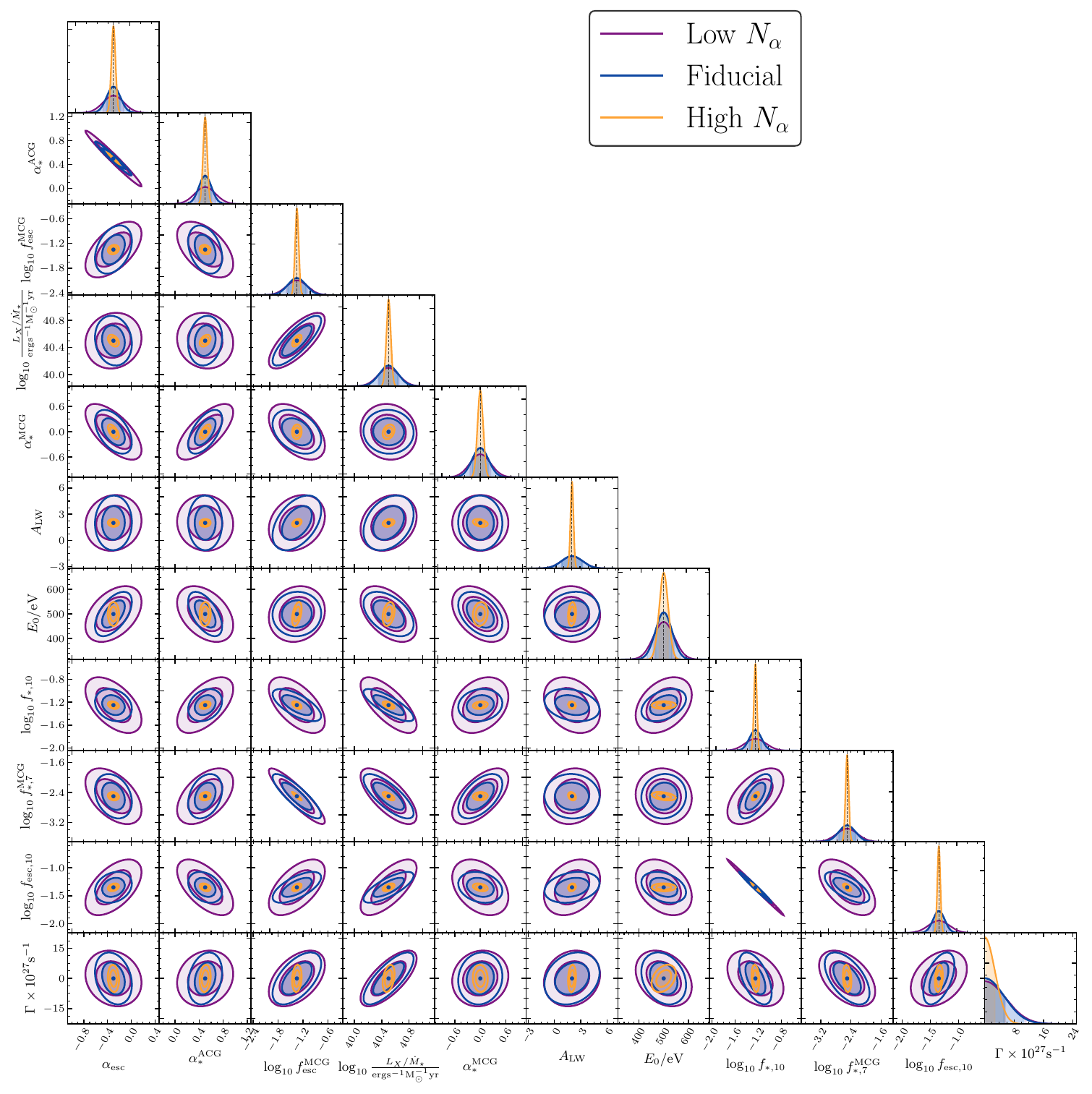}
    \caption{\textbf{Triangle plot comparing the covariance between the decay rate and the astrophysical parameters for three Lyman-$\boldsymbol{\alpha}$ scenarios.} Results are shown for a decaying DM particle of mass $m_\chi = 20.4~{\rm eV}$, decaying via $\chi \to \gamma \gamma$, using the projected sensitivity of {\tt SKA1-Low}. We compare the fiducial scenario to two alternative cases with enhanced and reduced Lyman-$\alpha$ photon production by the first stars (see Table~\ref{tab:fiducial_scenarios}).}
    \label{fig:Triangle_Na}
\end{figure*}

In this supplementary material we provide additional details on the energy deposition treatment, show the behavior of the power spectrum at different scales in Fig.~\ref{fig:panel_plot_different_k}, and provide full triangle plots  showing the correlations between the astrophysical parameters and the decay rate $\Gamma$, for several fiducial astrophysical models in Figs.~\ref{fig:Triangle_plot}, ~\ref{fig:Triangle_L_X}, ~\ref{fig:Triangle_Na}. 

\section{Corrected behavior compared to {\tt DarkHistory}}
In Fig.~\ref{fig:energy_deposition}, we showed the deposition of energy by channel, $f_c$, for decaying dark matter of masses 19--50 eV, at $z = 35$, computed using {\tt DarkHistory}~\cite{Liu:2019bbm}, and with lifetime $\tau = 10^{28}$~s. We remarked in the text, that there are two regimes, shown by the gray hashed regions, where the energy deposition treatment in \texttt{DM21cm} is inaccurate due to binning/discretization issues. This inaccuracy occurs because the spectrum of injected photons is mapped onto a (somewhat coarse) grid of photon energies in such a way as to preserve the number and total energy of the injected photons, which generically involves smearing the photons across neighboring energy bins. Where the functions describing how photons' energy is  partitioned between heating, ionization, and Lyman-$\alpha$ have sharp thresholds at specific energies, this smearing tends to smear out those thresholds.

However, the underlying treatment in \texttt{DarkHistory v1.0} \cite{Liu:2019bbm} is rather simple. Photons with energies modestly exceeding $13.6$ eV are assumed to photoionize hydrogen, contributing 13.6 eV of energy to the ``H ion'' channel, with the remaining energy being attributed to the kinetic energy of a secondary electron. In our mass range $m_\chi < 50$ eV, the kinetic energy is dissipated solely into heat as the electron thermalizes through Compton scattering. Consequently, for this mass range, the approximations made in \texttt{DarkHistory v1.0} (and inherited by \texttt{DM21cm}) give rise to the following behavior for the fractions of energy deposited into the various channels:
\begin{align}
f_{\rm Ly\alpha}(m_\chi) &=
\begin{cases}
1, & m_\chi \in [20.4,\,27.2]~\mathrm{eV},\\[3pt]
0, & \text{otherwise},
\end{cases} \\[8pt] \label{eq:dhapprox_lya}
f_{\rm Heat}(m_\chi) &=
\begin{cases}
0, & m_\chi \le 27.2~\mathrm{eV},\\[3pt]
1 - \frac{27.2}{m_\chi}, & m_\chi > 27.2~\mathrm{eV},
\end{cases} \\[8pt]
f_{\rm H\,ion}(m_\chi) &=
\begin{cases}
0, & m_\chi \leq 27.2~\mathrm{eV},\\[3pt]
\frac{27.2}{m_\chi} , & m_\chi > 27.2~\mathrm{eV},
\end{cases} \\[8pt]
f_{\rm continuum}(m_\chi) &=
\begin{cases}
1, & m_\chi < 20.4~\mathrm{eV},\\[3pt]
0, & \text{otherwise.}
\end{cases}\label{eq:dhapprox_cont}
\end{align}
The power into helium ionization is taken to be zero for the masses of interest, since photoionization of neutral helium requires a 24.6 eV photon, and so can be neglected for $m_\chi \lesssim 50$ eV. At the top end of this mass range, where  the secondary electron has kinetic energy exceeding 10.2 eV, it may in principle excite hydrogen atoms and thus contribute to the Lyman-$\alpha$ flux, but this only occurs for an initial photon energy exceeding 23.8 eV, and thus can be ignored for $m_\chi \lesssim 48$ eV. Furthermore, in the treatment of \texttt{DarkHistory v1.0}, collisional excitation/ionization by secondary electrons is only captured in detail for electron kinetic energies $\ge 14$ eV, corresponding to $m_\chi > 55$ eV. In the 10.2--13.6 eV range for electron kinetic energy, at the very top of our mass range, \texttt{DarkHistory v1.0} adopts an approximate treatment that in principle can give rise to a non-zero excitation contribution and modestly suppressed heating; however, we do not observe any significant change in the \texttt{DM21cm} results in this mass region, compared to slightly lower masses.

We overplot the curves given in Eqs.~(\ref{eq:dhapprox_lya}--\ref{eq:dhapprox_cont}), compared with the binned results from \texttt{DarkHistory}, in Fig.~\ref{fig:energy_deposition} in the main text. We see that this simple approximation agrees well with the numerical results from \texttt{DarkHistory} and resolves the unphysical thresholds associated with the bin edges. 

When computing our bounds for $20.4$ eV $< m_\chi < 27.2$ eV, we thus use the signal  computed with \texttt{DM21cm} for energies sufficiently far from the thresholds that $f_{\text{Ly}\alpha}=1$ in the \texttt{DarkHistory} calculation, and then modify the inferred limit according to the redshifting and $f_\text{recycle}$ factors as discussed in the main text (to account for the efficiency for which photons injected into the Lyman-series band actually convert to Lyman-$\alpha$, which is not captured in $f_{\text{Ly}\alpha}$). For $m_\chi > 27.2$ eV, we use \texttt{DM21cm} results for $m_\chi \gtrsim 31$ eV, and extrapolate the results into the $\sim27.2\text{--}31$ eV mass range where the default \texttt{DarkHistory} calculation underestimates the ionization contribution and overestimates the Lyman-$\alpha$ contribution (which should be zero) due to discretization/binning effects.

\section{Power spectrum at different scales}

\section{Triangle Plots}\label{sec:triangle}
\begin{figure*}[b]
    \includegraphics[width=1\columnwidth]{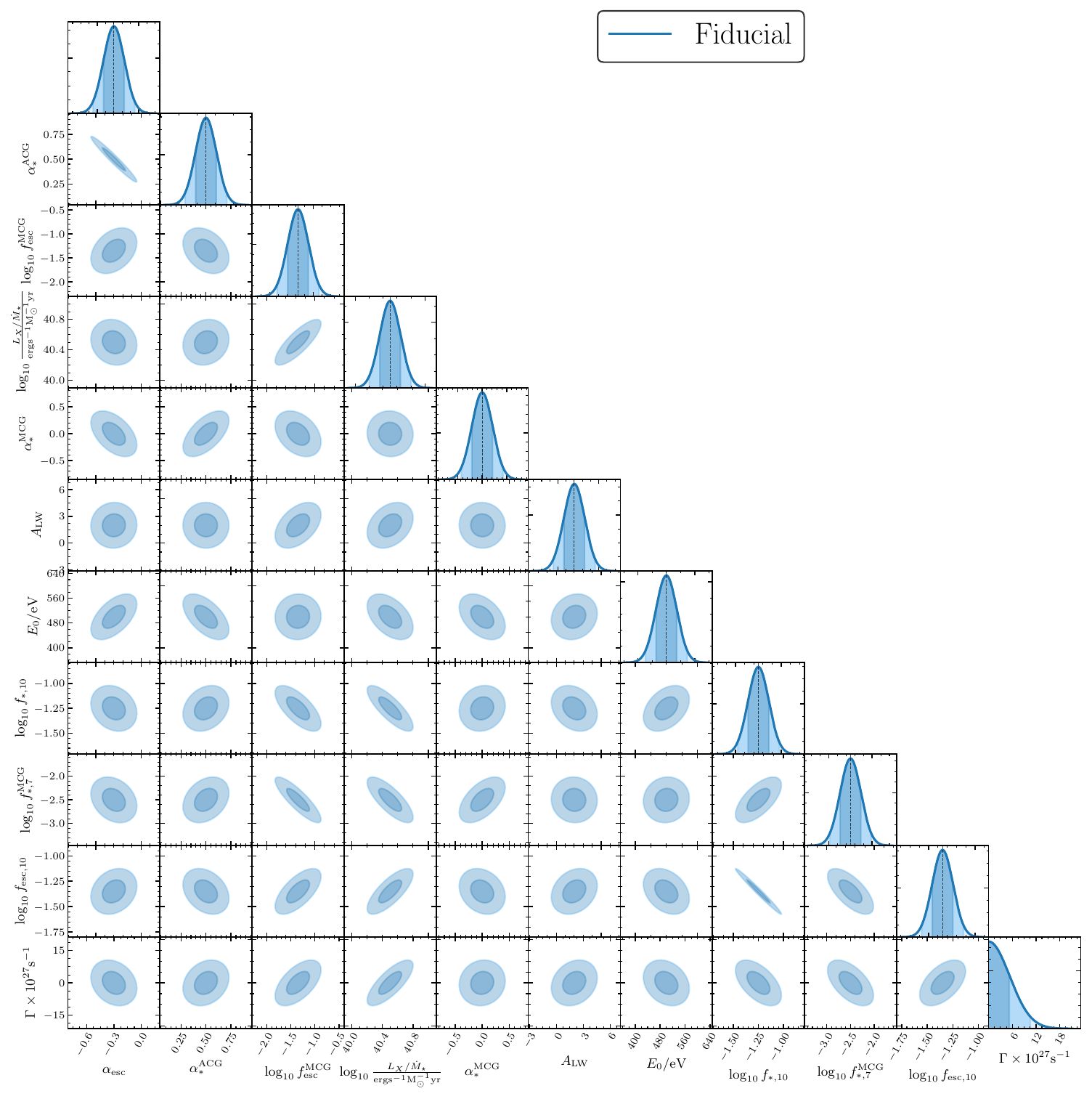}
    \caption{\textbf{Triangle plot showing the covariance between the decay rate and the astrophysical parameters.}  Results are shown for a decaying DM particle of mass $m_\chi = 20.4~{\rm eV}$, decaying via $\chi \to \gamma \gamma$, using the projected sensitivity of {\tt SKA1-Low}, with the fiducial astrophysical model.}
    \label{fig:Triangle_plot}
\end{figure*}

\end{document}